\documentclass[journal]{IEEEtran}

\usepackage [latin1]{inputenc}
\usepackage{mathrsfs}  
\usepackage{graphicx}
\usepackage{epstopdf}
\usepackage{subfloat}
\usepackage{float}
\usepackage{amsmath}
\usepackage{CJK}
\usepackage{amssymb}
\usepackage{indentfirst}
\usepackage{mathtools}
\usepackage{subfigure}
\usepackage{stfloats}
\usepackage{mathrsfs}
\usepackage{bm}
\usepackage{blindtext}
\usepackage{algorithm}
\usepackage{algpseudocode}
\usepackage{amsmath}
\usepackage{cite}
\usepackage{colortbl}
\definecolor{mygray}{gray}{.9}
\usepackage[table]{xcolor}
\usepackage{booktabs}
\usepackage{enumerate}

\usepackage{amsmath,amssymb}

\usepackage[Symbol]{upgreek}
\usepackage{booktabs}
\usepackage{multirow}

\usepackage{array}

\ifCLASSINFOpdf

\else

\fi

\newcolumntype{C}[1]{>{\centering\let\newline\\\arraybackslash\hspace{0pt}}m{#1}}

\begin{document}
\title{Reduced Reference Perceptual Quality Model and Application to Rate Control for 3D Point Cloud Compression}

\author{Qi~Liu, Hui~Yuan,~\IEEEmembership{Senior Member,~IEEE,}
        Raouf~Hamzaoui,~\IEEEmembership{Senior Member,~IEEE,}\\
        Honglei~Su, Junhui~Hou,~\IEEEmembership{Senior Member,~IEEE,} and Huan~Yang,~\IEEEmembership{Member,~IEEE}\\
\thanks{This work was supported in part by the National Natural Science Foundation of China under Grant 61871342; in part by the open project program of state key laboratory of virtual reality technology and systems, Beihang University, under Grant VRLAB2019B03; in part by the Hong Kong Research Grants Council under Grants 9042955 (CityU 11202320), and in part by Shandong Provincial Natural Science Foundation, China under Grants ZR2018PF002. \textit{Corresponding author: Hui Yuan.}}
\thanks{Q. Liu is with the School of Information Science and Engineering, Shandong University, Qingdao 266237, China. (Email: sdqi.liu@gmail.com).}
\thanks{H. Yuan is with the School of Control Science and Engineering, Shandong University, Ji'nan 250061, China. (Email: huiyuan@sdu.edu.cn).}
\thanks{R. Hamzaoui is with the School of Engineering and Sustainable Development, De Montfort University, Leicester, UK. (Email: rhamzaoui@dmu.ac.uk).}
\thanks{H. Su is with the School of Electronic Information, Qingdao University, Qingdao 266071, China. (Email: suhonglei@qdu.edu.cn).}
\thanks{J. Hou is with the Department of Computer Science, City University of Hong Kong, Kowloon, Hong Kong. (Email: jh.hou@cityu.edu.hk).}
\thanks{H. Yang is with the College of Computer Science and Technology, Qingdao University, Qingdao 266071, China. (Email: cathy\_huanyang@hotmail.com).}
}

\markboth{Submitted to IEEE Transactions on Image Processing}%
{Shell \MakeLowercase{\textit{et al.}}: Bare Demo of IEEEtran.cls
for Journals}

\maketitle

\begin{abstract}

In rate-distortion optimization, the encoder settings are determined
by maximizing a reconstruction quality measure subject to a
constraint on the bit rate. One of the main challenges of this
approach is to define a quality measure that can be computed with
low computational cost and which correlates well with perceptual
quality. While several quality measures that fulfil these two
criteria have been developed for images and video, no such one
exists for 3D point clouds. We address this limitation for the
video-based point cloud compression (V-PCC) standard by proposing a
linear perceptual quality model whose variables are the V-PCC
geometry and color quantization parameters and whose coefficients
can easily be computed from two features extracted from the original
3D point cloud. Subjective quality tests with 400 compressed 3D
point clouds show that the proposed model correlates well with the
mean opinion score, outperforming state-of-the-art full reference
objective measures in terms of Spearman rank-order and Pearson¡¯s
linear correlation coefficient. Moreover, we show that for the same
target bit rate, rate-distortion optimization based on the proposed
model offers higher perceptual quality than rate-distortion
optimization based on exhaustive search with a point-to-point
objective quality metric.

\end{abstract}


\IEEEpeerreviewmaketitle

%
\section{Introduction}
\IEEEPARstart{W}{ith} the rapid development of 3D data acquisition
technologies, point clouds are now readily available and popular. A
3D point cloud (3DPC) comprises a set of points with geometric
coordinates and associated attributes, such as color, reflectance,
normal vectors, and so on. These points can be stored, transmitted,
and rendered in a variety of ways. There are already many 3DPC
applications in the fields of virtual reality, immersive
communication, architecture, and automatic driving,
etc.~\cite{gu20193d}.
\begin{figure*}[t!]
  \centering
  \includegraphics[width=2\columnwidth]{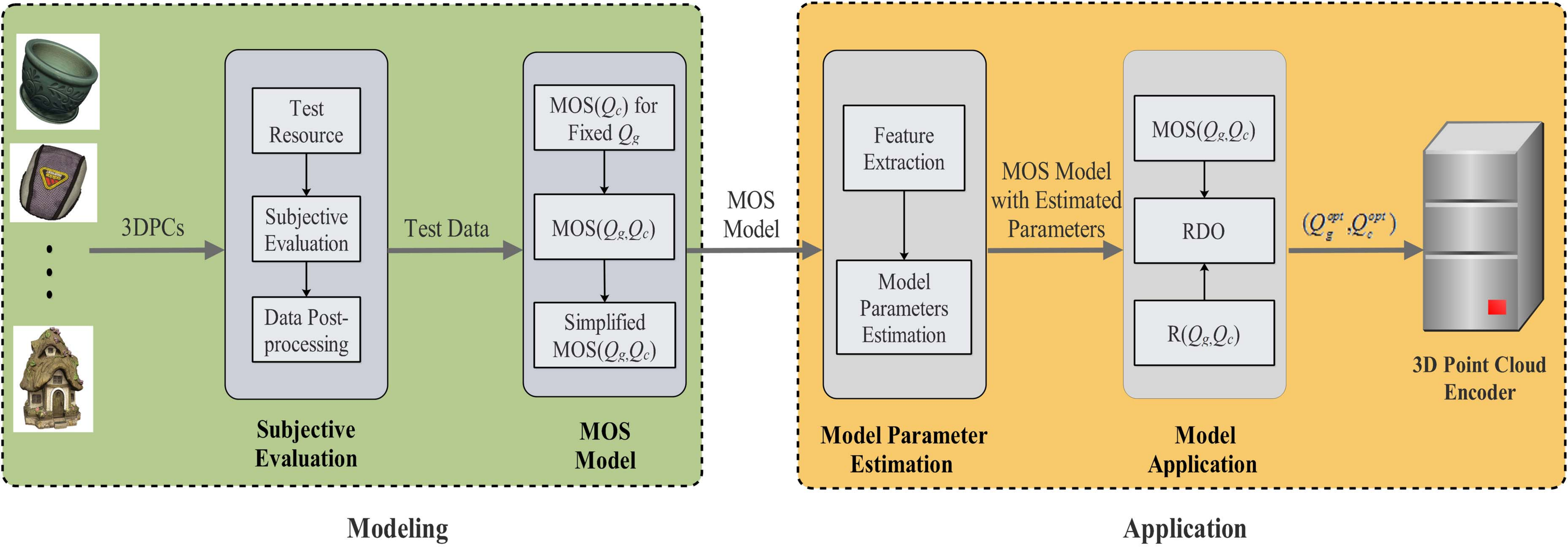}
  \caption{Compressed 3DPC perceptual quality modeling and its
  application. $Q_g$ and $Q_c$ denote the geometry quantization step and color
quantization step, respectively, $(Q^{opt}_g,Q^{opt}_c)$ is the
optimal geometry and color quantization step pair, $R(\cdot)$ and
$MOS(\cdot)$ are the rate and MOS functions, respectively.}
  \label{framework}
\end{figure*}
3DPCs can be classified into objects and scenes. Each class can
consist of static or dynamic 3DPCs. In this paper, we mainly focus
on static 3DPC objects~\cite{8786879}.

To represent the surface of an object with high fidelity, a 3DPC
usually contains millions, even billions of points, which results in
a large amount of data that needs to be efficiently stored and
transmitted~\cite{gu20203d}. Recently, the Moving Picture Experts
Group (MPEG) standardized two compression platforms for Point Cloud
Compression (PCC): Geometry-based Point Cloud Compression
(G-PCC)~\cite{gpcc} and Video-based Point Cloud Compression
(V-PCC)~\cite{vpcc}. In these platforms, both geometry and color
information are compressed~\cite{liu2019comprehensive}. Therefore,
the distortion of geometry and color will inevitably influence the
perceived quality of the reconstructed 3DPCs.

Similar to image/video quality assessment methods, point cloud
quality assessment methods can be classified into three categories:
Full Reference (FR), Reduced Reference (RR), and No Reference (NR).
To evaluate the quality of a distorted point cloud, FR methods use
the pristine uncompressed point cloud as a reference, while RR
methods only require statistical features that are extracted from
the reference point cloud. On the other hand, NR methods evaluate
the quality of the distorted point cloud in the absence of the
reference one.

FR 3DPC objective quality assessment techniques can be based on the
point-to-point~\cite{girardeau2005change},
point-to-plane~\cite{tian2017geometric} or
point-to-mesh~\cite{mekuria2016evaluation} distortion metric. The
point-to-point metric uses geometric distances between points in the
reference and distorted 3DPC, but it does not consider the fact that
points in a 3DPC usually represent surfaces on the object. The
point-to-plane metric is based on the projected error along the
normal of a reference point. This method depends on the calculation
of the normal and, essentially, larger costs are assigned to points
deviated from the underlying surface. The
point-to-mesh~\cite{mekuria2016evaluation} metric requires
construction of 3D meshes and is therefore hard to deploy in real
time applications. Beyond that, there is an angular similarity-based
FR metric~\cite{alexiouangular} and a local curvature analysis-based
FR metric~\cite{meynet2019pc} for 3DPC quality assessment. Both of
them are limited by the high complexity of searching for the
neighboring points to construct the normal or curvature. In
addition, the above objective quality metrics cannot predict the
visual quality of 3DPCs accurately, especially when the coding
distortion is
involved~\cite{alexiou2017subjective}~\cite{alexiou2018point}.

In this paper, we propose a reduced reference model to accurately
predict the mean opinion score (MOS) of V-PCC compressed 3DPCs from
the quantization parameters of the geometry and color encoders. The
proposed model is analytically simple and can be used for
rate-distortion optimized (RDO) rate control, as shown in
Fig.~\ref{framework}. The main contributions of this paper are as
follows:
\begin{enumerate}[1)]
\item We conduct comprehensive subjective tests to obtain MOSs
of V-PCC compressed 3DPCs with different combinations of geometry
and color quantization steps.
\item We develop a simple yet effective analytical model to
 predict the MOS from the geometry and color
 quantization steps.
\item We study the dependent factors of the model parameters and propose two features to estimate them.
\item We propose a perceptually optimal rate control method based on the proposed analytical model.
\end{enumerate}

The remainder of this paper is organized as follows. Section II
briefly reviews related work. The subjective test and the test
results are described in Section III. In Section IV, we present the
proposed perceptual quality model and validate its accuracy by using
the subjective test results. The dependent factors of the model
parameters are studied in Section V. Based on the study, we propose
an efficient model parameter estimation method by extracting two
features from the original 3DPCs. Subsequently, the subjective
quality-based rate control method is presented and evaluated in
Section VI. Finally, Section VII concludes the paper.

\section{Related work}
To develop an accurate perceptual quality model for 3DPCs,
subjective experiments are necessary. In recent years, some datasets
were provided to study the impact of compression on the subjective
quality of the reconstructed point clouds. Alexiou~\textit{et
al.}~\cite{alexiou2019comprehensive} provided a database which has
eight reference point clouds and the tested point clouds are
compressed by G-PCC and V-PCC. Zerman~\textit{et
al.}~\cite{zerman2019subjective} used V-PCC to generate a dataset of
3DPCs showing two people playing football. The remaining
datasets~\cite{alexiou2018impact}~\cite{sjtu-pcqa} study the impact
of multiple degradations types on point cloud subjective quality,
without focusing on the compression degradation type. Usually, the
number of raw 3DPCs limits the accuracy of the subjective quality
test. Therefore, we need to build a new subjective test dataset that
contains sufficient reference content and various encoding
degradation levels.

Generally, subjective quality assessment tests involve the
participation of subjects in experiments in which distorted objects
are visualized and rated.
In~\cite{alexiou2018point,8122239,alexiou2017performance}, the
geometry distortion was evaluated, while the effect of color
distortion was ignored. Torlig~\textit{et
al.}~\cite{torlig2018novel} considered the geometry and color
distortion jointly when doing the subjective assessment. However,
only six 3DPCs and their related degradations were assessed.
Su~\textit{et al.}~\cite{su2019perceptual} proposed a complete point
cloud data sets with various quality levels and made preliminary
verification on the performance of the existing objective quality
evaluation model. As reported in~\cite{su2019perceptual}, the visual
information fidelity in pixel domain (VIFP) achieves the best
performance compared to other assessment models. However, the PLCC
and SRCC of VIFP is only 0.77 which means the accuracy of 3DPC
quality assessment model still needs to be improved. Inspired by the
human visual system (HVS), eyes are not directly sensing the
individual point intensity, but rather the connected local neighbor
structures due to the low-pass spread functionality of our eye
optics~\cite{thibos1989image}. Yang~\textit{et
al.}~\cite{yang2020inferring} proposed a graph-based objective
metric instead of a point-based one. Although the metric can predict
the MOS more accurately than point-wise metrics, the resampling and
local graph construction operations greatly increase its complexity,
which limits its applications. Moreover, the existing FR objective
3DPC quality model is hard to be satisfied in some applications. For
example, in 3DPC streaming, a 3DPC is often requested by users with
diverse sustainable channel bandwidth. To address this diversity, it
can be coded into a scalable stream with several geometry and color
quantization parameters ($QP$s) combinations. Given a particular
target bitrate, the encoder needs to determine appropriate geometry
and color $QP$s to achieve the best perceptual quality. When there
are only FR metrics, time consuming exhaustive pre-coding must be
conducted to evaluate the performance of different $QP$
combinations~\cite{liu2018model}~\cite{liu2020rate}~\cite{liu2020coarse}.
In~\cite{liu2018model}, a model-based technique was developed to
efficiently determine the optimal maximum octree level (geometric
distortion) and JPEG\_VALUE (color distortion) for point cloud
library-based point cloud compression (PCL-PCC) platform. However,
only the color difference between the original point cloud and the
reconstructed point cloud was considered in the bit allocation
problem. In~\cite{liu2020rate}, a linear combination of the geometry
and color distortions was used to represent the point-to-point
distortion of 3DPCs. In~\cite{liu2020coarse}, a coarse to fine rate
control algorithm was proposed, in which the point-to-point
distortion metric was also adopted.  In all those methods, the
perceptual quality of the reconstructed 3DPCs was not considered,
which may limit their performance to some extent.

\section{Subjective quality assessment}\label{sec:2}
\subsection{Subjective test dataset}  
\begin{figure}[t!]
  \centering
  \includegraphics[width=0.85\columnwidth]{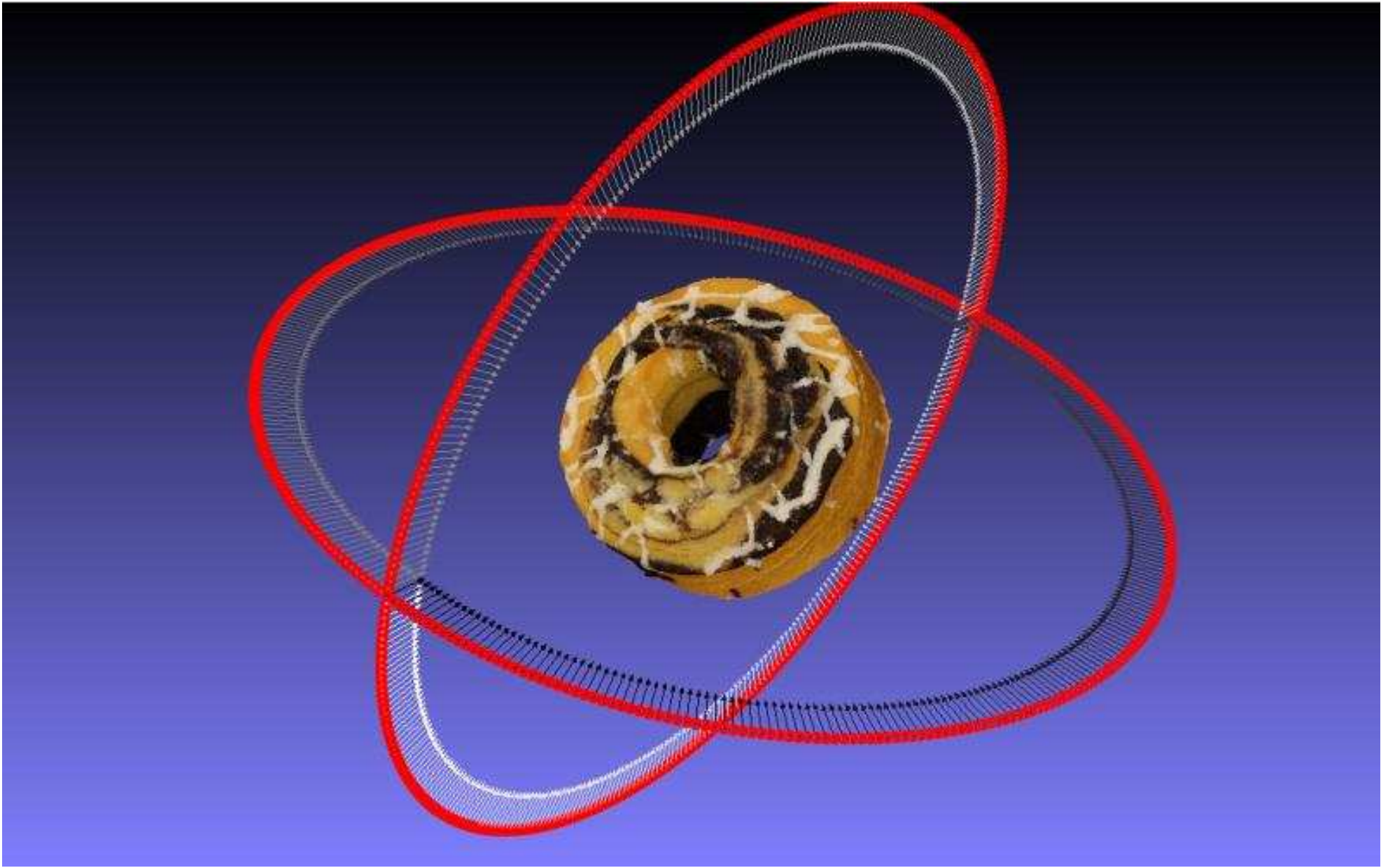}
  \caption{Illustration of the generation of pictures from 360 viewpoints of a 3DPC.}
  \label{differentviews}
\end{figure}
It is hard for an observer to distinguish the quality degradation of
3DPC with intrinsic distortion~\cite{8122239}. In the early stage of
the MPEG standardization for  point cloud compression(PCC), there
are not enough high quality raw 3DPCs. Therefore, sixteen high
quality point clouds, i.e., $Bag$ (1267845 points), $Banana$ (807184
points), $Biscuits$ (952579 points), $Cake$ (2486566 points),
$Cauliflower$ (1936627 points), $Flowerpot$ (2407154 points),
$House$ (1568490 points), $Litchi$ (1039942 points), $Mushroom$
(1144603 points), $Ping-pong\_bat$ (703879 points), $Puer\_tea$
(412009 points), $Pumpkin$ (1340343 points), $Ship$ (684617 points),
$Statue$ (1637577 points), $Stone$ (1086453 points), and $Tool\_box$
(1054211 points) were chosen from the Waterloo Point Cloud (WPC)
dataset ~\cite{su2019perceptual} in the subjective evaluation. These
3DPCs have various geometric and textural complexity. Since the MPEG
V-PCC platform achieves almost the best
performance~\cite{liu2019comprehensive} in all the existing public
encoders for both static and dynamic 3DPCs, all the 3DPCs were coded
by the V-PCC test model v7~\cite{tmc2}. For each 3DPC, there are 25
degraded versions with five geometry $QP$s (26, 32, 38, 44, and 50)
and five color $QP$s (26, 32, 38, 44, and 50). The corresponding
quantization steps range from 12.75 to 204. As a result, we have $16
\times 5 \times 5=400$ 3DPCs in the subjective evaluation. To show a
3DPC as fully as possible, we generated 180 pictures along the
horizontal and vertical directions with a step of two degrees
separately, for each 3DPC, as shown in Fig.~\ref{differentviews}.
Afterwards, the degraded and the original pictures were concatenated
to generate a 10-second video sequence with 360 frames.

A total of 30 subjects, consisting of 15 males and 15 females aged
between 20 and 35, were recruited in the subjective evaluation. All
subjects had normal or corrected-to-normal vision.

\subsection{Subjective evaluation}  
\begin{figure}[t!]
  \centering
  \includegraphics[width=0.85\columnwidth]{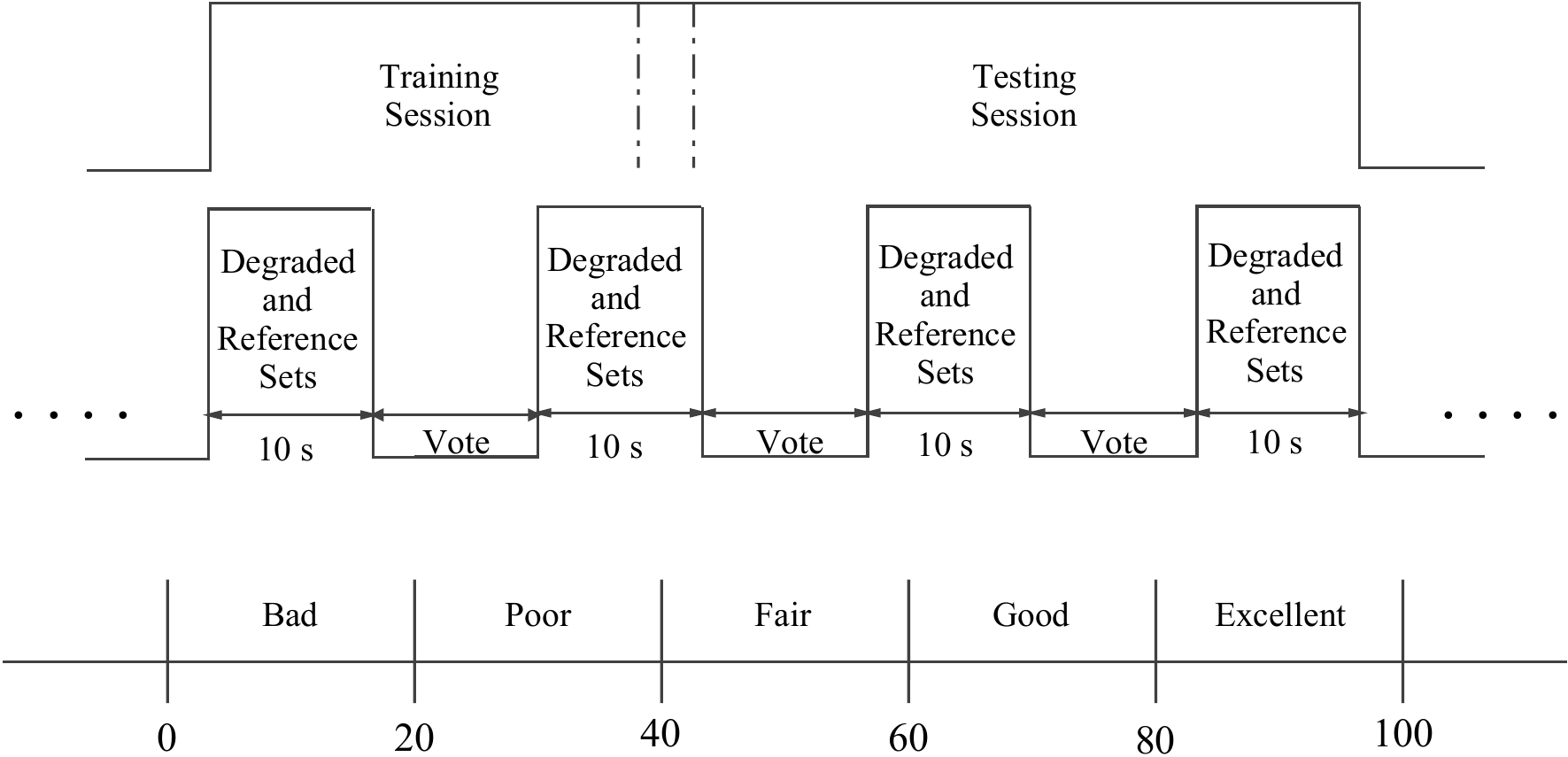}
  \caption{Schematic diagram of the subjective experiment.}
  \label{subjective}
\end{figure}

\begin{figure}[t!]
  \centering
  \includegraphics[width=0.85\columnwidth]{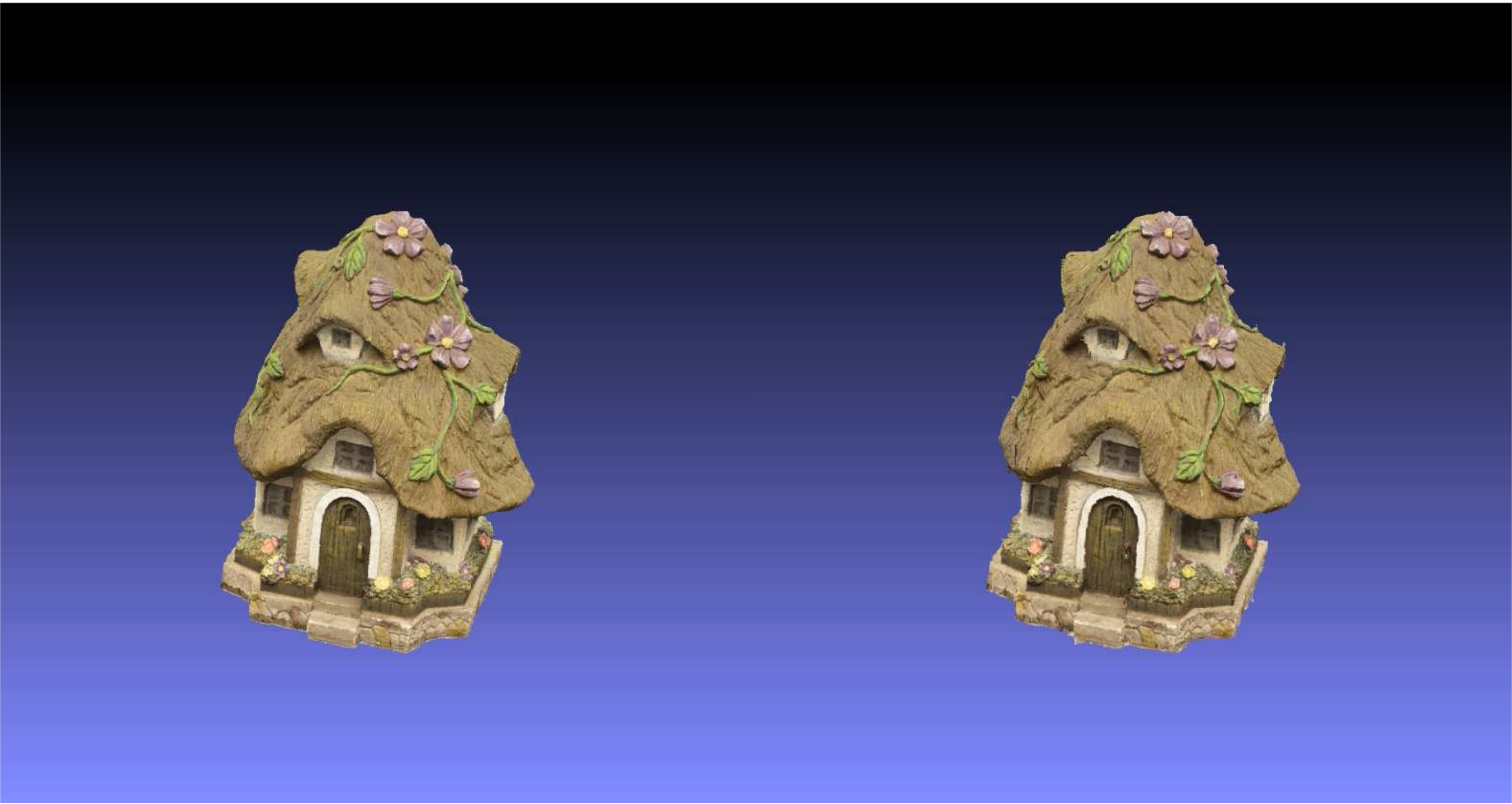}
  \caption{Example of a subjective evaluation.}
  \label{subexample}
\end{figure}

The Double-Stimulus Impairment Scale (DSIS)
methodology~\cite{nehme2019comparison} was adopted in the subjective
evaluation. As normal operation, to expand the rating range and
obtain finer distinctions, the DSIS method was adopted with 100
points continuous scale instead of 5 levels rating, as shown in
Fig.~\ref{subjective}.
To display the stimuli, a DELL E2417H displayer with an In-Plane
Switching Display of 23.8 inch (res. 1920 $\times$ 1080) was used.
Both the original and the distorted videos generated from a 3DPC
were simultaneously shown to the observer side-by-side, as shown in
Fig.~\ref{subexample}.
The observer viewed these videos from a distance equal to twice the
screen height and rated them through a customized interface after
the playback finished by keyboard input to guarantee there is no
time restriction.

At the beginning of each evaluation, a training session was
conducted to make the observers familiar with the artifacts in the
assessment. The 3DPCs used for training were different from those
used for the evaluation. Therefore, the observers were familiar with
the distortion types and the quality levels, but not familiar with
the content. The duration of each test for a given subject was about
two hours, divided into four sections, with three five-minute breaks
in-between to minimize the effect of fatigue.

\subsection{Data post-processing}  
Since the ratings range from 0 to 100, the scores given by different
observers tend to fall in fairly small subranges. Therefore, we need
to convert the subjective scores to Z-scores~\cite{van1995quality}
based on the mean and standard deviation of all the scores of each
observer. The Z-score of the $m$-th 3DPC at the $j$-th degraded
level from the $i$-th viewer is
\begin{equation}
\label{eq:zcore} Z_{mij} =\frac{X_{mij}-\mu_{X_i}}{\delta_{X_i}},
\end{equation}
where $X_{mij}$ denotes the raw rating, and $\mu_{X_i}$ and
$\delta_{X_i}$ represent the mean and the standard deviation of the
ratings of the $i$-th viewer, respectively. Besides, we adopted the
outlier removal technique suggested in~\cite{bt50013} to remove
outliers. No participants were removed but outlier ratings from each
participant were discarded. The obtained Z-scores lie in the range
[0, 100]. The average of the Z-scores from all valid subjects were
calculated to be the MOS of each degraded 3DPC. By taking the MOS as
the ``ground truth", the PLCC and SRCC between each viewer's scores
and MOSs were calculated to verify the performance of individual
subjects~\cite{su2019perceptual}. Both the mean PLCC and SRCC
between each observer scores and the calculated MOS were as high as
0.84, indicating substantial agreement between individual subjects.
\begin{figure}[t!]
  \centering
  \includegraphics[width=0.7\columnwidth]{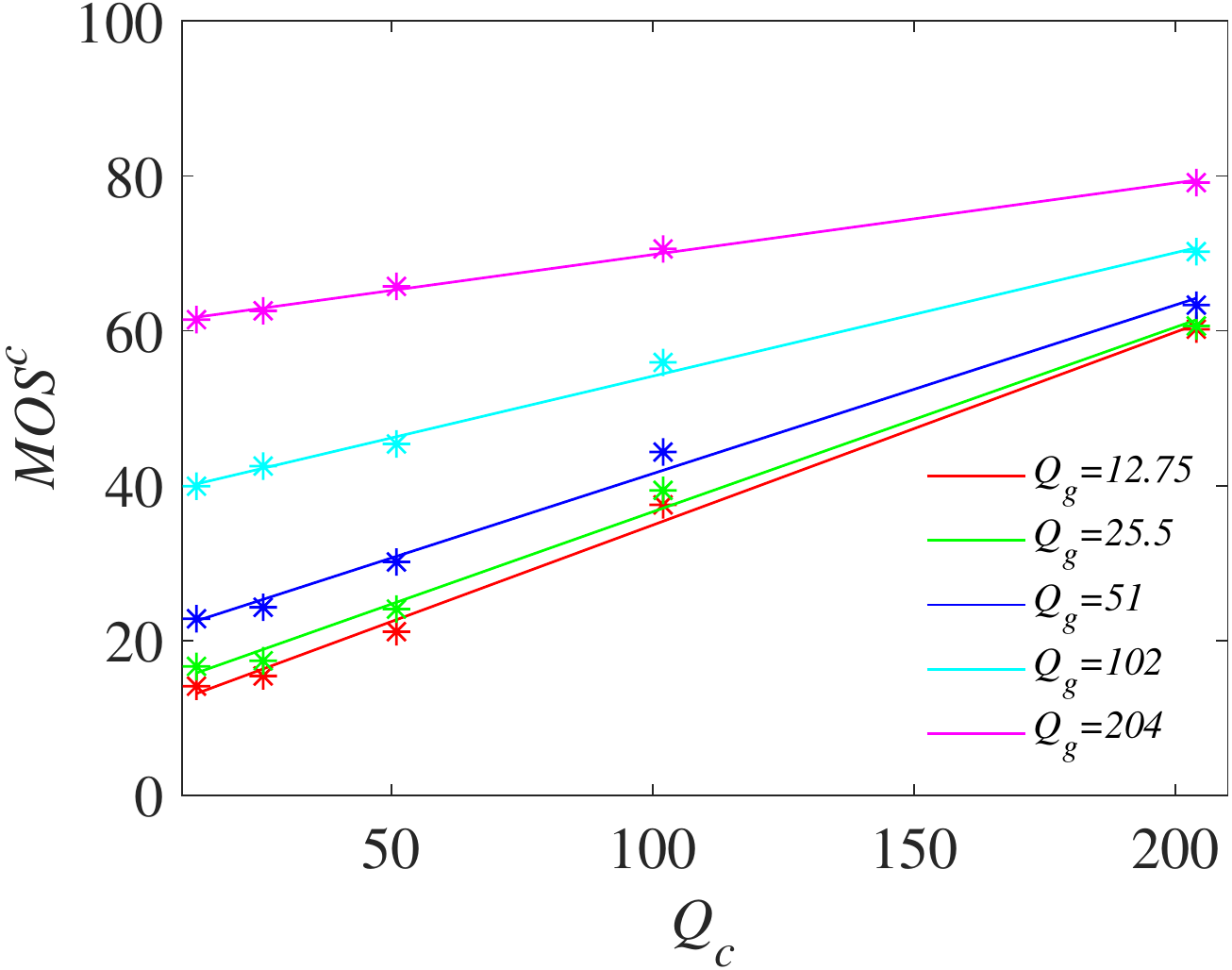}
  \caption{Relationships between $MOS^c=100-MOS$ and $Q_c$ for different $Q_g$s.}
  \label{aveMosTqstep}
\end{figure}
\section{Proposed quality metric model}\label{sec:3}
To determine the relationship between the perceived quality and the
quantization steps of the geometry and color, the distorted 3DPCs
with different geometry and color quantization steps were rated, as
shown in Fig.~\ref{aveMosTqstep}.
\begin{figure}[t!] \centering \subfigure[]{
\label{fig2:subfig:a}
\includegraphics[width=0.48\columnwidth]{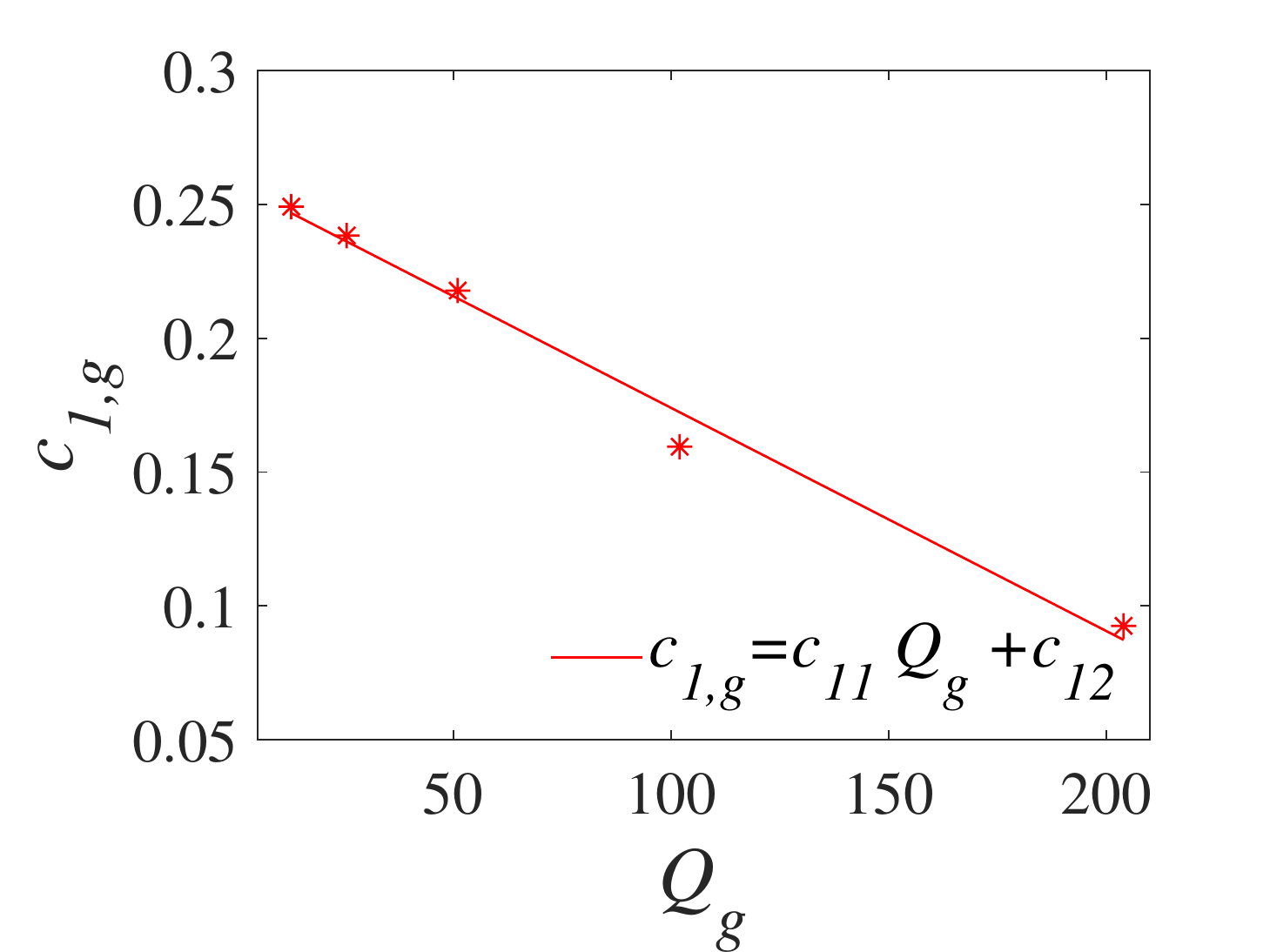}}
\subfigure[]{ \label{fig2:subfig:b}
\includegraphics[width=0.48\columnwidth]{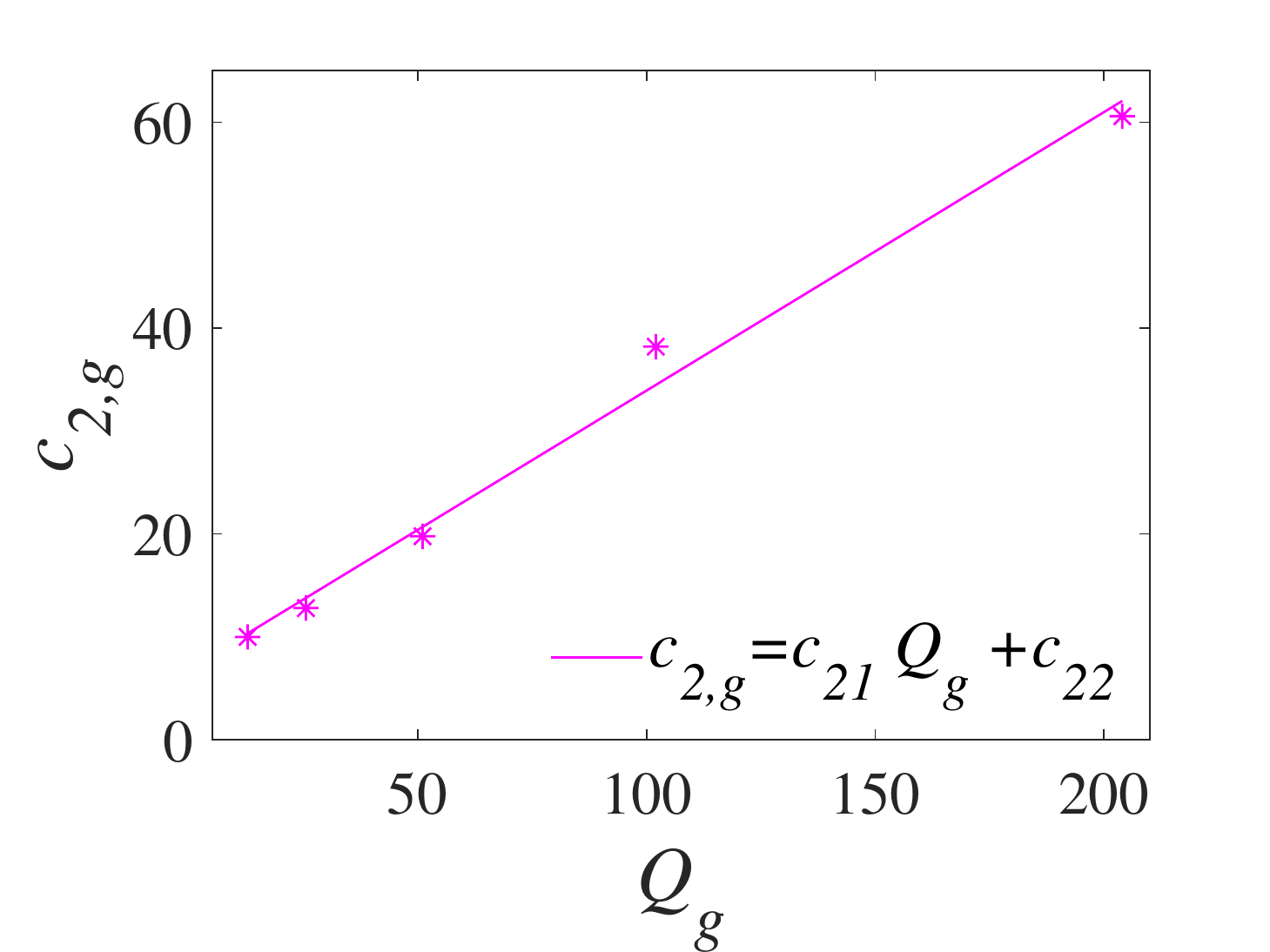}}
\caption{Relationship between the slope $c_{1,g}$ and intercept
$c_{2,g}$ in~\eqref{eq:mos100colorqs} and $Q_g$. (a) $c_{1,g}$ vs.
$Q_g$, (b) $c_{2,g}$ vs. $Q_g$.} \label{fig2}
\end{figure}
We can observe that there is a linear relationship between
$MOS^c=100-MOS$ and the the color quantization step $Q_c$ for a
fixed geometry quantization step $Q_g$, that is,
\begin{equation}
\label{eq:mos100colorqs} MOS^c =c_{1,g}  Q_c + c_{2,g},
\end{equation}
where $c_{1,g}$ and $c_{2,g}$ are the model parameters. Here, we use
$MOS^c$ to represent the perceptual distortion for the standard
mathematical expression used in rate-distortion optimization. From
Table~\ref{tab:mos100cqs}, we can also see that the squared
correlation coefficient (SCC) between $MOS^c$ and $Q_c$ with
different $Q_g$s is larger than or equal to 0.993, while the root
mean squared error (RMSE) is smaller than or equal to 1.785.
\begin{table}[t!]
\centering \caption{Accuracy of the linear
relationship~\eqref{eq:mos100colorqs}} \label{tab:mos100cqs}
  \begin{tabular}{ccccc}
      \toprule
      \midrule
      $Q_g$  &$c_{1,g}$   &$c_{2,g}$    &SCC      &RMSE \\\hline
      12.75  &0.249 &9.986   &0.994    &1.731 \\
      25.5   &0.238 &12.782  &0.993    &1.785 \\
      51     &0.218 &19.765  &0.993    &1.634 \\
      102    &0.159 &38.187  &0.994    &1.070 \\
      204    &0.093 &60.571  &0.996    &0.525 \\\hline
      \bottomrule
  \end{tabular}
\end{table}
Moreover, as shown in Fig.~\ref{fig2}, the relationship between the
slope $c_{1,g}$ (respectively the intercept $c_{2,g}$) and $Q_g$ can
be represented by the linear models
\begin{equation}
\label{eq:p1gqs} c_{1,g} =  c_{11} Q_g + c_{12},
\end{equation}
\begin{equation}
\label{eq:p2gqs} c_{2,g} =  c_{21} Q_g + c_{22},
\end{equation}
where the SCCs of $Q_g$ and $c_{1,g}$, and $Q_g$ and $c_{2,g}$ are
0.988 and 0.990, respectively. Accordingly, the quality model can be
rewritten as
\begin{equation}
\label{eq:mos100com} MOS^c =a Q_g Q_c + b Q_g + c Q_c +d,
\end{equation}
where $a=c_{11}$, $b=c_{21}$, $c=c_{12}$, and $d=c_{22}$ are model
parameters. The accuracy of~\eqref{eq:mos100com} for each 3DPC is
given in Table~\ref{tab:MOS100sequencescom}.
\begin{table}[t!]
\centering \caption{Accuracy of~\eqref{eq:mos100com} for each 3DPC}
\label{tab:MOS100sequencescom}
  \begin{tabular}{ccccccc}
      \toprule
      \midrule
      Point Cloud          &$a$     &$b$    &$c$    &$d$   &SCC    &RMSE \\\hline
      \emph{Bag}           &-0.0005 &0.263 &0.223 &3.192 &0.963 &4.317 \\
      \emph{Banana}        &-0.0006 &0.294 &0.127 &19.860 &0.925 &5.663 \\
      \emph{Biscuits}      &-0.0006 &0.190 &0.204 &8.293 &0.964 &3.158 \\
      \emph{Cake}          &-0.0008 &0.303 &0.188 &5.519 &0.977 &3.192 \\
      \emph{Cauliflower}   &-0.0010 &0.327 &0.258 &3.389 &0.967 &4.372 \\
      \emph{Flowerpot}     &-0.0005 &0.332 &0.115 &13.016 &0.889 &8.097 \\
      \emph{House}         &-0.0012 &0.311 &0.361 &-3.666 &0.981 &3.814 \\
      \emph{Litchi}        &-0.0012 &0.288 &0.359 &-3.440 &0.970 &4.536 \\
      \emph{Mushroom}      &-0.0010 &0.244 &0.304 &12.295 &0.946 &5.203 \\
      \emph{Ping-pong\_bat}&-0.0014 &0.351 &0.332 &5.463 &0.951 &5.875 \\
      \emph{Puer\_tea}     &-0.0009 &0.192 &0.366 &6.488 &0.982 &3.379 \\
      \emph{Pumpkin}       &-0.0007 &0.184 &0.276 &3.242 &0.969 &3.557 \\
      \emph{Ship}          &-0.0006 &0.312 &0.112 &13.296 &0.928 &5.905 \\
      \emph{Statue}        &-0.0007 &0.308 &0.196 &14.527 &0.874 &8.496 \\
      \emph{Stone}         &-0.0010 &0.245 &0.366 &-1.385 &0.981 &3.588 \\
      \emph{Tool\_box}     &-0.0008 &0.184 &0.333 &9.886 &0.951 &5.124 \\\hline
      \bottomrule
  \end{tabular}
\end{table}
By further considering the fact that the fitting parameter $a$ is
very small
(Table~\ref{tab:MOS100sequencescom}),~\eqref{eq:mos100com} can be
further simplified by removing the impact of $Q_g \cdot Q_c$ on the
perceptual quality. This makes the model convex, which is useful in
many applications such as rate-distortion optimization.
\begin{figure*}[t!]
\centering \subfigure[]{ \label{fig3:subfig:a}
\includegraphics[width=0.48\columnwidth, height=3.5cm]{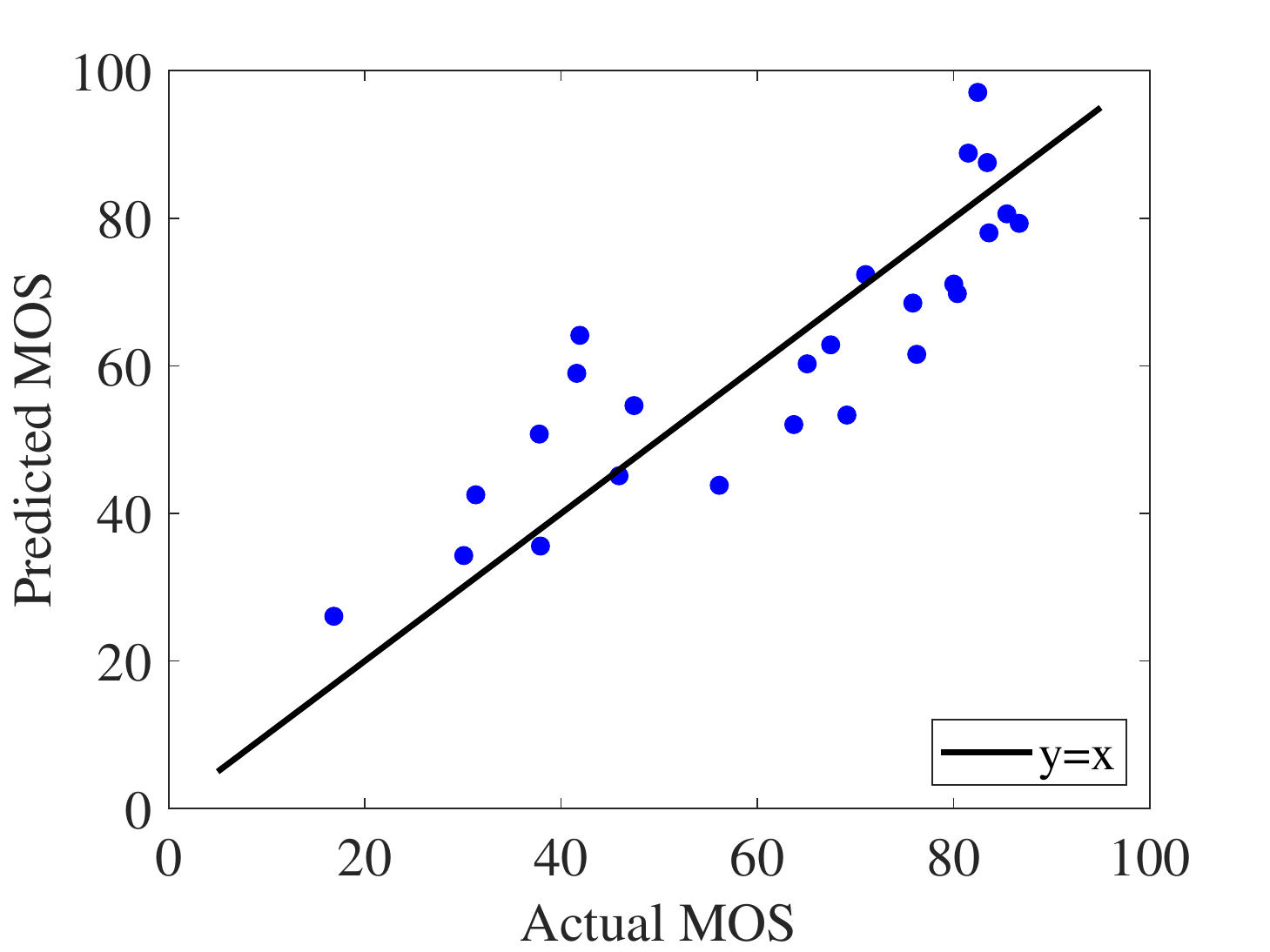}}
\subfigure[]{ \label{fig3:subfig:b}
\includegraphics[width=0.48\columnwidth, height=3.5cm]{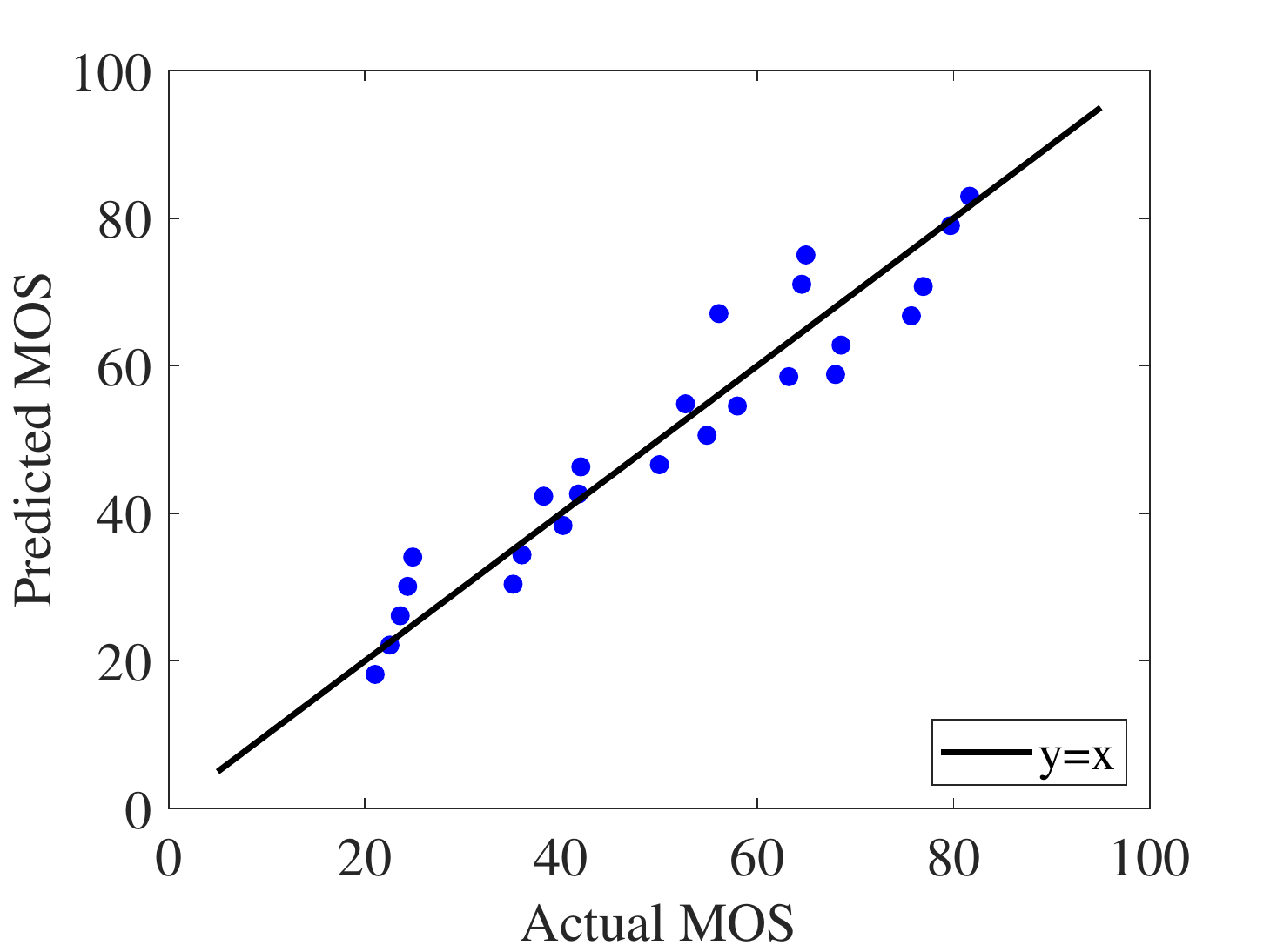}}
\subfigure[]{ \label{fig3:subfig:c}
\includegraphics[width=0.48\columnwidth, height=3.5cm]{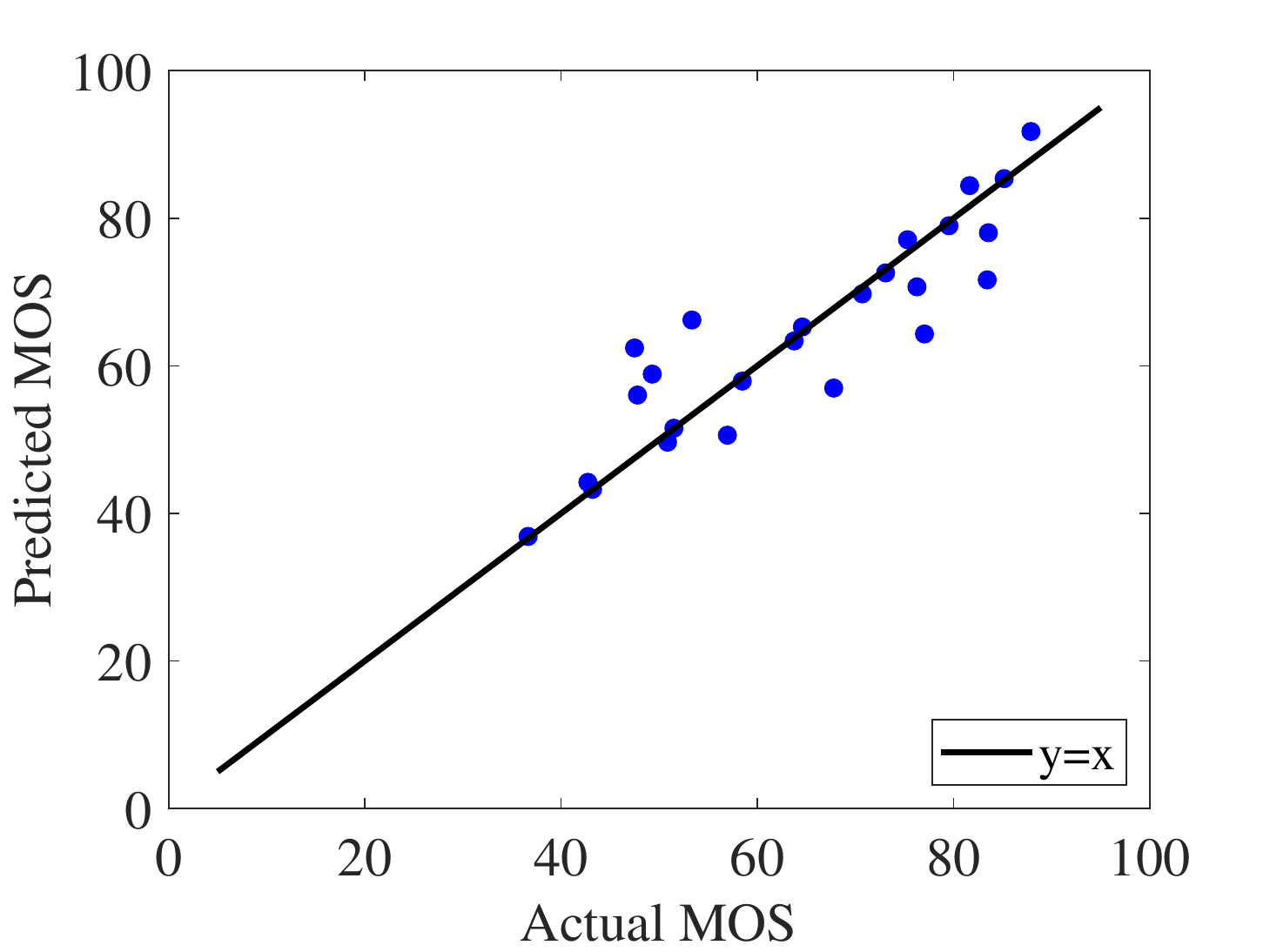}}
\subfigure[]{ \label{fig3:subfig:d}
\includegraphics[width=0.48\columnwidth, height=3.5cm]{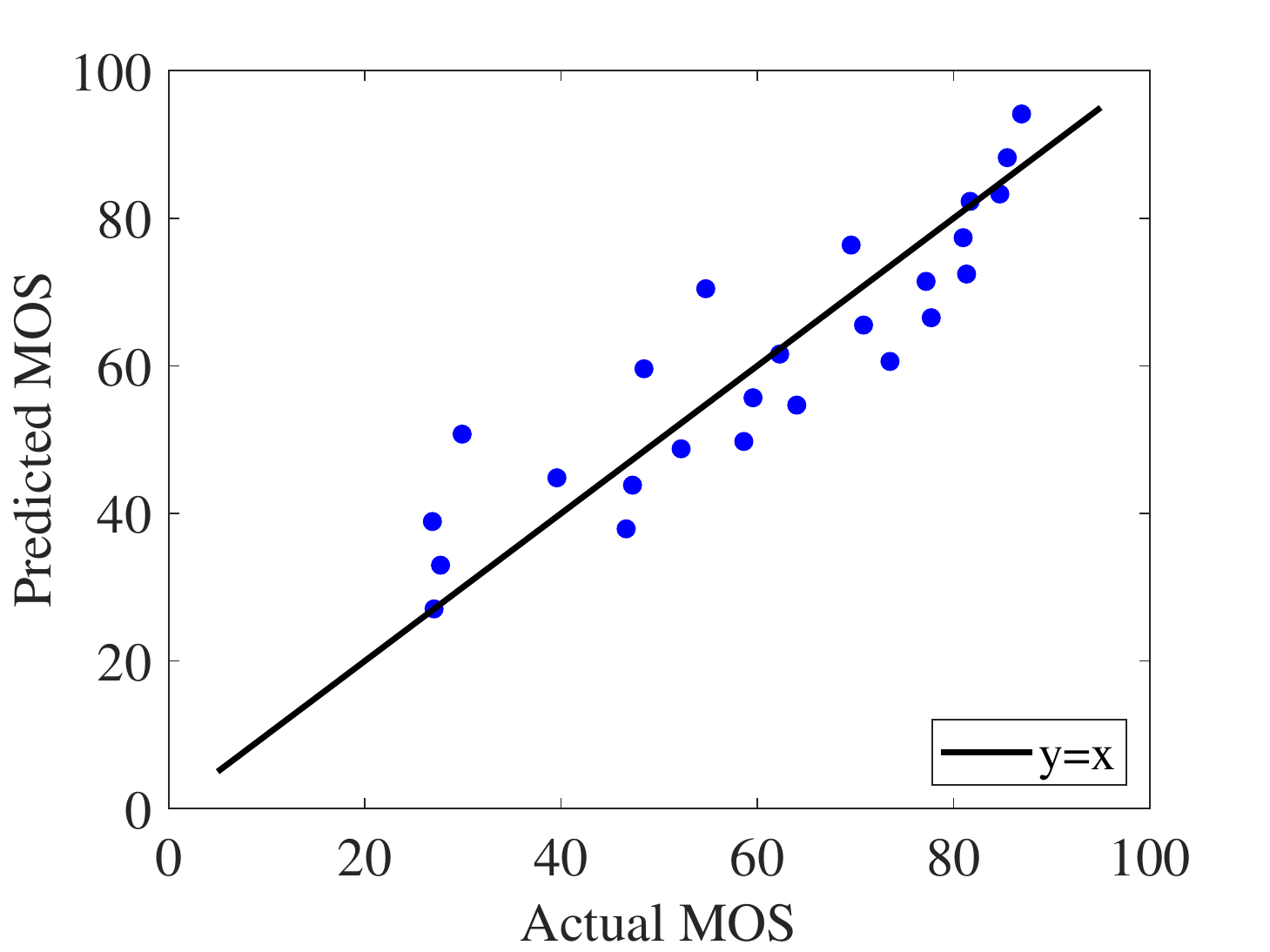}}
\subfigure[]{ \label{fig3:subfig:e}
\includegraphics[width=0.48\columnwidth, height=3.5cm]{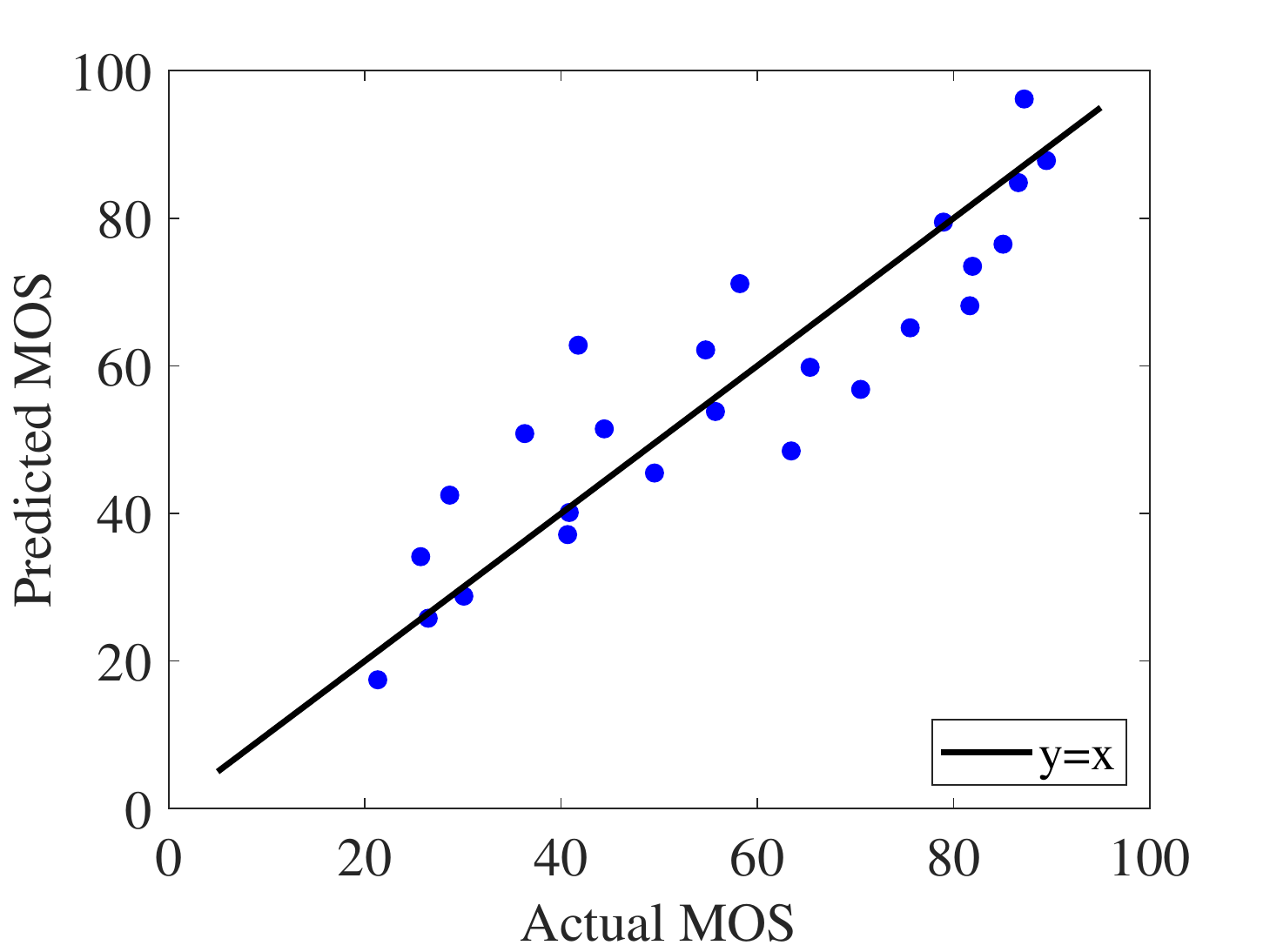}}
\subfigure[]{ \label{fig3:subfig:f}
\includegraphics[width=0.48\columnwidth, height=3.5cm]{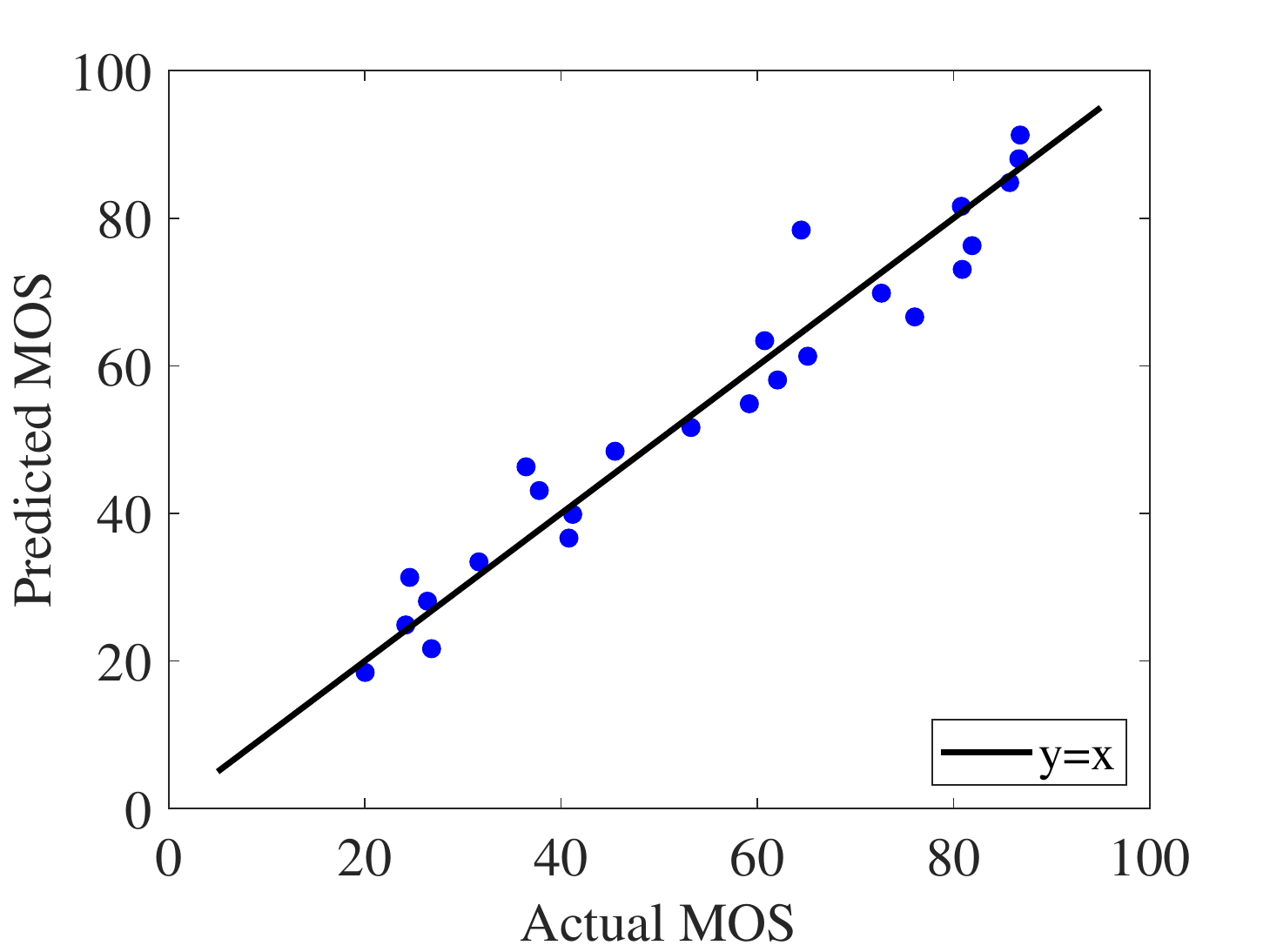}}
\subfigure[]{ \label{fig3:subfig:g}
\includegraphics[width=0.48\columnwidth, height=3.5cm]{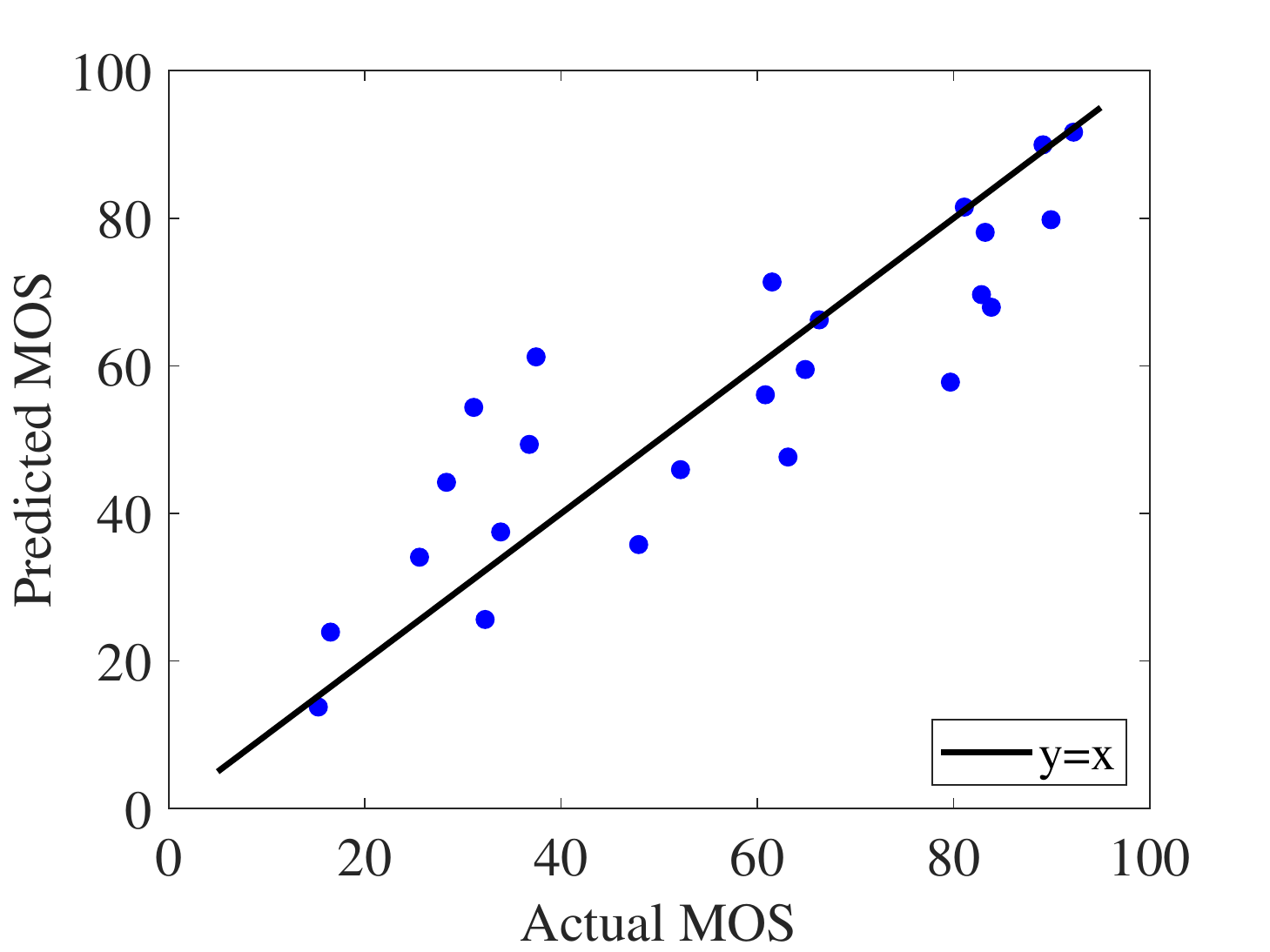}}
\subfigure[]{ \label{fig3:subfig:h}
\includegraphics[width=0.48\columnwidth, height=3.5cm]{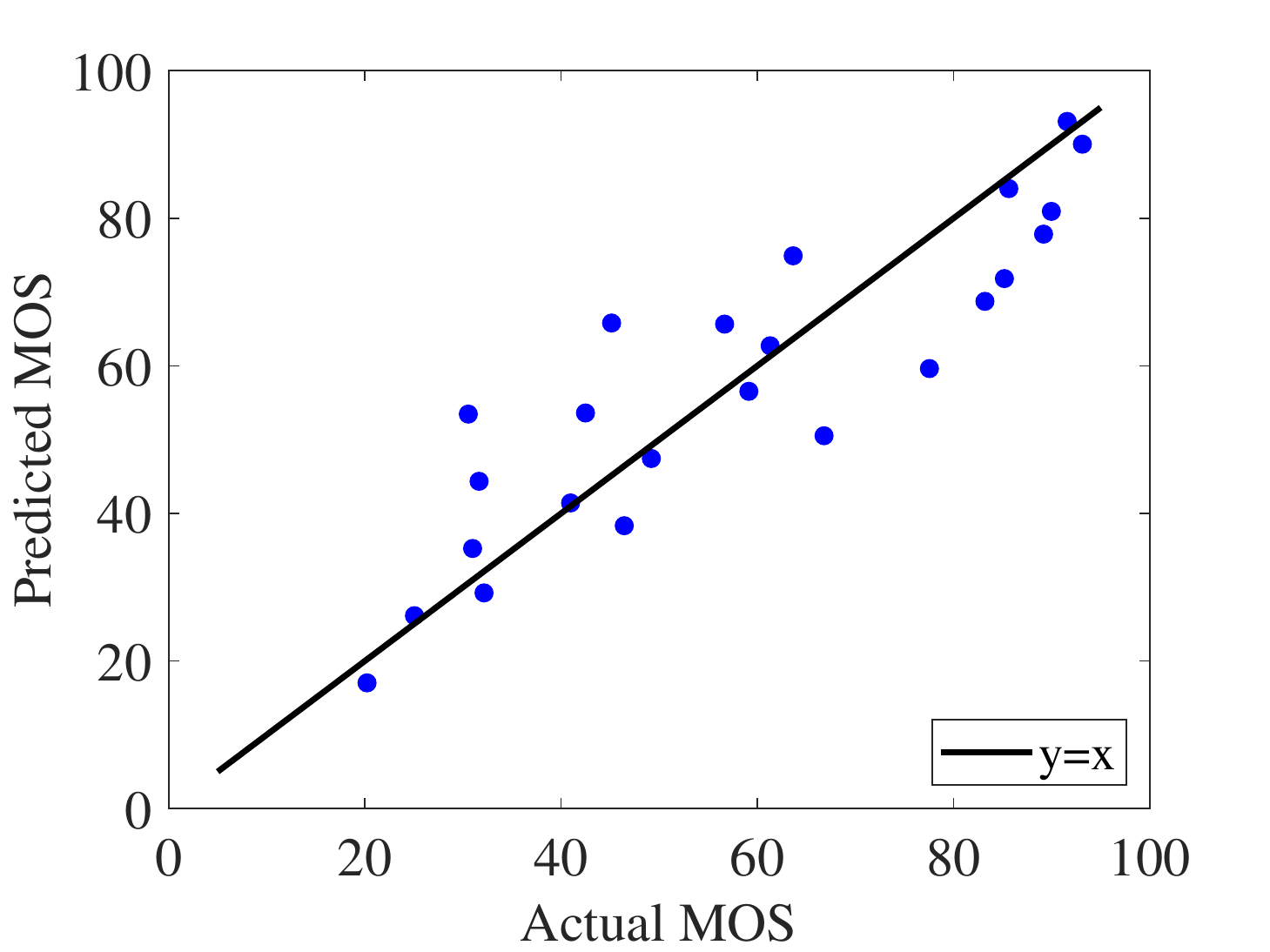}}
\subfigure[]{ \label{fig3:subfig:i}
\includegraphics[width=0.48\columnwidth, height=3.5cm]{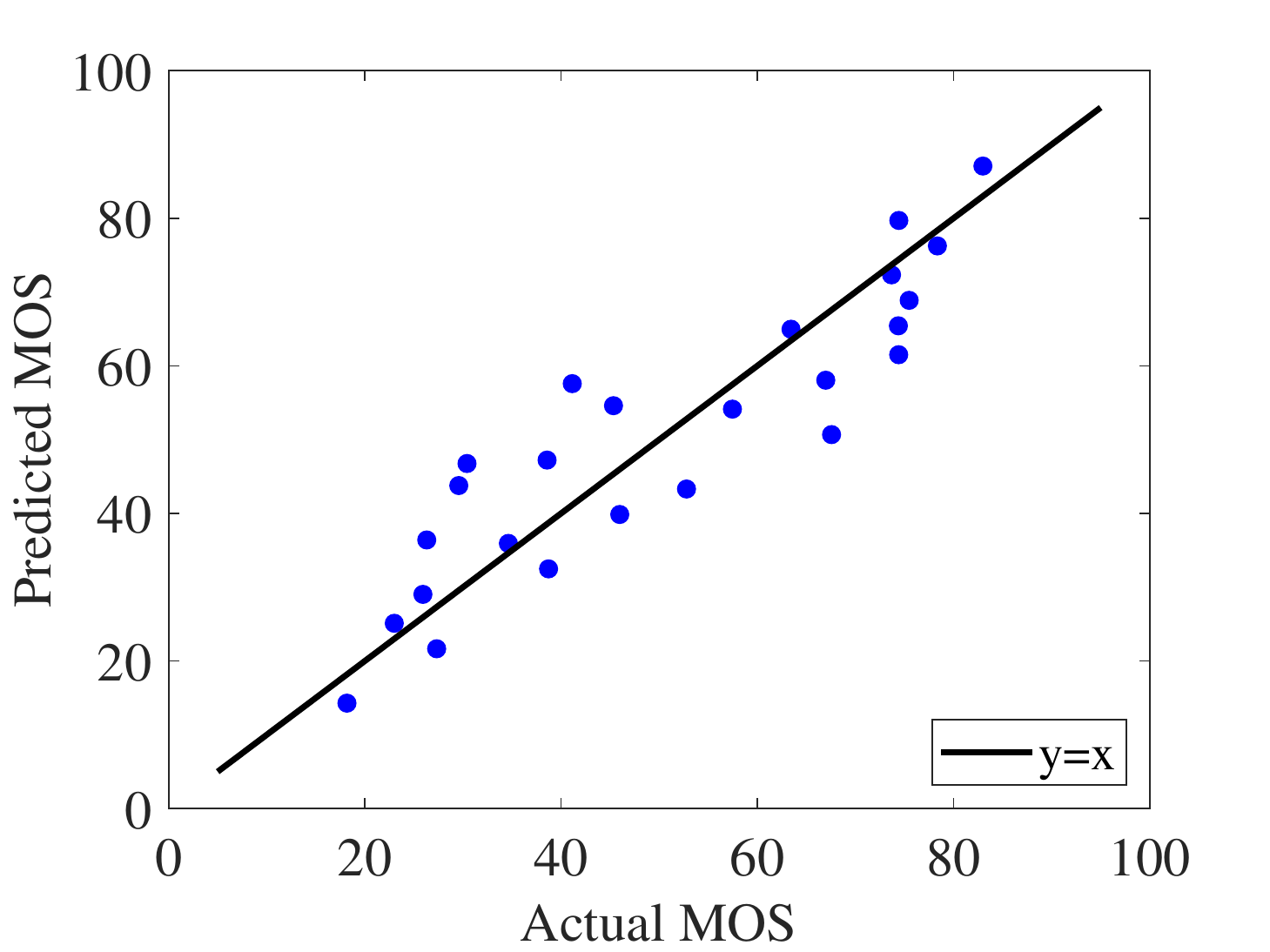}}
\subfigure[]{ \label{fig3:subfig:j}
\includegraphics[width=0.48\columnwidth, height=3.5cm]{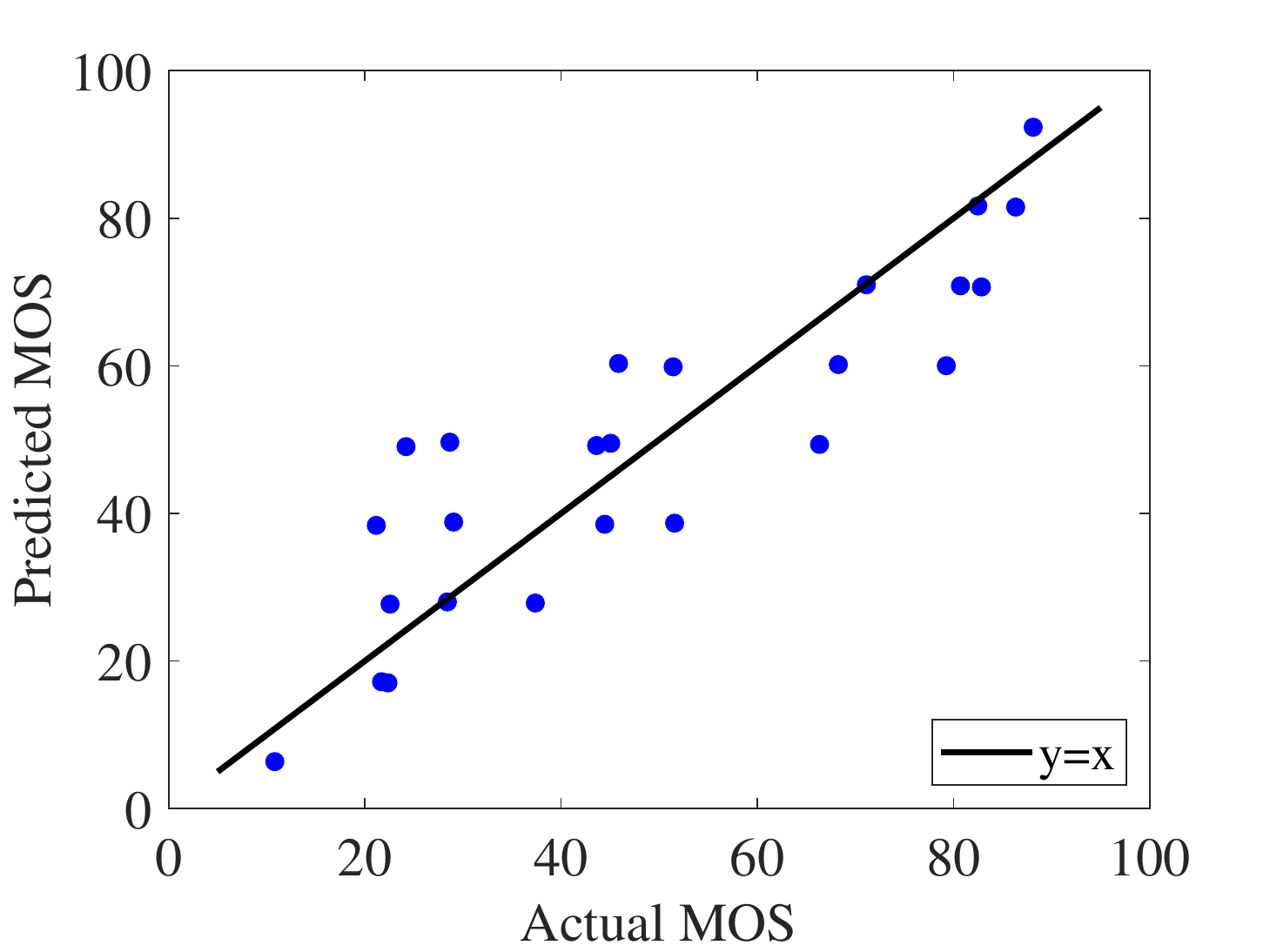}}
\subfigure[]{ \label{fig3:subfig:k}
\includegraphics[width=0.48\columnwidth, height=3.5cm]{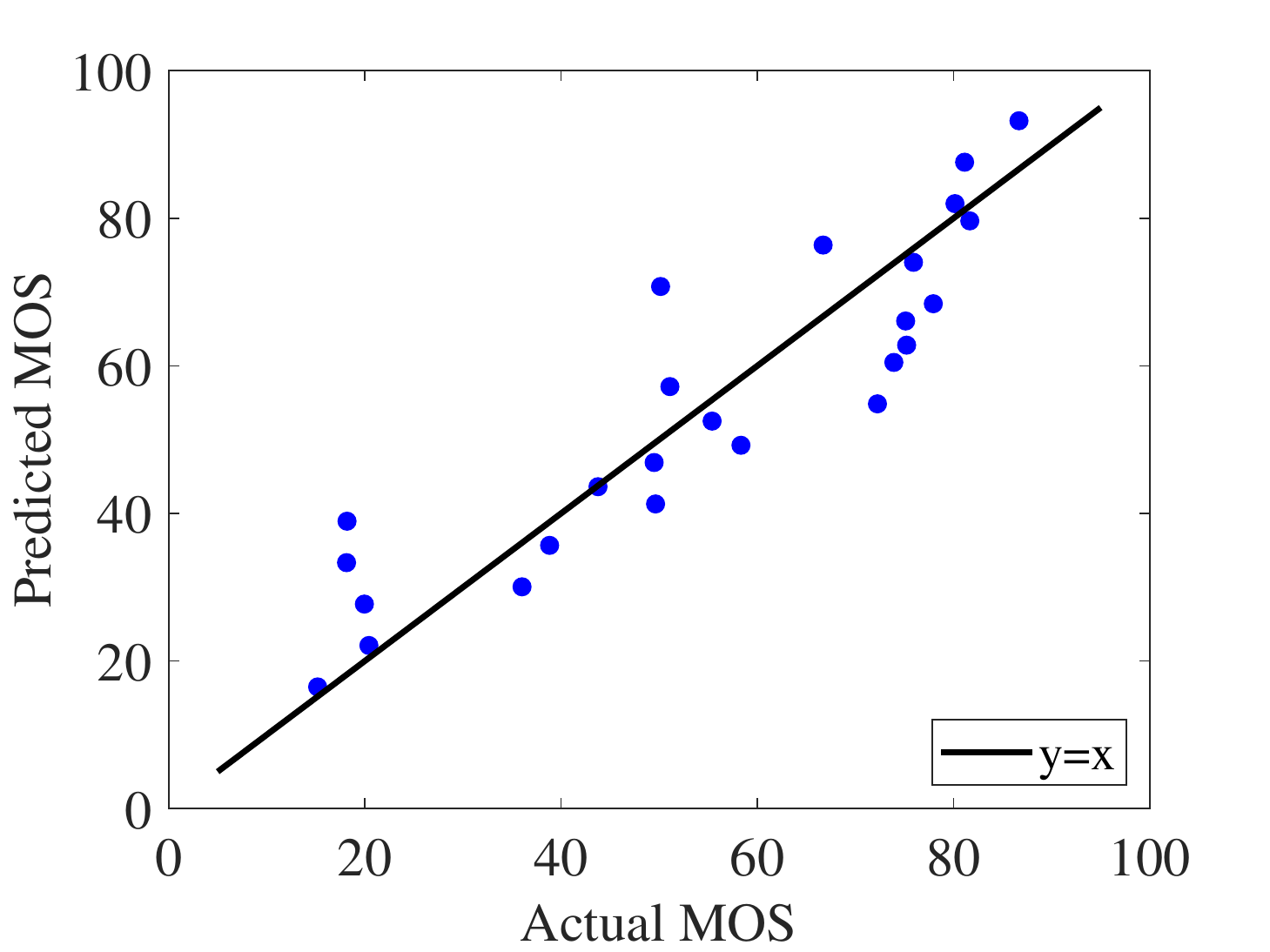}}
\subfigure[]{ \label{fig3:subfig:l}
\includegraphics[width=0.48\columnwidth, height=3.5cm]{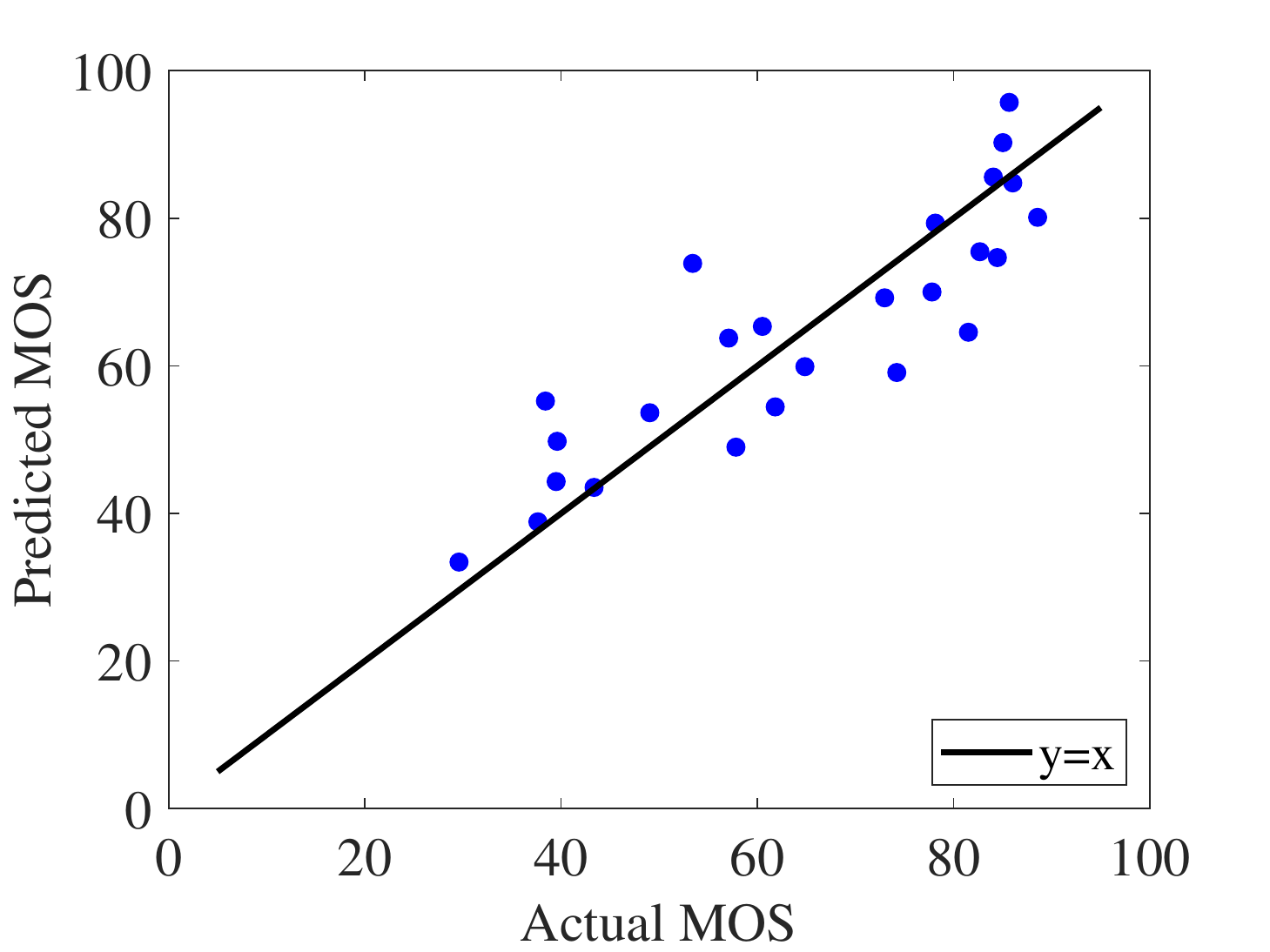}}
\subfigure[]{ \label{fig3:subfig:m}
\includegraphics[width=0.48\columnwidth, height=3.5cm]{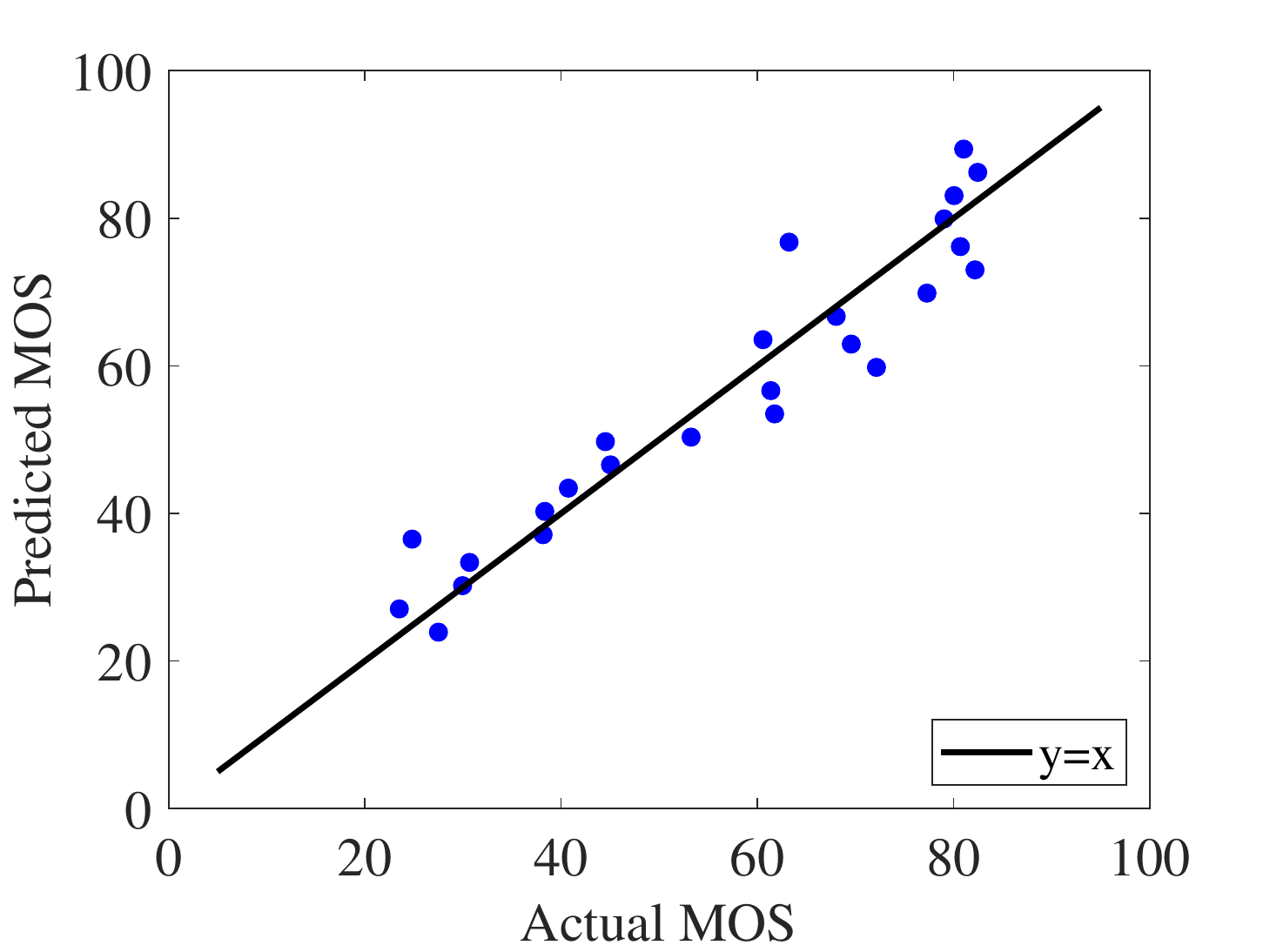}}
\subfigure[]{ \label{fig3:subfig:n}
\includegraphics[width=0.48\columnwidth, height=3.5cm]{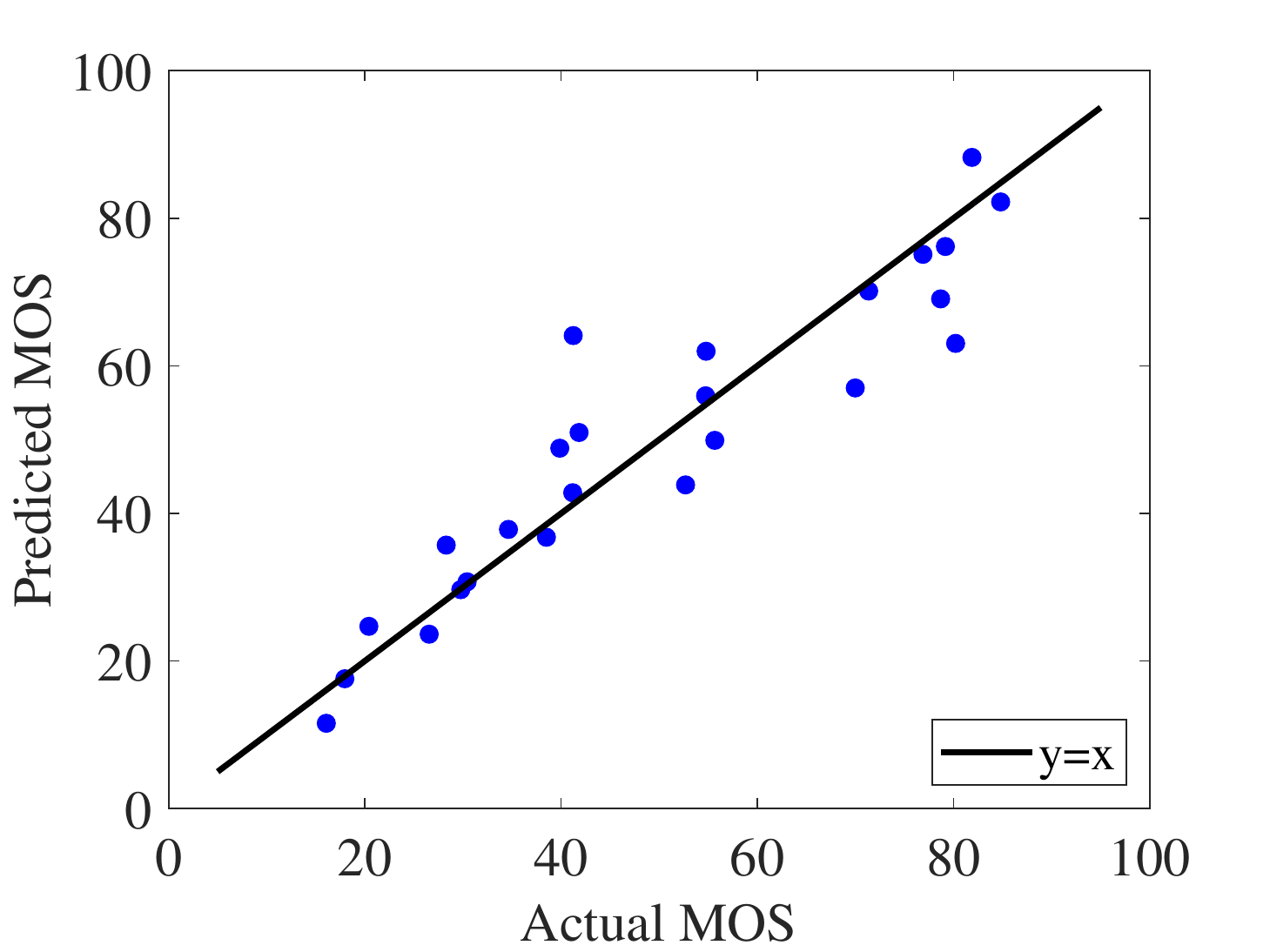}}
\subfigure[]{ \label{fig3:subfig:o}
\includegraphics[width=0.48\columnwidth, height=3.5cm]{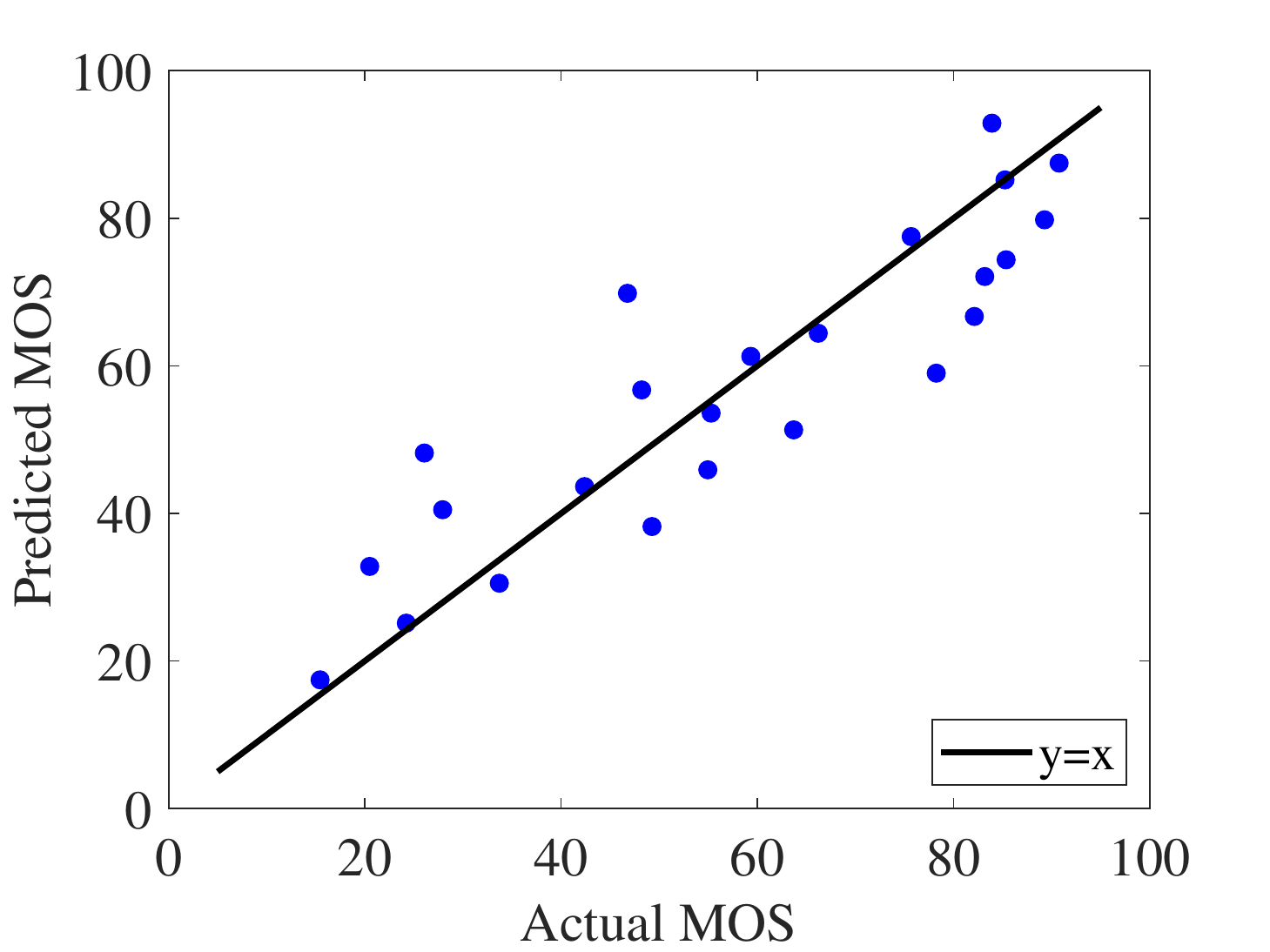}}
\subfigure[]{ \label{fig3:subfig:p}
\includegraphics[width=0.48\columnwidth, height=3.5cm]{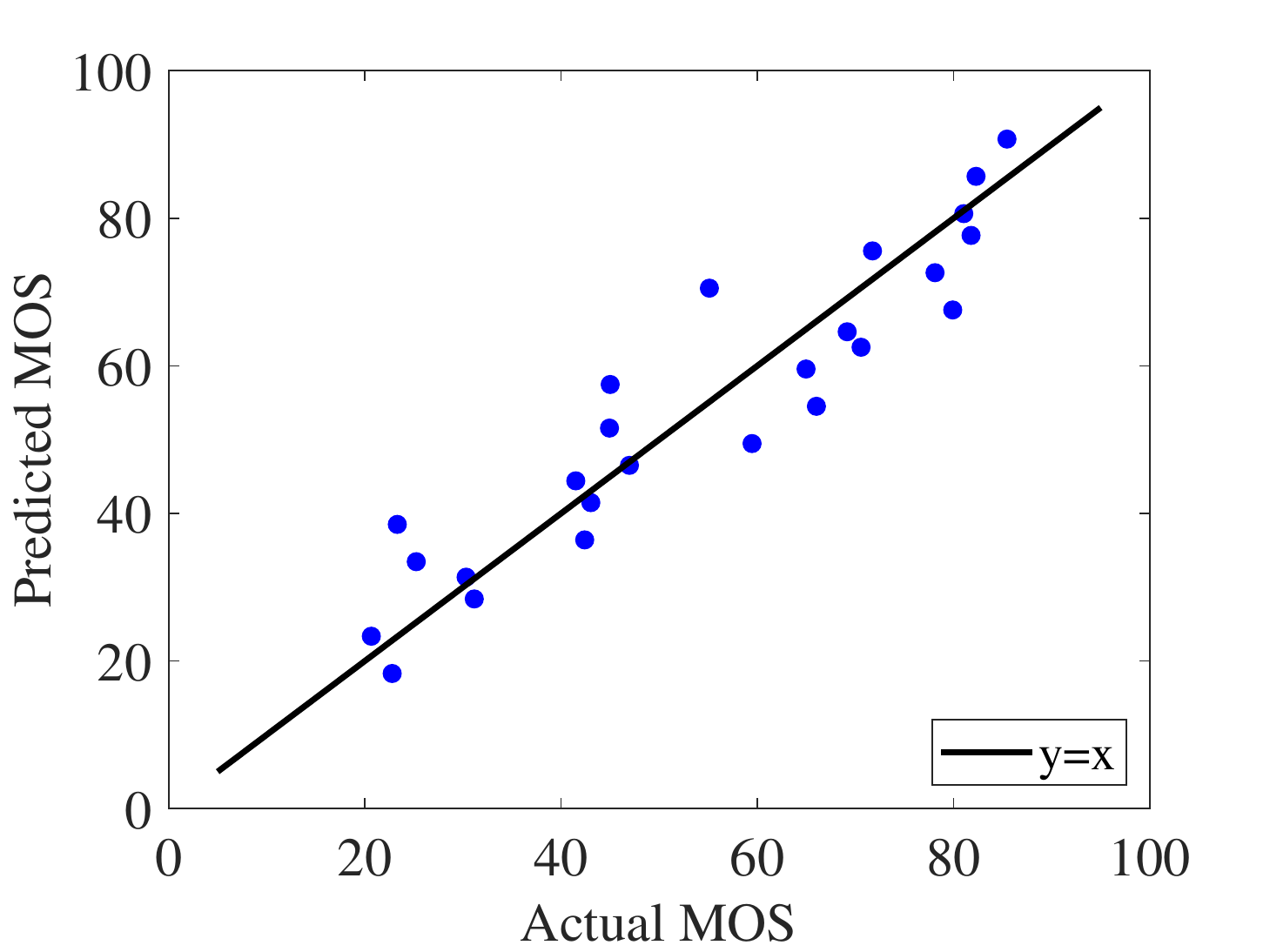}}
\caption{Accuracy of model~\eqref{eq:mos100com}. (a)-(d):
\emph{Bag}, \emph{Banana}, \emph{Biscuits}, and \emph{Cake},
(e)-(h): \emph{Cauliflower}, \emph{Flowerpot}, \emph{House}, and
\emph{Litchi}, (i)-(l): \emph{Mushroom}, \emph{Ping-pong\_bat},
\emph{Puer\_tea}, and \emph{Pumpkin}, (m)-(p): \emph{Ship},
\emph{Statue}, \emph{Stone}, and \emph{Tool\_box}.} \label{fig3}
\end{figure*}
Therefore, we also examined the statistical significance of the
three parts in~\eqref{eq:mos100com}, i.e., $Q_g \cdot Q_c$, $Q_g$,
and $Q_c$ using a two-way ANOVA test~\cite{girden1992anova}. In the
test, the $F$-values are based on the ratio of mean squares (MS) of
the test factor group and the error group. The MS is the mean of the
square of the standard deviation (SS) that accounts for the degrees
of freedom (DF). Therefore the $F$-value can be calculated as
\begin{equation}
\label{eq:fvalue}
\begin{aligned}
F &= MS_{t}}/{MS_{e}\\
&= (\frac{SS_{t}}{DF_{t}}) / (\frac{SS_{e}}{DF_{e}})\\
\end{aligned}
\end{equation}
where $MS_{t}$ and $MS_{e}$ represent the mean sum of squares of
deviations of the test factor group and the error group,
respectively. They can be calculated as $\frac{SS_{t}}{DF_{t}}$ and
$\frac{SS_{e}}{DF_{e}}$, respectively, where $SS_{t}$ and $SS_{e}$
represent the sum of squared deviations of the test factor group and
the error group, respectively, while $DF_{t}$ and $DF_{e}$ represent
the degrees of freedom of the test factor group and the error group,
respectively. Specifically, $SS_t \in \left\{SS_{Q_g}, SS_{Q_c},
SS_{Q_g \cdot Q_c}\right\}$, where $SS_{Q_g}, SS_{Q_c}, SS_{Q_g
\cdot Q_c}$ denote the SS of the test factors $Q_g, Q_c$, and $Q_g
\cdot Q_c$, respectively. Here $SS_{Q_g}, SS_{Q_c}, SS_{Q_g \cdot
Q_c}$, and $SS_{e}$ can be calculated as follows
\begin{equation}
\label{eq:fvaluedetailSS}
\begin{aligned}
\begin{small}
\begin{cases}
SS_{Q_g}  =  J L \sum_{i=1}^{I}(\overline{MOS^c_{i..}}-\overline{MOS^c})^2 \\
SS_{Q_c} =  I L \sum_{j=1}^{J}(\overline{MOS^c_{.j.}}-\overline{MOS^c})^2 \\
SS_{Q_g\! \cdot\! Q_c}\!
=\!L\!\sum_{i=1}^{I}\!\sum_{j=1}^{J}\!(\overline{MOS^c_{ij.}}\!-\!\overline{MOS^c_{i..}}\!-\!\overline{MOS^c_{.j.}}\!+\!\overline{MOS^c})^2
\\
SS_{e}= \sum_{i=1}^{I}\sum_{j=1}^{J}\sum_{l=1}^{L}(MOS^c_{ijl}-\overline{MOS^c_{ij.}})^2 \\
\end{cases}
\end{small}
\end{aligned}
\end{equation}
where $I$ denotes the number of possible $Q_g$ levels, $J$ denotes
the number of possible $Q_c$ levels, $L$ denotes the number of
tested 3DPCs, $MOS^c_{ijl}$ denotes the $MOS^c$ value of the $i$-th
$Q_g$ level ($i=1,2,...,I$) and $j$-th $Q_c$ level ($j=1,2,...,J$)
for the $l$-th 3DPC ($l=1,2,...,L$), $MOS^c_{i..}$ denotes the
$MOS^c$ value of the $i$-th $Q_g$ level with all the possible $Q_c$
levels for all the 3DPCs, $MOS^c_{.j.}$ denotes the $MOS^c$ value of
the $j$-th $Q_c$ level with all the possible $Q_g$ levels for all
the 3DPCs, $MOS^c_{ij.}$ denotes the $MOS^c$ value of the $i$-th
$Q_g$ and the $j$-th $Q_c$ level for all the 3DPCs, and
$\overline{MOS^c}$ is the mean of different combinations of $Q_g$
level, $Q_c$ level, and the tested 3DPCs. The degree of freedom of
the test factor group $DF_t \in \left\{DF_{Q_g}, DF_{Q_c}, DF_{Q_g
\cdot Q_c}\right\}$ and the values of $DF_{Q_g}, DF_{Q_c},$ and
$DF_{Q_g \cdot Q_c}$ are $I-1$, $J-1$, and $(I-1)(J-1)$,
respectively. Finally, $DF_{e}= I J (L-1)$.
Through~\eqref{eq:fvalue}, we can calculate the corresponding $F$
values, i.e., the $MOS^c$ variations over $Q_g \cdot Q_c$, $Q_g$,
and $Q_c$, as shown in Table~\ref{tab:anova}.
\begin{table}[t!]
\centering \caption{Two-way ANOVA on $MOS^c$} \label{tab:anova}
  \begin{tabular}{cccc}
      \toprule
      \midrule
      Factors &$Q_g$    &$Q_c$     &$Q_g \cdot Q_c$ \\\hline
      $F$-value &226.802  &197.838   &4.660 \\\hline
      \bottomrule
  \end{tabular}
\end{table}
The larger the $F$-value is, the more significant the corresponding
parameter is. From Table~\ref{tab:anova}, we can see that the
statistical significance of $Q_g \cdot Q_c$ is much smaller than
that of $Q_g$ and $Q_c$.
Therefore,~\eqref{eq:mos100com} is further simplified to
\begin{equation}
\label{eq:mos100sim} MOS^c =p_1 Q_g + p_2 Q_c +p_3,
\end{equation}
where $p_1$, $p_2$, and $p_3$ are model parameters. By
using~\eqref{eq:mos100sim}, the SCC between the fitted $MOS^c$ and
the actual one is up to 0.949. The model parameters $p_1$, $p_2$,
and $p_3$ in~\eqref{eq:mos100sim}, the SCCs, and the RMSEs between
the actual $MOS^c$s and the fitted values of all the evaluated 3DPCs
are given in Table~\ref{tab:MOS100sequencesillustrate}.
\begin{table}[t!]
\centering \caption{Parameters and accuracy of the perceptual
quality model} \label{tab:MOS100sequencesillustrate}
  \begin{tabular}{cccccc}
      \toprule
      \midrule
      Point Cloud  &$p_1$    &$p_2$    &$p_3$    &SCC    &RMSE \\\hline
      \emph{Bag}  &0.223 &0.183 &6.342 &0.949 &4.954 \\
      \emph{Banana}  &0.247 &0.080 &23.601 &0.902 &6.336  \\
      \emph{Biscuits}  &0.143 &0.156 &12.072 &0.927 &4.387  \\
      \emph{Cake}  &0.241 &0.125 &10.489 &0.938 &5.153 \\
      \emph{Cauliflower}  &0.246 &0.177 &9.773 &0.916 &6.782 \\
      \emph{Flowerpot}  &0.291 &0.075 &16.212 &0.877 &8.339 \\\
      \emph{House}  &0.220 &0.269 &3.597 &0.930 &7.059 \\
      \emph{Litchi}  &0.195 &0.266 &3.874 &0.914 &7.488 \\
      \emph{Mushroom}  &0.164 &0.225 &18.579 &0.890 &7.262 \\
      \emph{Ping-pong\_bat}  &0.240 &0.221 &14.240 &0.872 &9.243 \\
      \emph{Puer\_tea}  &0.124 &0.297 &11.921 &0.948 &5.568 \\
      \emph{Pumpkin}  &0.131 &0.223 &7.424 &0.939 &4.898 \\
      \emph{Ship}  &0.268 &0.068 &16.756 &0.910 &6.438 \\
      \emph{Statue}  &0.254 &0.142 &18.777 &0.852  &9.011 \\
      \emph{Stone}  &0.170 &0.291 &4.555 &0.945 &6.026 \\
      \emph{Tool\_box}  &0.117 &0.266 &15.152 &0.914 &6.630 \\\hline
      Average  &- &- &- &\textbf{0.914}    &\textbf{6.598}\\\hline
      \bottomrule
  \end{tabular}
\end{table}
We can see that the average SCC is 0.914, indicating that the
derived simplified perceptual quality model is accurate.
Fig.~\ref{fig3} illustrates the accuracy of~\eqref{eq:mos100sim}.

\section{Model parameter prediction using content features}\label{sec:4}
\begin{figure}[t!] \centering
\subfigure[]{ \label{fig4:subfig:a}
\includegraphics[width=0.35\columnwidth, height=3cm]{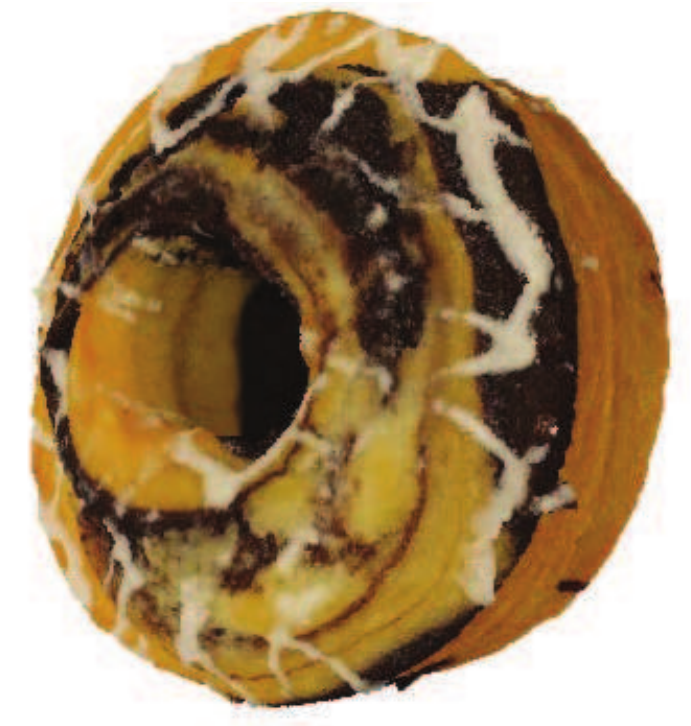}}
\subfigure[]{ \label{fig4:subfig:b}
\includegraphics[width=0.35\columnwidth, height=3cm]{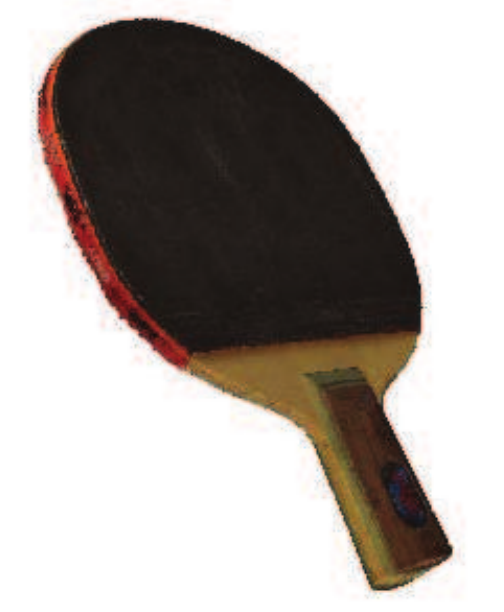}}
\subfigure[]{ \label{fig4:subfig:c}
\includegraphics[width=0.70\columnwidth]{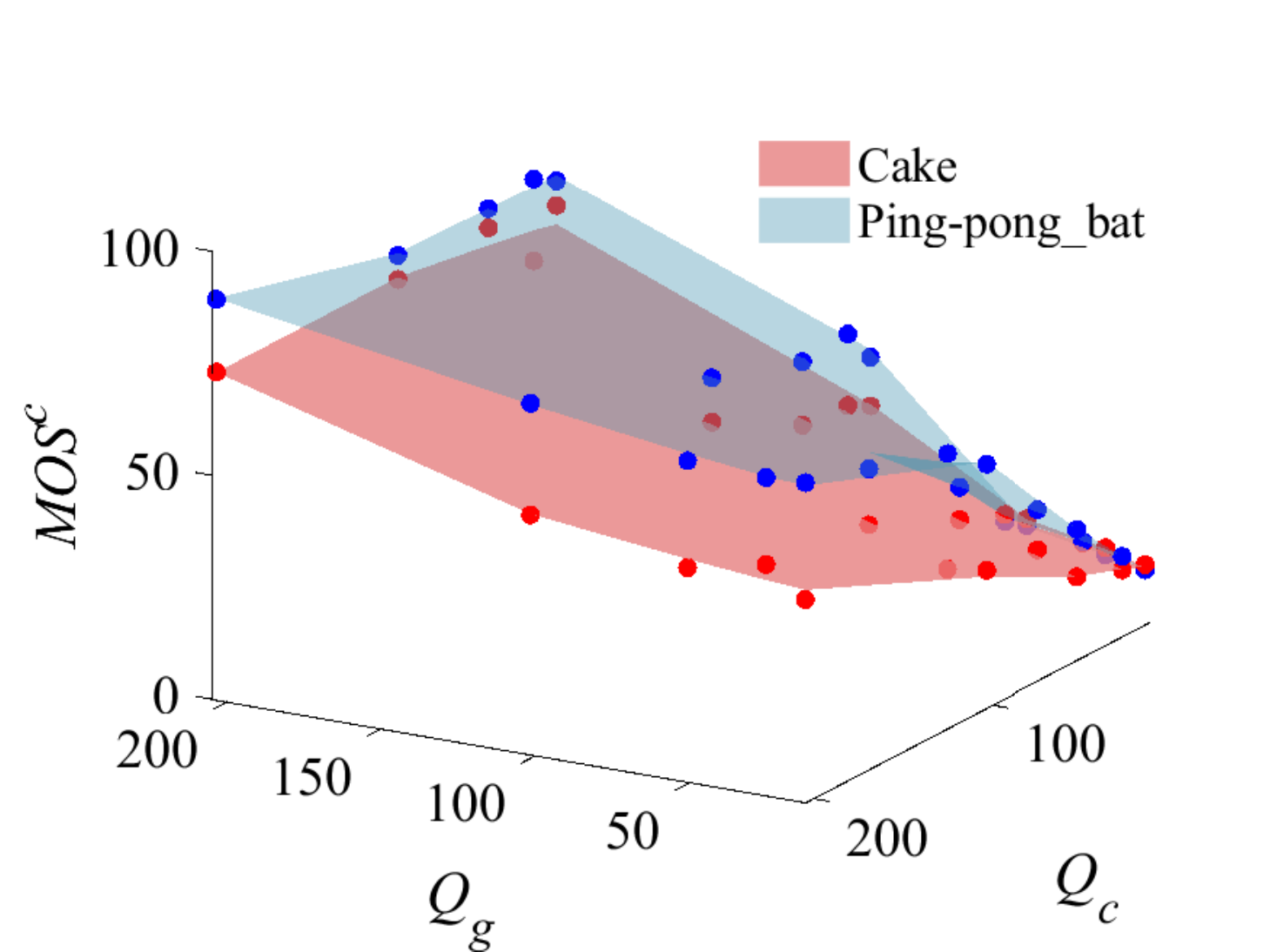}}
\caption{Relationship between content complexity and $MOS^c$. (a)
\emph{Cake} , (b) \emph{Ping-pong\_bat}, (c) mesh curve of $MOS^c$,
$Q_g$, and $Q_c$ for \emph{Cake} and \emph{Ping-pong\_bat}.}
\label{fig4}
\end{figure}

As shown in Fig.~\ref{fig4}, 3DPCs with rich texture characteristics
(e.g., \emph{Cake}) usually have lower $MOS^c$ (corresponding to
higher $MOS$) for the same quantization steps. In contrast, 3DPCs
with simple texture characteristics (e.g., \emph{Ping-pong\_bat})
have higher $MOS^c$ (corresponding to lower $MOS$) for the same
quantization steps. This is because the content has a concealing
effect on the coding distortion, which is consistent with the
characteristics of the human visual
system~\cite{karunasekera1995distortion}. That is to say, the model
parameters are highly content dependent. In this section, we propose
two features to predict the model parameters efficiently.
The perceptual quality of a 3DPC depends on both the geometry and
color distortion. But the influence of geometry and color distortion
are different~\cite{javaheri2019point}. By analyzing the local
topological and color consistencies, Alexiou and
Ebrahimi~\cite{alexiou2020towards} and Meynet~\textit{et
al.}~\cite{meynet2020pcqm} reported that color-based features
achieve the best performance in predicting the perceptual quality.
Accordingly, we extracted two novel texture features (a local
feature and a global feature) to predict the model parameters
effectively. The local feature represents the color fluctuation over
a geometric distance (CFGD), while the global feature is the color
block mean variance (CBMV).
\subsection{Color fluctuation over geometric distance (CFGD)} 
\begin{figure}[t!]
  \centering
  \includegraphics[width=0.75\columnwidth]{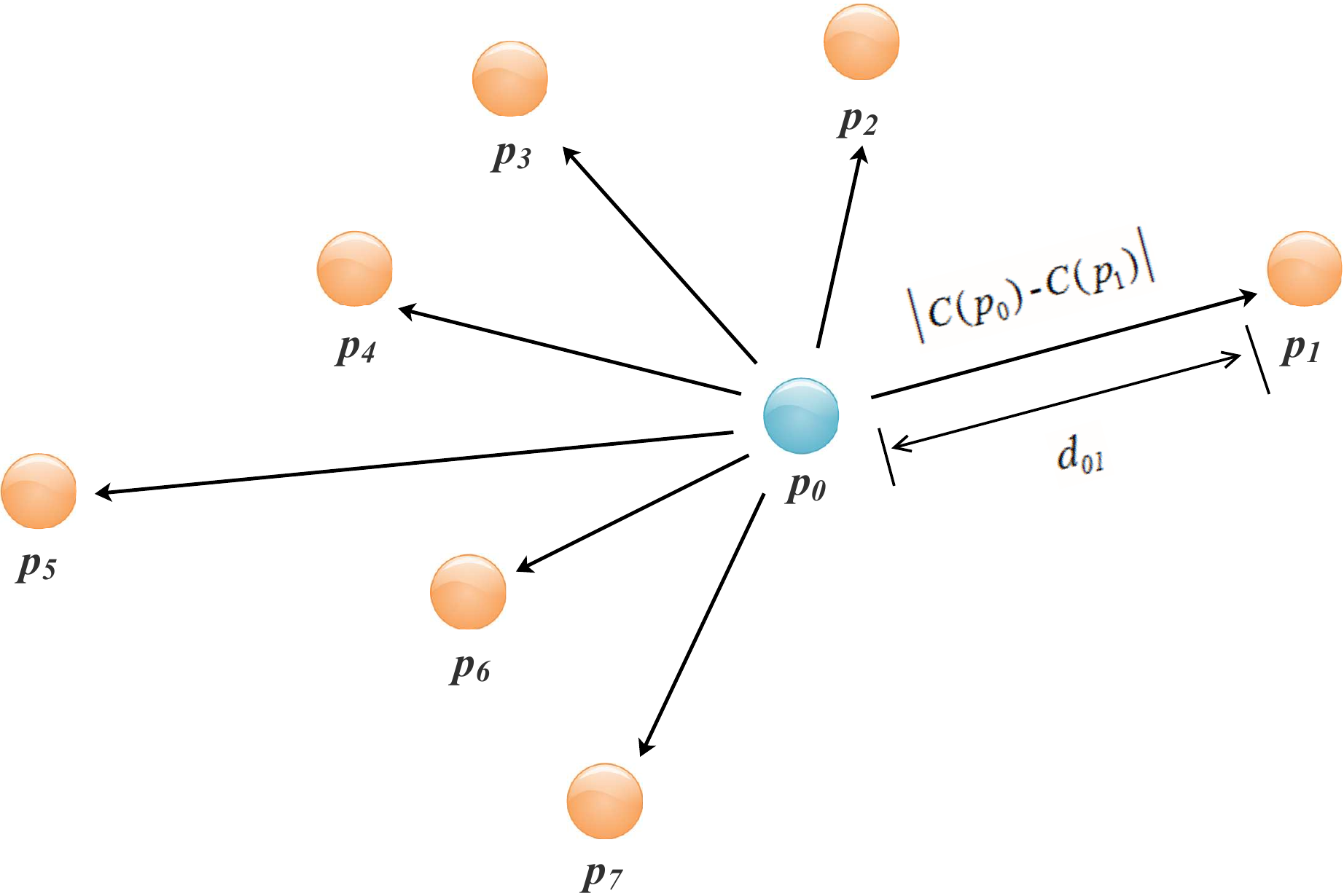}
  \caption{Illustration of the calculation of CFGD.}
  \label{CGFeature}
\end{figure}

Color gradient appropriately describes local texture variation,
therefore, we define the CFGD to describe the local content
characteristic for a 3DPC. As shown in Fig.~\ref{CGFeature}, the
mean value of the neighboring color intensity differences of the
current point is calculated to be the CFGD feature of the point:
\begin{equation}
\label{eq:CG} CFGD_{i}=\frac{1}{N_i}\sum_{p_j\in
\mathbb{S}_i}\frac{|C(p_i)-C(p_j)|}{d_{i,j}},
\end{equation}
where $CFGD_{i}$ denotes the value of CFGD for point $p_i$,
$C(\cdot)$ denotes the color attribute of a point, $d_{i,j}$ denotes
the distance between points $p_i$ and $p_j$, $\mathbb{S}_i$ is the
set of the $K$ nearest neighbors of point $p_i$, and $N_i$ is the
number of points in $\mathbb{S}_i$. For simplicity, we only consider
the $Y$ (luminance) component~\cite{mekuria2017performance} in this
paper. Then, the CFGD of all the points is defined as
\begin{equation}
\label{eq:CGtotal} CFGD=\frac{1}{T}\sum_{i\in \mathbb{P}}CFGD_{i},
\end{equation}
where $T$ is the number of points in the 3DPC $\mathbb{P}$.

\subsection{Color block mean variance (CBMV)} 
\begin{figure}[t!]
  \centering
  \includegraphics[width=0.8\columnwidth]{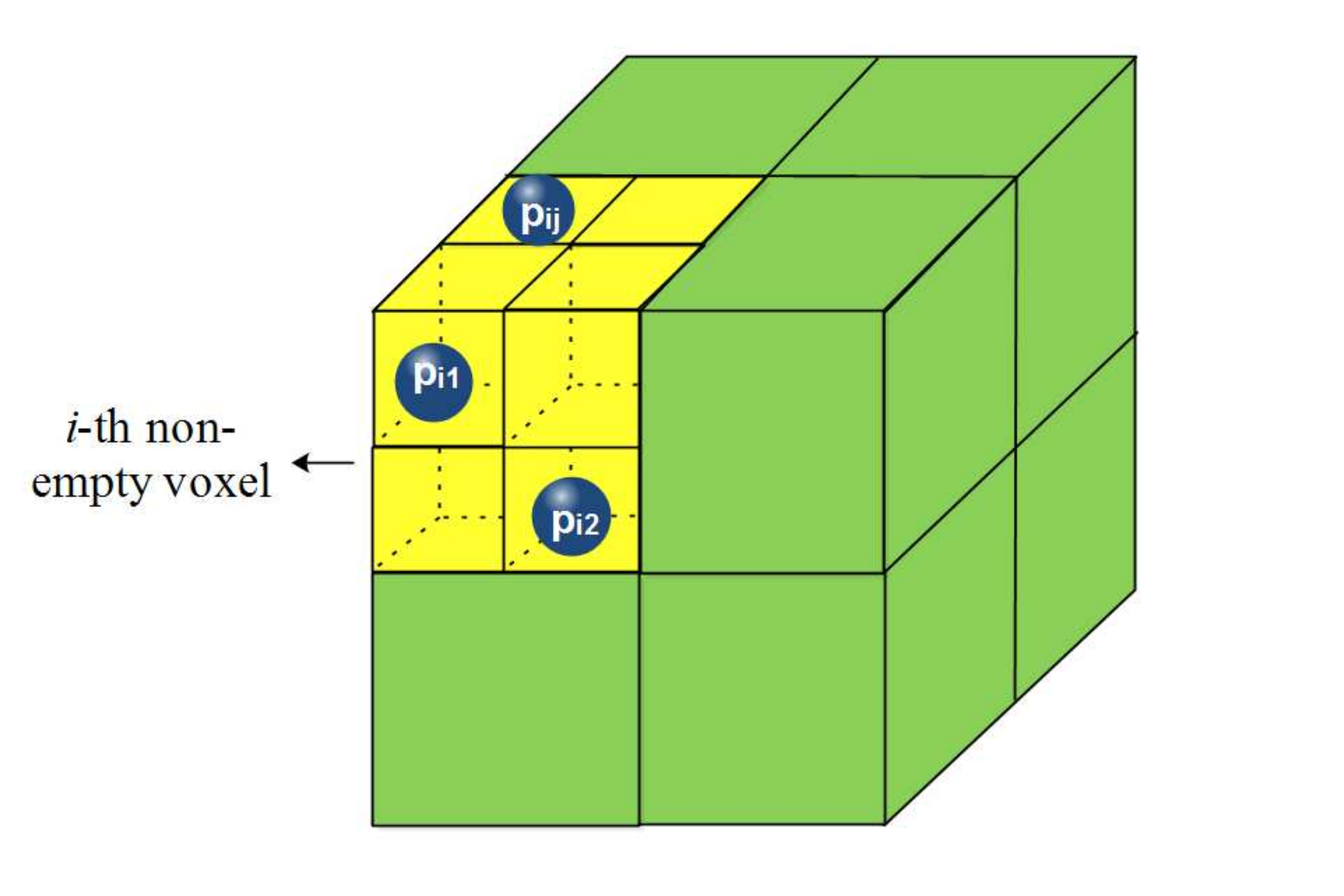}
  \caption{Voxelized 3DPC. The voxel size can be $8^3$, $16^3$, $32^3$, or $64^3$}
  \label{CDFeature}
\end{figure}

The standard deviation is commonly used as a global feature for
image/video quality
assessment~\cite{fang2014no}~\cite{rimac2009spatial}~\cite{xue2013gradient}.
Similarly, we use it to build a global feature for 3DPCs. Assuming
that the 3DPCs are voxelized~\cite{thanou2016graph}
(Fig.~\ref{CDFeature}), the CBMV is computed as
\begin{equation}
\label{eq:CD} CBMV=\frac{1}{B}\sum_{i=1}^B
\sqrt{\frac{1}{D}{\sum_{j=1}^D}(C(p_{ij})-\mu_i)^2},
\end{equation}
where $B$ denotes the number of non-empty voxels, $D$ denotes the
number of points in the $i$-th non-empty voxel, $C(p_{ij})$ is the
color of the $j$-th point in the $i$-th non-empty voxel, and $\mu_i$
is the color mean value of the $i$-th non-empty voxel.

\subsection{Model parameter estimation}
\label{sec:4-3}
We used a generalized linear model
(GLM)~\cite{dobson2018introduction} to predict the model parameters
from the extracted two features. Let $p_{m,j}$ denote the $j$-th
parameter in~\eqref{eq:mos100sim} of the $m$-th 3DPC, $m=1,2,...,M$,
where $M$ is the number of 3DPCs. Let $f_{m,k}$ denotes the value of
the $k$-th feature for the $m$-th 3DPC, $k=1,2,...,K$, and $K$ is
the number of extracted features (in this paper, $K=2$). Then, the
parameter $p_{m,j}$ is estimated by a generalized linear predictor
\begin{equation}
\label{ppredictor} p_{m,j}=h_{j,0}+\sum_{k=1}^K f_{m,k}h_{j,k},
\end{equation}
where $h_{j,k}$ is the weight of the $j$-th parameter in
~\eqref{eq:mos100sim} of the $k$-th feature, $j=$1, 2, and 3. The
$h_{j,0}$ is the constant weight of the $j$-th parameter. The
generalized linear predictor can be described using the vector form
$\bf{\hat{P}_m}=\bf{F_m}\bf{H}$, where $\bf{\hat{P}_m}$ is a
three-dimensional vector, representing the model parameters $[p_1,
p_2, p_3]$ in~\eqref{eq:mos100sim} of the $m$-th 3DPC, and
$\bf{F_m}$=$[1, f_{m,1},f_{m,2}]$, where $f_{m,1}$ and $f_{m,2}$
represent the two feature values of the $m$-th 3DPC. $\bf{H}$ is a
3$\times(K+1)$ coefficients matrix with elements $h_{j,k}$. The aim
is to find a matrix $\bf{H}$ that minimizes the prediction error
$\varepsilon$.

\begin{figure}[t!]
  \centering
  \includegraphics[width=0.8\columnwidth]{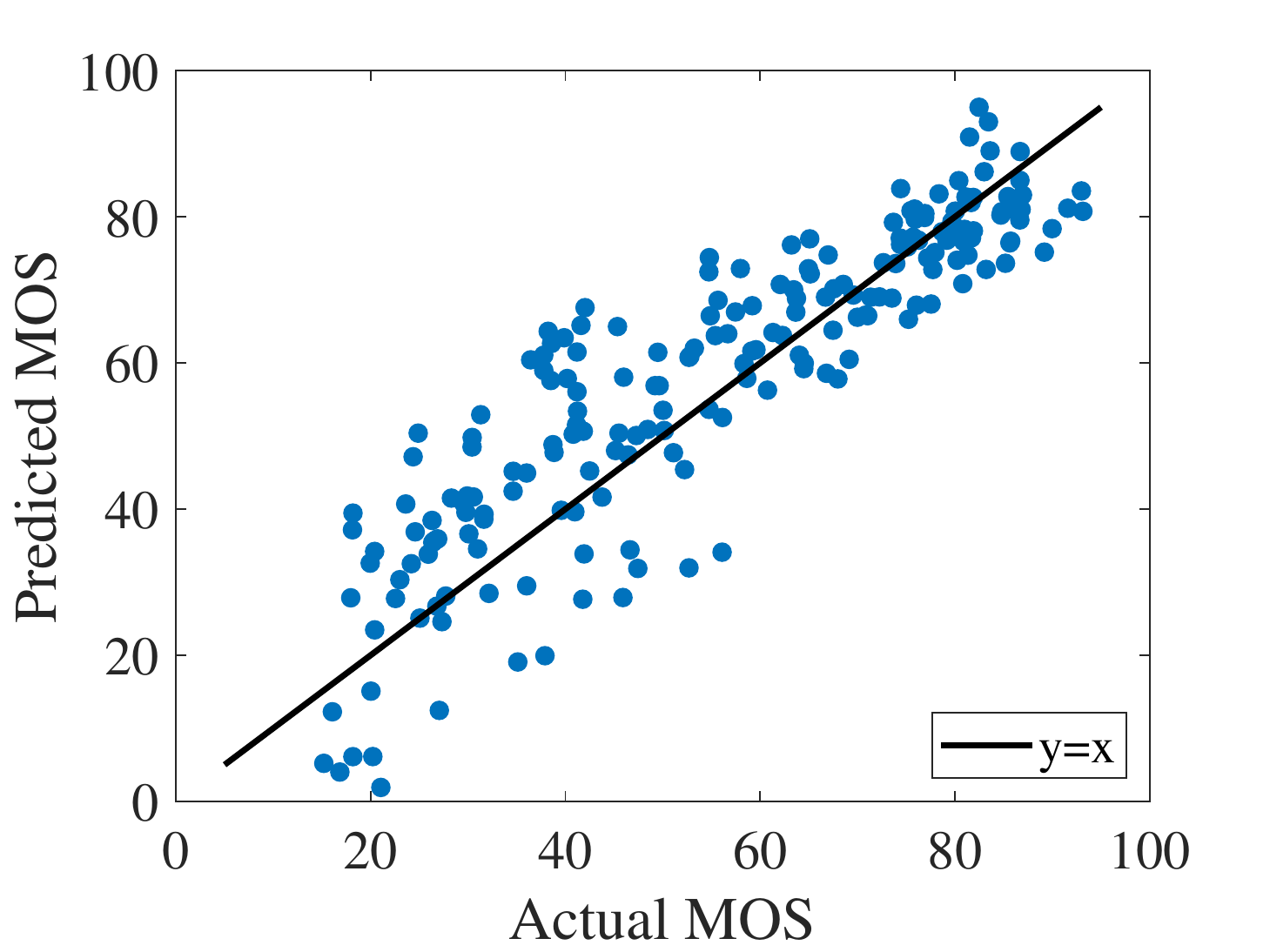}
  \caption{Scatter plot of the actual MOS and the MOS predicted by the proposed quality model for the test set.}
  \label{mosandpredictmos}
\end{figure}

\begin{figure*}[t!] \centering
\subfigure[]{ \label{comparedFR:subfig:a}
\includegraphics[width=0.6\columnwidth]{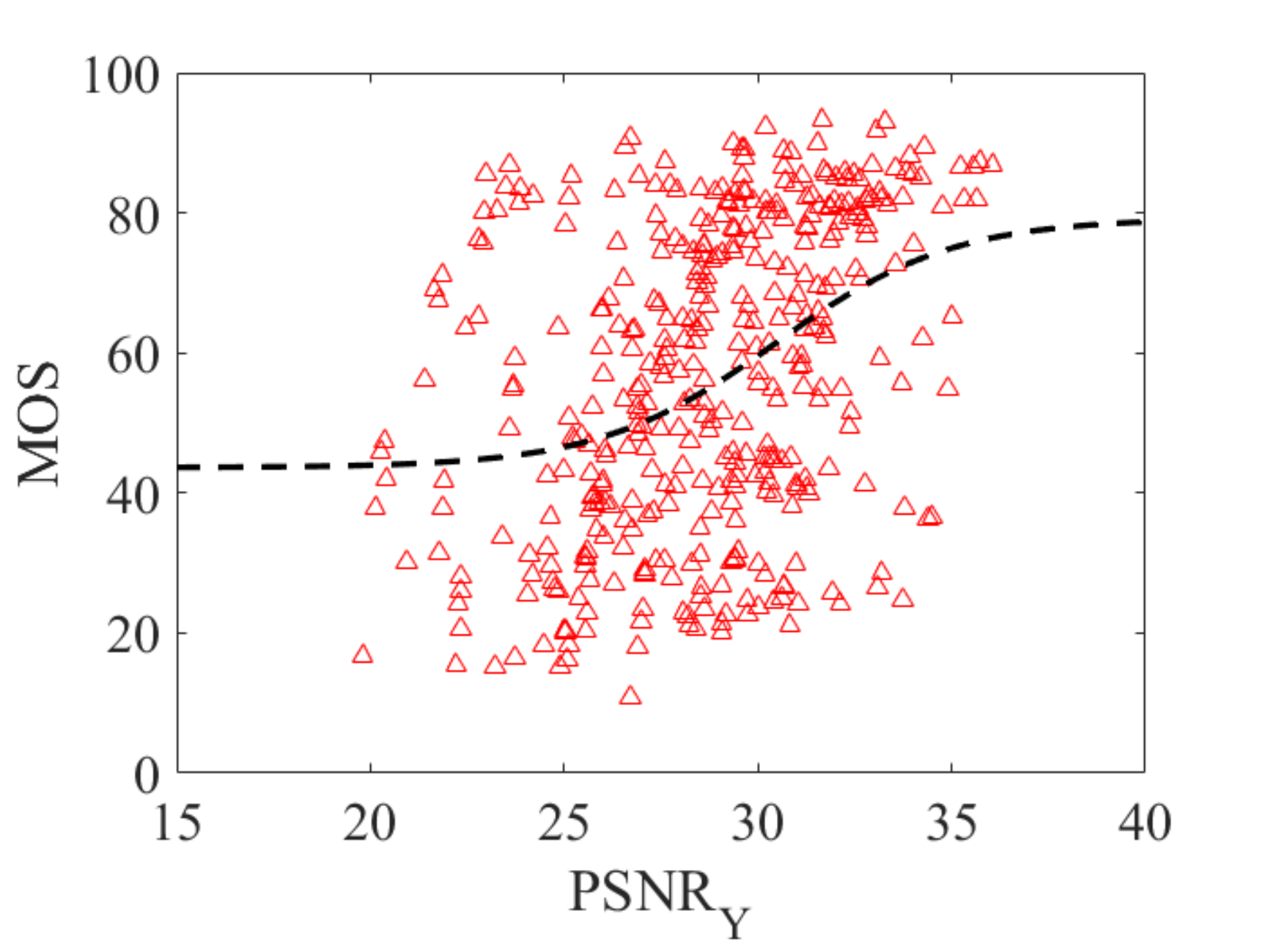}}
\subfigure[]{ \label{comparedFR:subfig:b}
\includegraphics[width=0.6\columnwidth]{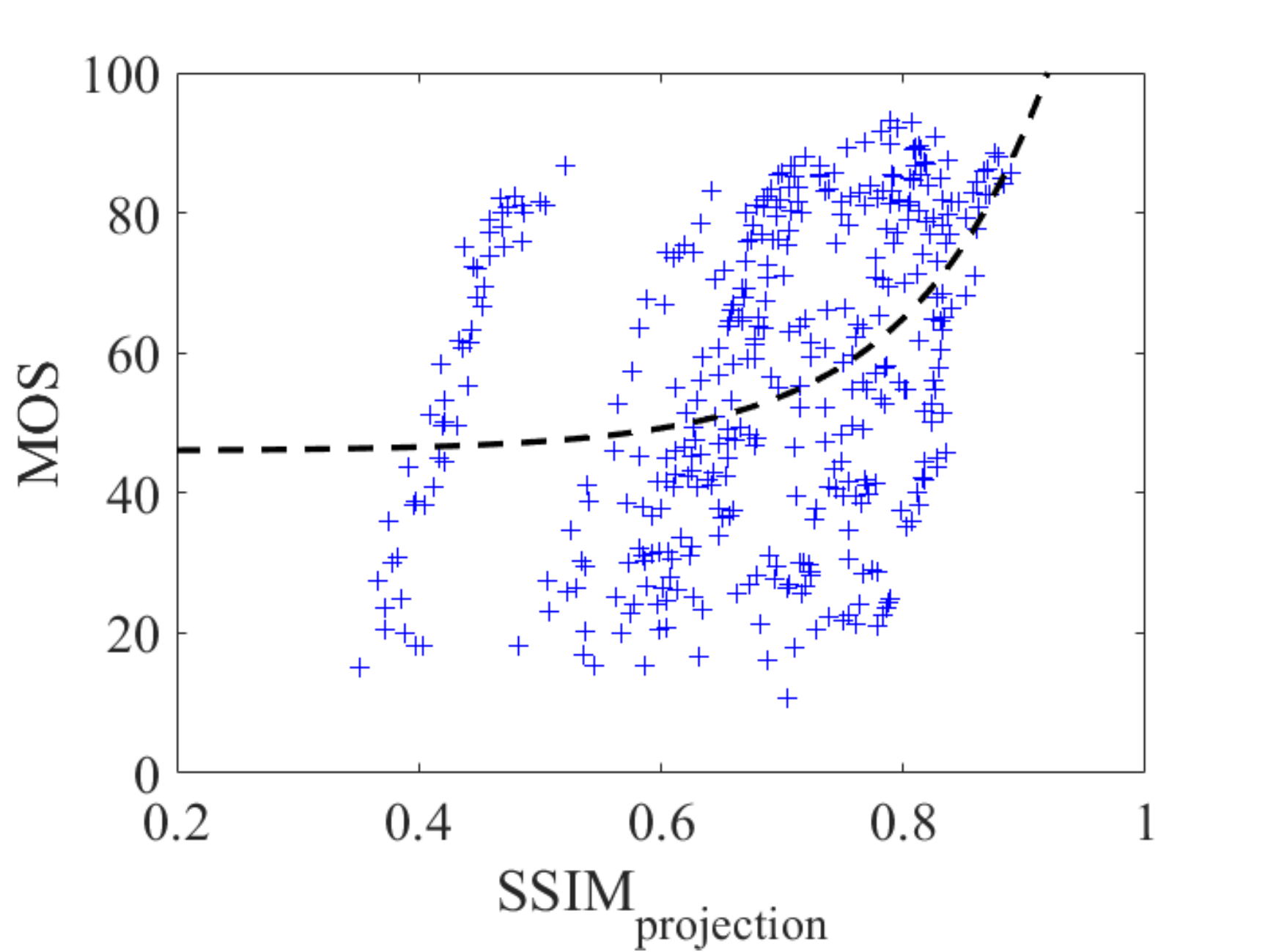}}
\subfigure[]{ \label{comparedFR:subfig:c}
\includegraphics[width=0.6\columnwidth]{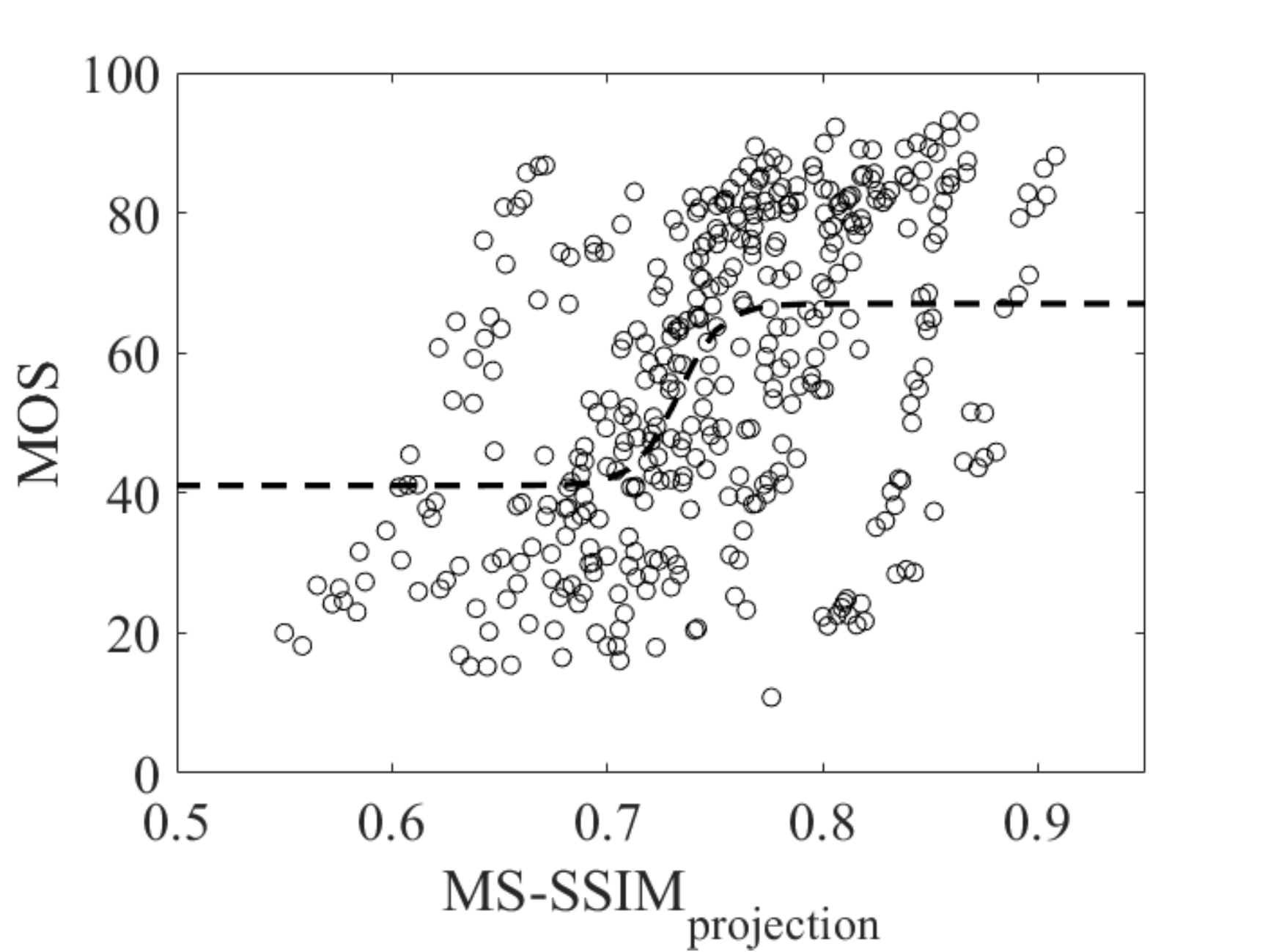}}
\subfigure[]{ \label{comparedFR:subfig:d}
\includegraphics[width=0.6\columnwidth]{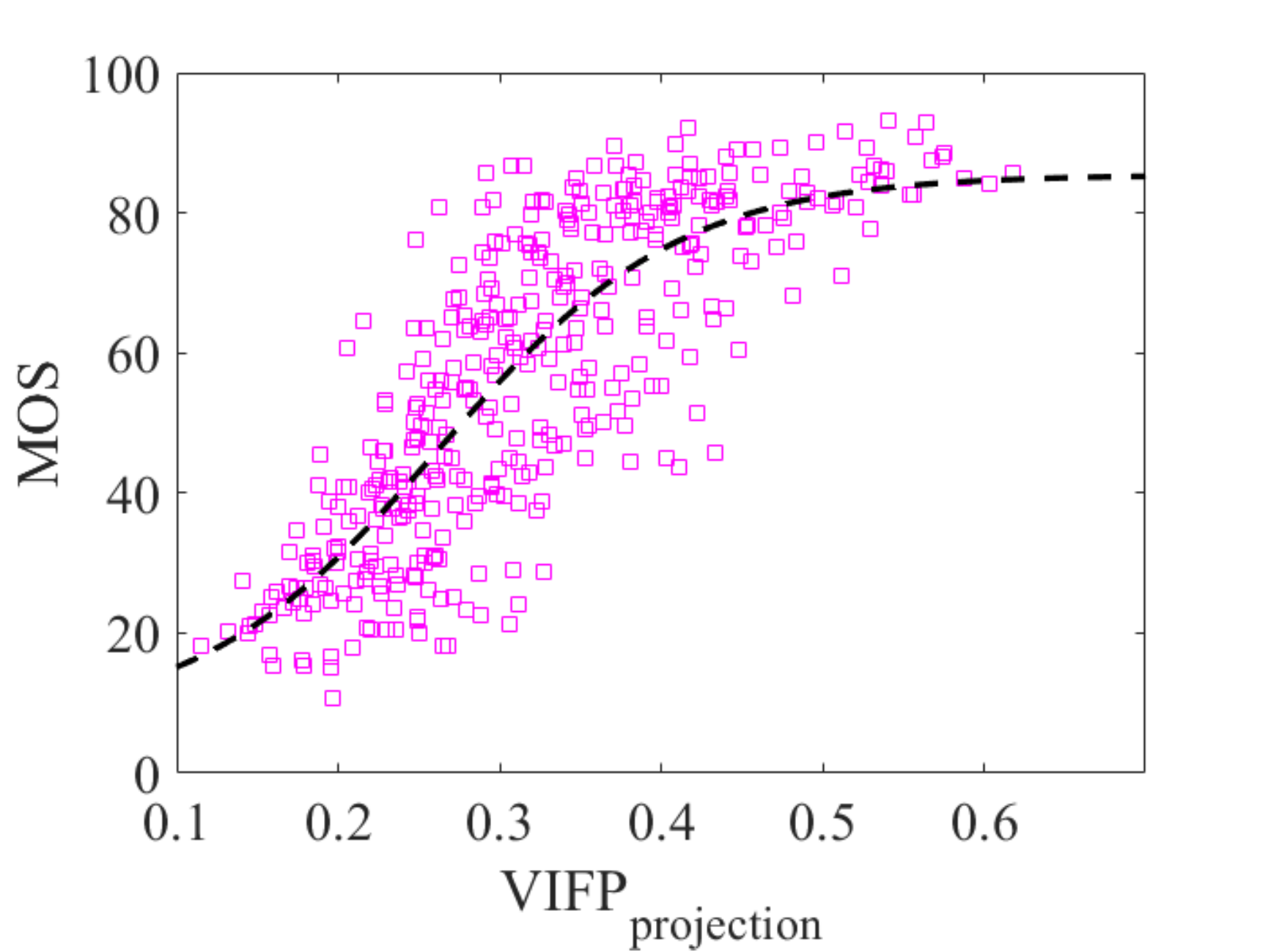}}
\subfigure[]{ \label{comparedFR:subfig:e}
\includegraphics[width=0.6\columnwidth]{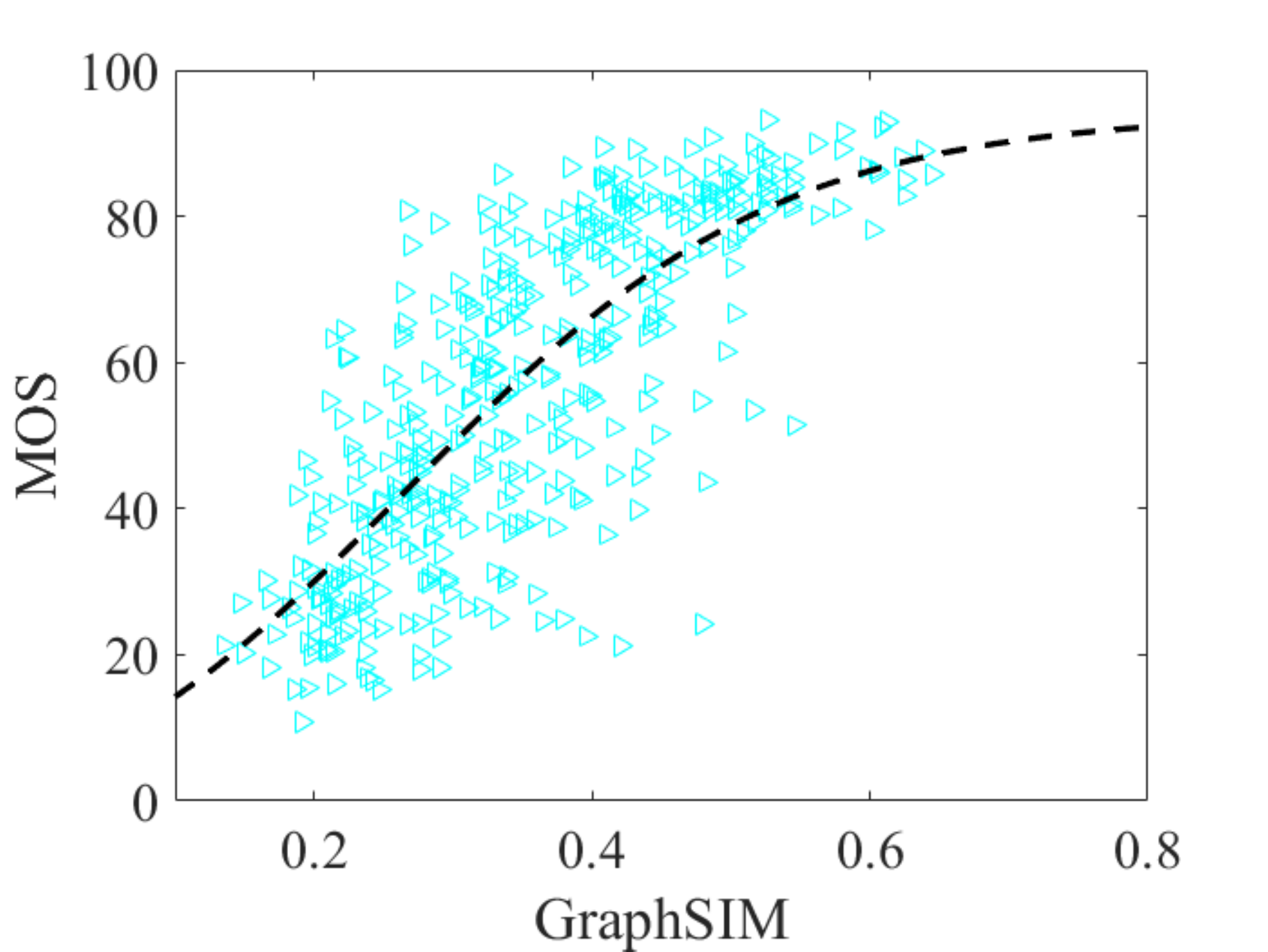}}
\subfigure[]{ \label{comparedFR:subfig:f}
\includegraphics[width=0.6\columnwidth]{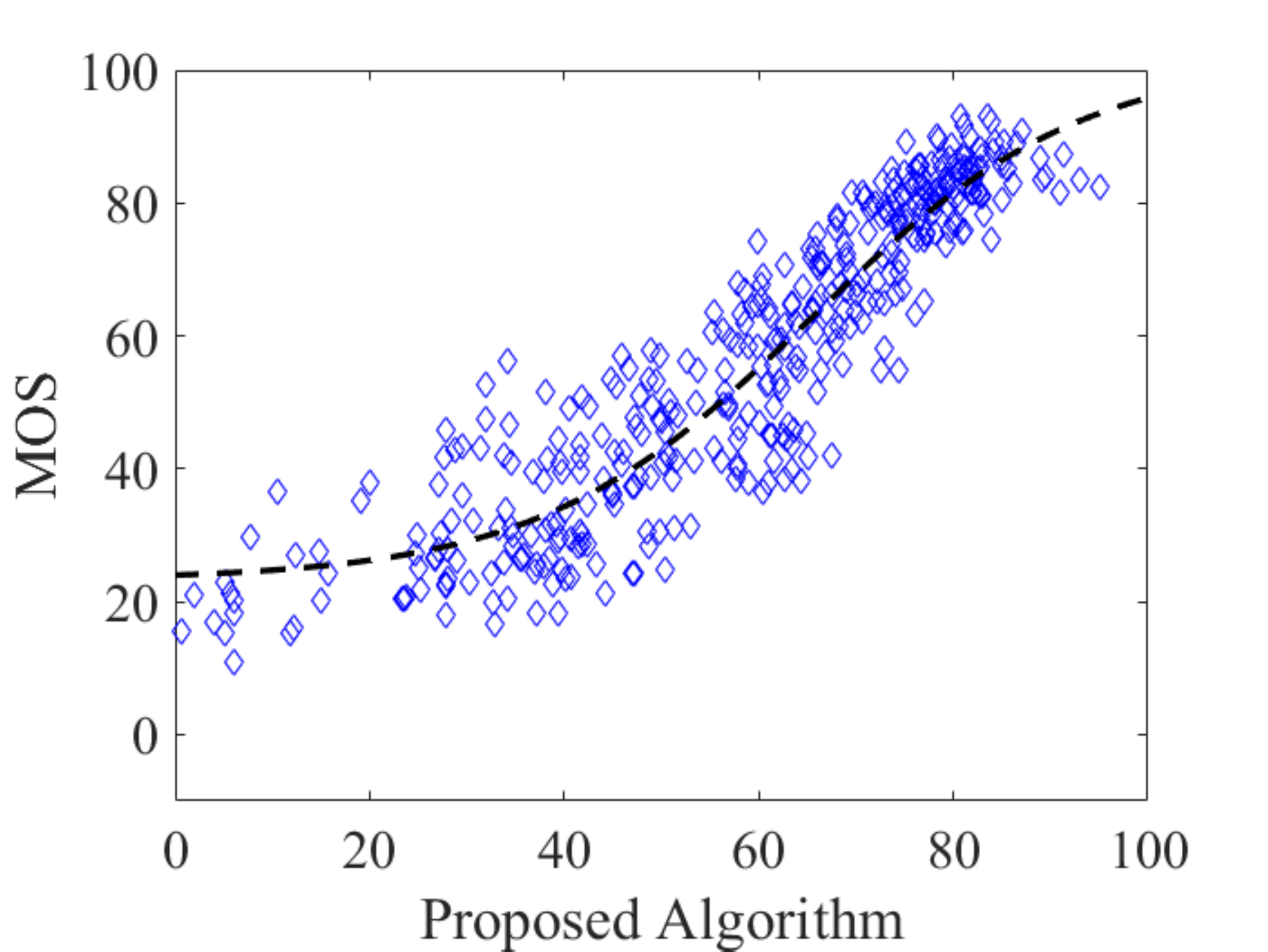}}
\caption{Scatter plots of objective score vs. MOS. The dashed curves
correspond to the best fitting logistic functions.}
\label{comparedFR}
\end{figure*}

In this paper, $\bf{H}$ is obtained by training, and we set the
voxel size equal to $64^3$ as an example for the CBMV. Eight 3DPCs:
$Cauliflower$, $Stone$, $House$, $Ship$, $Tool\_box$, $Pumpkin$,
$Biscuits$ and $Ping-pong\_bat$ that cover a wide range of content
characteristics were used for training. The remaining 3DPCs, i.e.,
$Litchi$, $Puer\_tea$, $Flowerpot$, $Bag$, $Cake$, $Statue$,
$Banana$ and $Mushroom$ were used for testing. We determined the
optimal $\bf{H}$ by minimizing the fitting error $\varepsilon$ for
the training 3DPCs set, defined as
\begin{equation}
\varepsilon=\sum_{m=1}^8 {\lVert \bf{\hat{P}_m}-\bf{P_m} \rVert}^2.
\end{equation}
where $\bf{\hat{P}_m}$ and $\bf{P_m}$ are the predicted model
parameter vector and the model parameter vector of the $m$-th 3DPC,
respectively. The optimal $\bf{H}$ is then calculated to be
\begin{equation}
\label{predH} \bf{H}^{opt}=\left[
  \begin{array}{ccc}
    0.1817 & 0.2058 & 18.4528\\
    0.0034 & -0.0070 & -0.0199\\
    -0.0116 & 0.0292 &-1.5427\\
  \end{array}
\right].
\end{equation}
By using $\bf{H}^{opt}$ and the extracted feature vector $\bf{F_m}$,
the model parameter vector $\bf{\hat{P}_m}$ can be calculated
directly. Furthermore, based on the estimated model parameters, we
can obtain the $MOS^c$ through~\eqref{eq:mos100sim}. We use PLCC,
SRCC~\cite{antkowiak2000final}, and RMSE between the actual $MOS^c$s
and the predicted ones to evaluate the accuracy of the proposed
model with the estimated parameters.
Table~\ref{tab:mosmodelaccuracy} shows that the PLCC and SRCC of the
proposed perceptual quality model of the test set are as high as
0.9133 and 0.9095, respectively, and RMSE is as small as 8.9090
(noting that the maximum MOS is 100). The accuracy of the model is
also illustrated in Fig.~\ref{mosandpredictmos} which shows the
relationship between the actual MOSs and the estimated ones.
\begin{table}[t!]
\centering \caption{Performance of the perceptual quality model on
the Training and Test Sets. $V_{size}$ is the voxel size for CBMV}
\label{tab:mosmodelaccuracy}
  \begin{tabular}{ccccc}
      \toprule
      \midrule
      \cellcolor{mygray}DataSet  &\cellcolor{mygray}$V_{size}$ &\cellcolor{mygray}PLCC     &\cellcolor{mygray}SRCC     &\cellcolor{mygray}RMSE  \\\hline
      \multirow{4}*{Training Set} & $8^3$ &0.9291   &0.9358  &8.1530  \\
                                  &\cellcolor{mygray}$16^3$ &\cellcolor{mygray}0.9335   &\cellcolor{mygray}0.9377   &\cellcolor{mygray}7.9047  \\
                                  & $32^3$ &0.9369   &0.9402   &7.7078  \\
                                  &\cellcolor{mygray}$64^3$ &\cellcolor{mygray}0.9377   &\cellcolor{mygray}0.9409   &\cellcolor{mygray}7.6597
                                  \\\hline
      \multirow{4}*{Test Set}     & $8^3$  &\textbf{0.8963}   &\textbf{0.8922}   &\textbf{9.7016}  \\
                                  &\cellcolor{mygray}$16^3$ &\cellcolor{mygray}0.8998   &\cellcolor{mygray}0.8972   &\cellcolor{mygray}9.5457  \\
                                  & $32^3$ &0.9080   &0.9053   &9.1651  \\
                                  &\cellcolor{mygray}$64^3$ &\cellcolor{mygray}\textbf{0.9133}   &\cellcolor{mygray}\textbf{0.9095}
                                  &\cellcolor{mygray}\textbf{8.9090}
                                  \\\hline
      \bottomrule
  \end{tabular}
\end{table}
To further validate the accuracy of the proposed RR quality metric
model, we compared it to three the representative FR objective
metric models: a point-based model~\cite{mekuria2016evaluation}, a
projection-based
model~\cite{torlig2018novel}~\cite{sheikh2006image}~\cite{wang2004image}~\cite{wang2003multiscale},
and a graph-based model~\cite{yang2020inferring}. The point-based
method captures the difference between the points in the reference
and the tested 3DPC, and we name it as $PSNR_Y$. Currently, the
point-based method is adopted by MPEG. For the projection-based
approaches, a 3DPC is mapped onto six conventional two-dimensional
image planes by orthographic projection. After obtaining the
projected image planes, the 2D image quality metrics structural
similarity (SSIM)~\cite{wang2004image}, multi-scale structural
similarity (MS-SSIM)~\cite{wang2003multiscale}, and visual
information fidelity in pixel domain (VIFP)~\cite{sheikh2006image}
are used to evaluate the six projection image quality, finally, the
average image quality of these six projection is mapped to MOS by
the best fitting logistic function, the mapped MOS is taken as the
quality of the 3DPC. We call these projection-based methods
SSIM$_{projection}$, MS-SSIM$_{projection}$, and
VIFP$_{projection}$, respectively. For the graph-based
method~\cite{yang2020inferring}, local graphs centered at the key
points were used to calculate the similarity between the original
and the distorted 3DPC. We call this method GraphSIM.
Table~\ref{tab:comparedfullreference} shows the comparison results
with the point-based and projection-based methods. We can see that
the point-based $PSNR_Y$ model does not seem to provide enough
accuracy due to a lack of overall perception. GraphSIM improves the
prediction accuracy to some extent; however, it is more complex and
requires many parameters to be determined. In contrast, the
projection-based models perform better among which VIFP achieves the
best performance compared to PSNR, SSIM and MS-SSIM. Nevertheless,
the quality prediction accuracy is only moderate when compared with
their performance on 2D images~\cite{su2019perceptual}.
Table~\ref{tab:comparedfullreference} shows that the PLCC of FR
quality metrics is in the range 0.4027 to 0.8199. In contrast, the
PLCC of the proposed RR quality metric is as high as 0.9133. In
addition to the PLCC, the SRCC of the worst and best FR quality
metrics are 0.3926 and 0.8187 respectively, whereas the SRCC of the
proposed RR quality metric is 0.9095. Beyond that, the RMSE of the
proposed quality metric is also much smaller than those compared
metrics.
\begin{table}[t!]
\newcommand{\tabincell}[2]{\begin{tabular}{@{}#1@{}}#2\end{tabular}}
\centering \caption{Performance of point cloud quality assessment
models. } \label{tab:comparedfullreference}
  \begin{tabular}{ccccc}
      \toprule
      \midrule
      \cellcolor{mygray}Model Type       &\cellcolor{mygray}Model &\cellcolor{mygray}PLCC     &\cellcolor{mygray}SRCC
      &\cellcolor{mygray}RMSE\\\hline
      \multirow{5}*{FR} &PSNR$_Y$  &0.3956 &0.3926 &20.2058\\
                        &\cellcolor{mygray}SSIM$_{projection}$ &\cellcolor{mygray}0.4027 &\cellcolor{mygray}0.4014 &\cellcolor{mygray}20.1382\\
                        &MS-SSIM$_{projection}$ &0.5126 &0.5025 &18.8910\\
                        &\cellcolor{mygray}VIFP$_{projection}$ &\cellcolor{mygray}0.8199 &\cellcolor{mygray}0.8187 &\cellcolor{mygray}12.5964\\
                        &GraphSIM &0.7748 &0.7786 &13.9095\\\hline
      \cellcolor{mygray}RR &\cellcolor{mygray}proposed ($V_{size}=64^3$) &\cellcolor{mygray}\textbf{0.9133} &\cellcolor{mygray}\textbf{0.9095} &\cellcolor{mygray}\textbf{8.9090}\\\hline
      \bottomrule
  \end{tabular}
\end{table}
Fig.~\ref{comparedFR} shows scatter plots of MOS vs. objective
scores for all models. The plots illustrate the superiority of the
proposed RR quality metric over the other models.
\section{Application}\label{sec:5}
\begin{figure*}[t!]
\centering \subfigure[]{ \label{fig13:subfig:a}
\includegraphics[width=0.485\columnwidth]{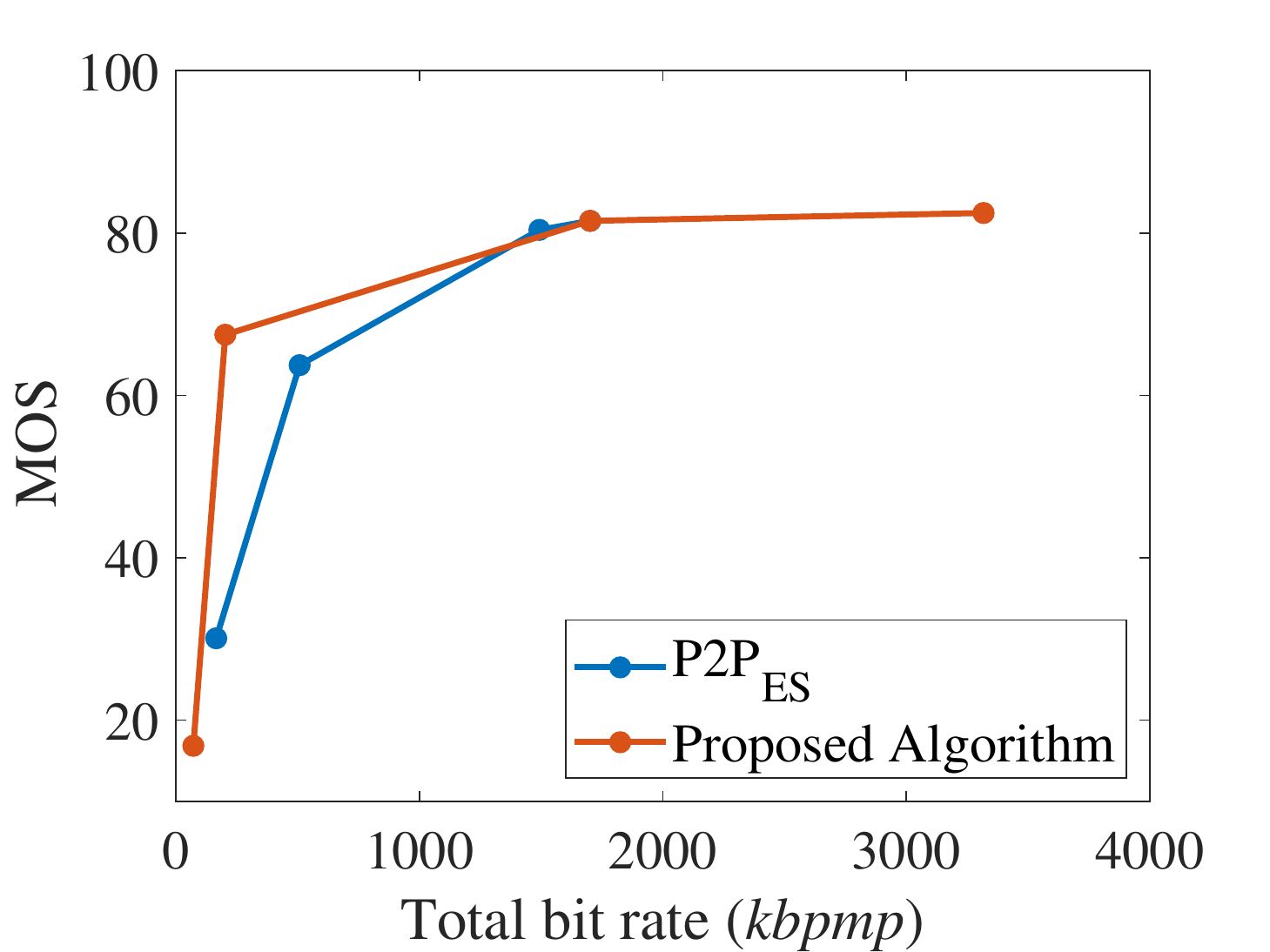}} 
\subfigure[]{ \label{fig13:subfig:b}
\includegraphics[width=0.485\columnwidth]{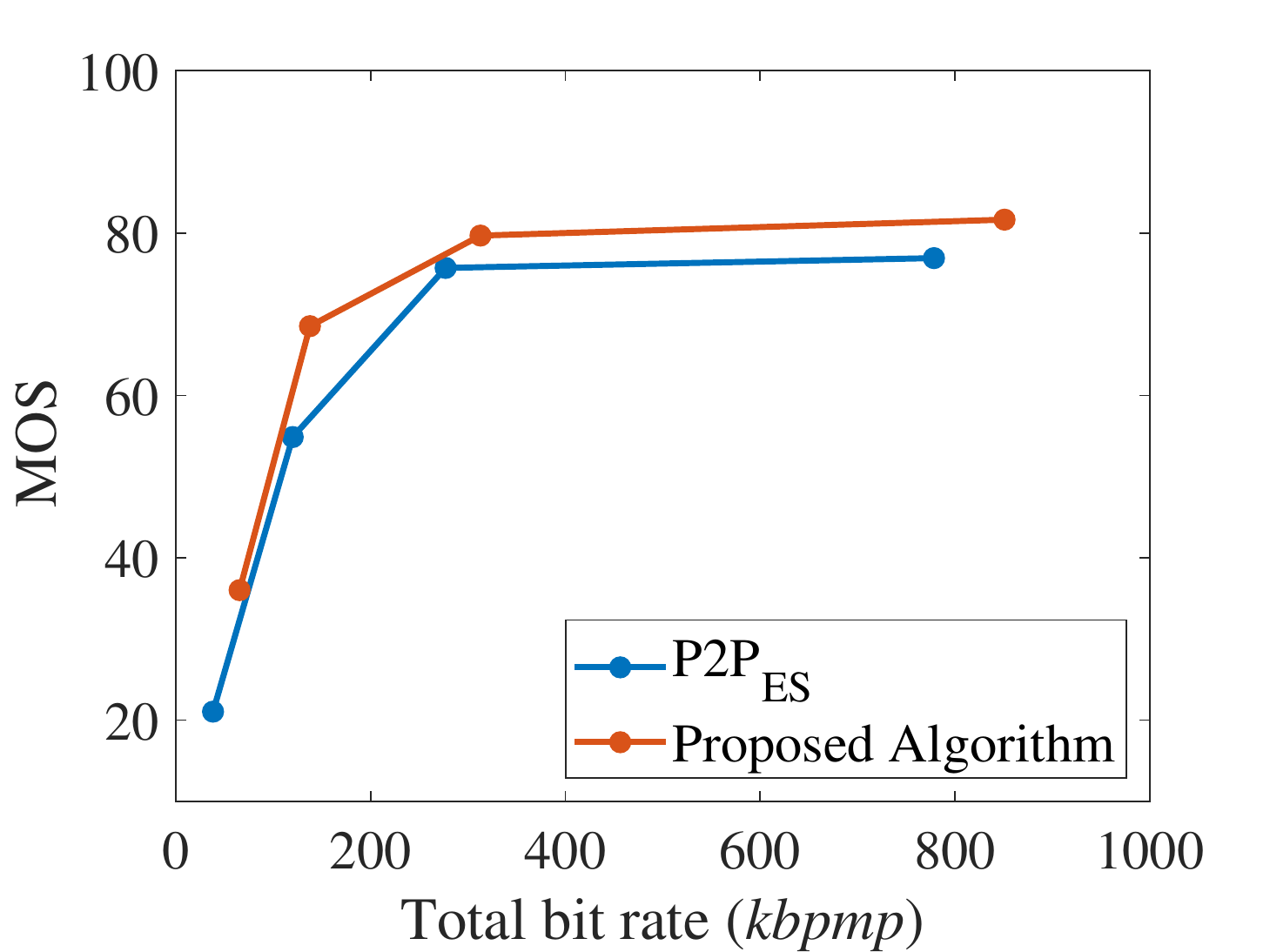}}
\subfigure[]{ \label{fig13:subfig:c}
\includegraphics[width=0.485\columnwidth]{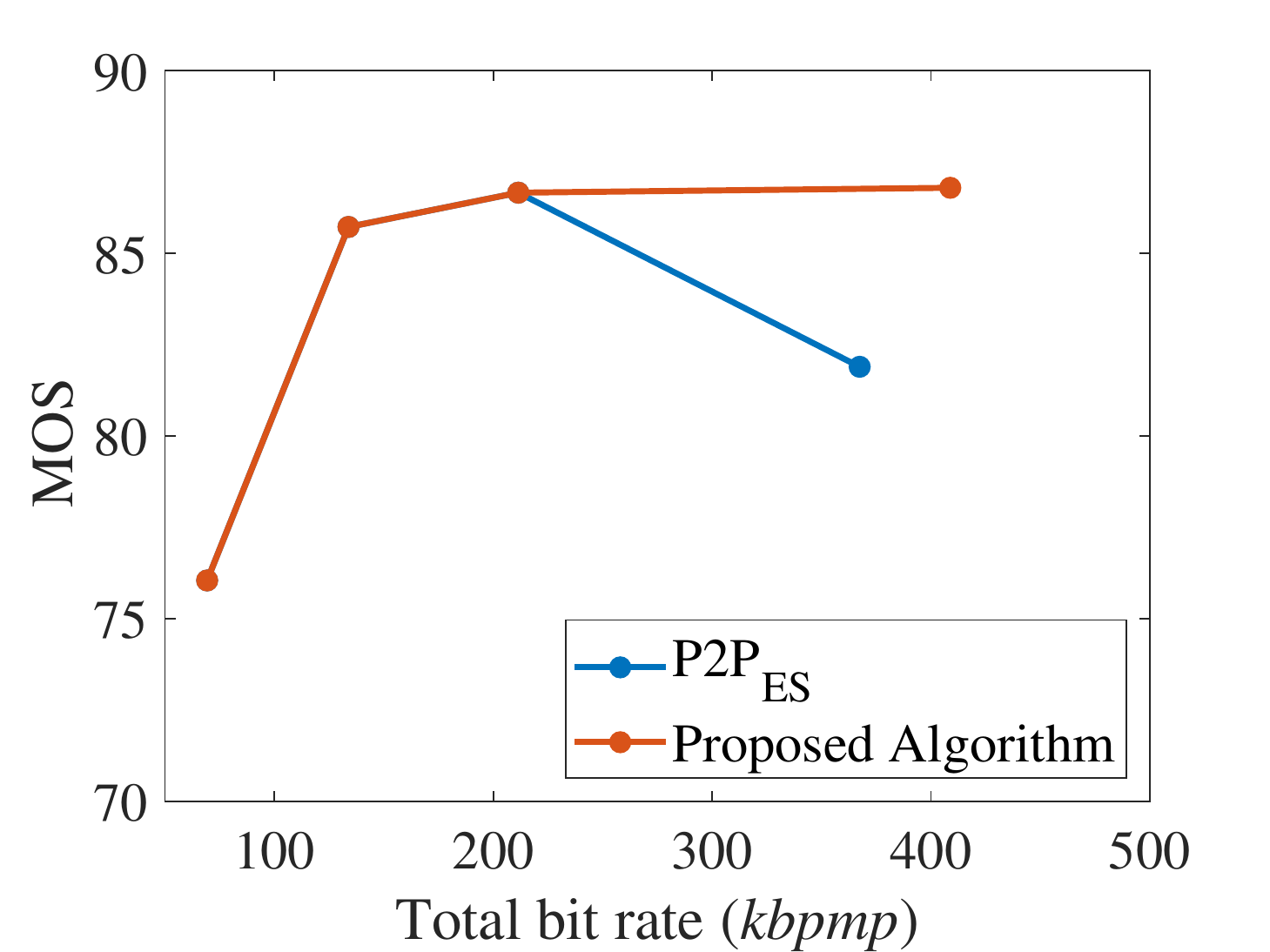}}
\subfigure[]{ \label{fig13:subfig:d}
\includegraphics[width=0.485\columnwidth]{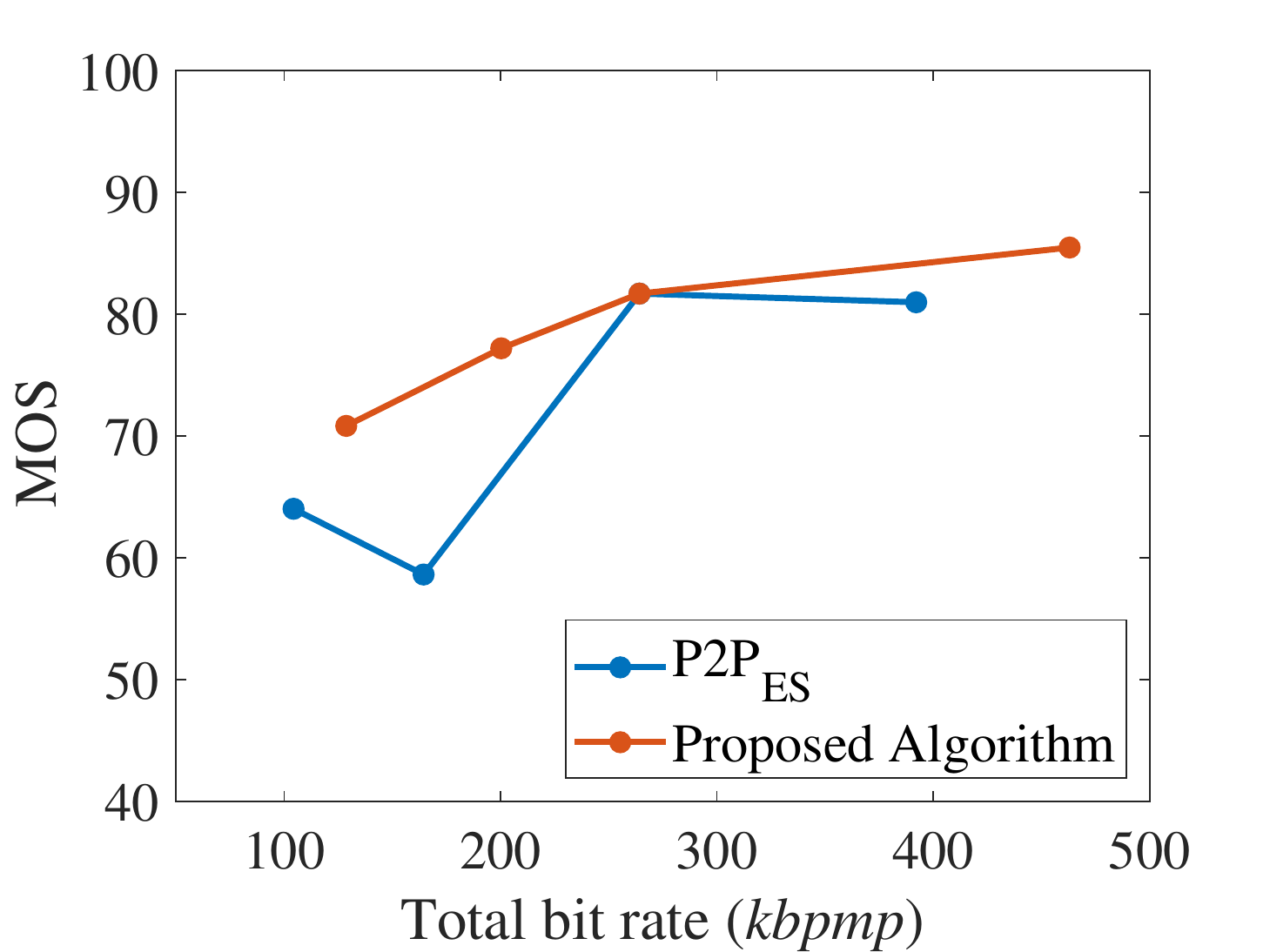}}
\subfigure[]{ \label{fig13:subfig:e}
\includegraphics[width=0.485\columnwidth]{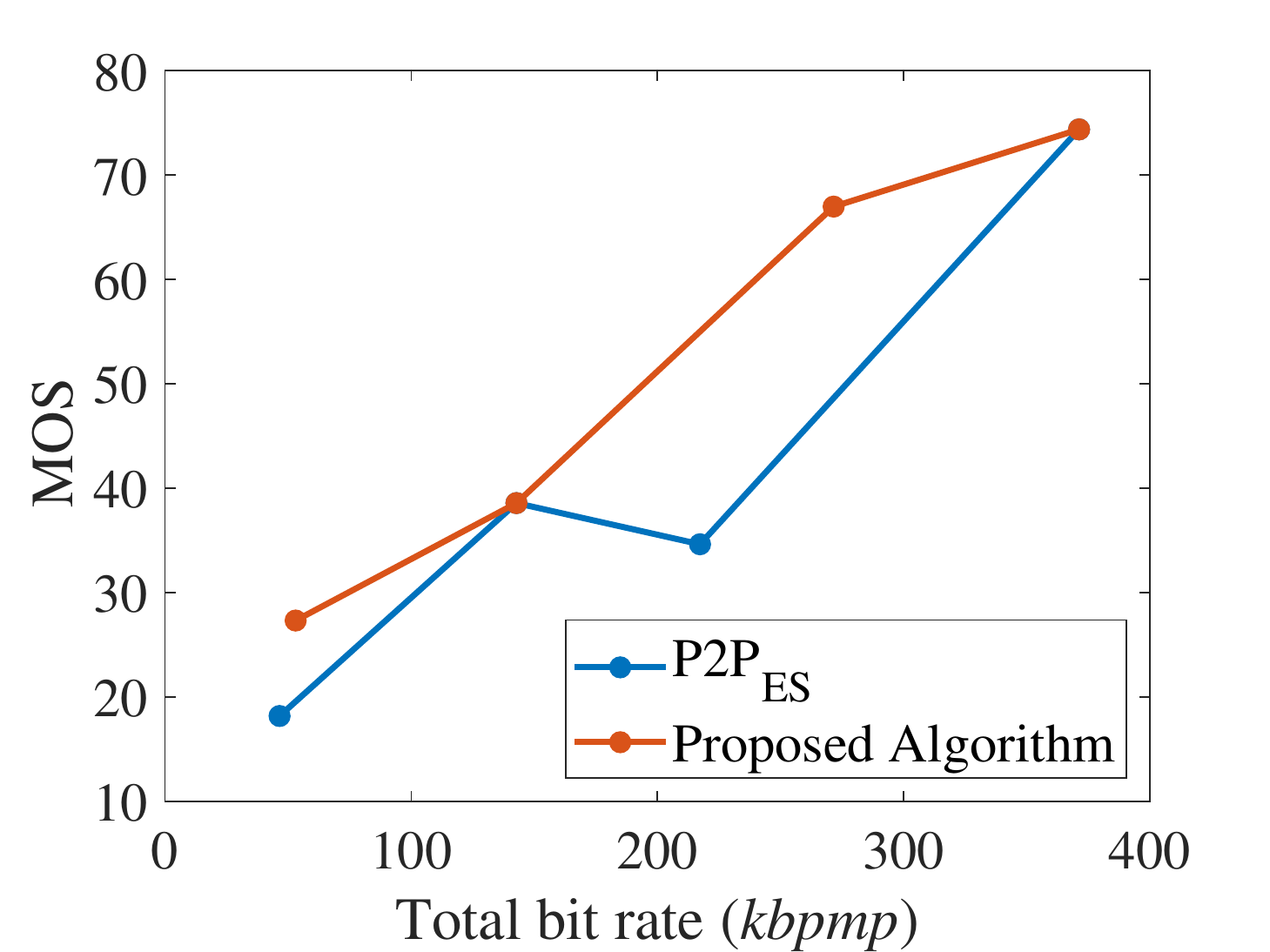}}
\subfigure[]{ \label{fig13:subfig:f}
\includegraphics[width=0.485\columnwidth]{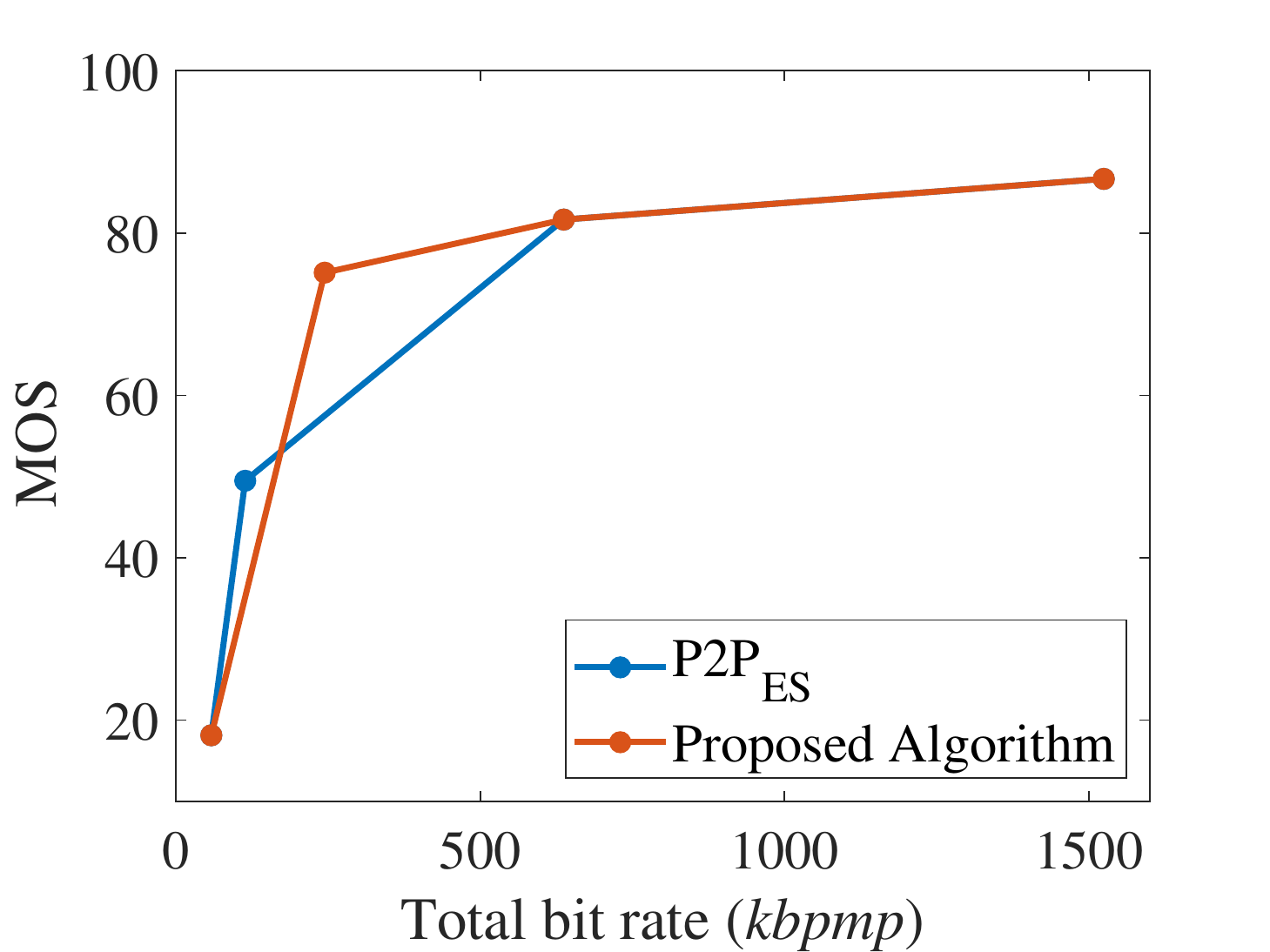}}
\subfigure[]{ \label{fig13:subfig:g}
\includegraphics[width=0.485\columnwidth]{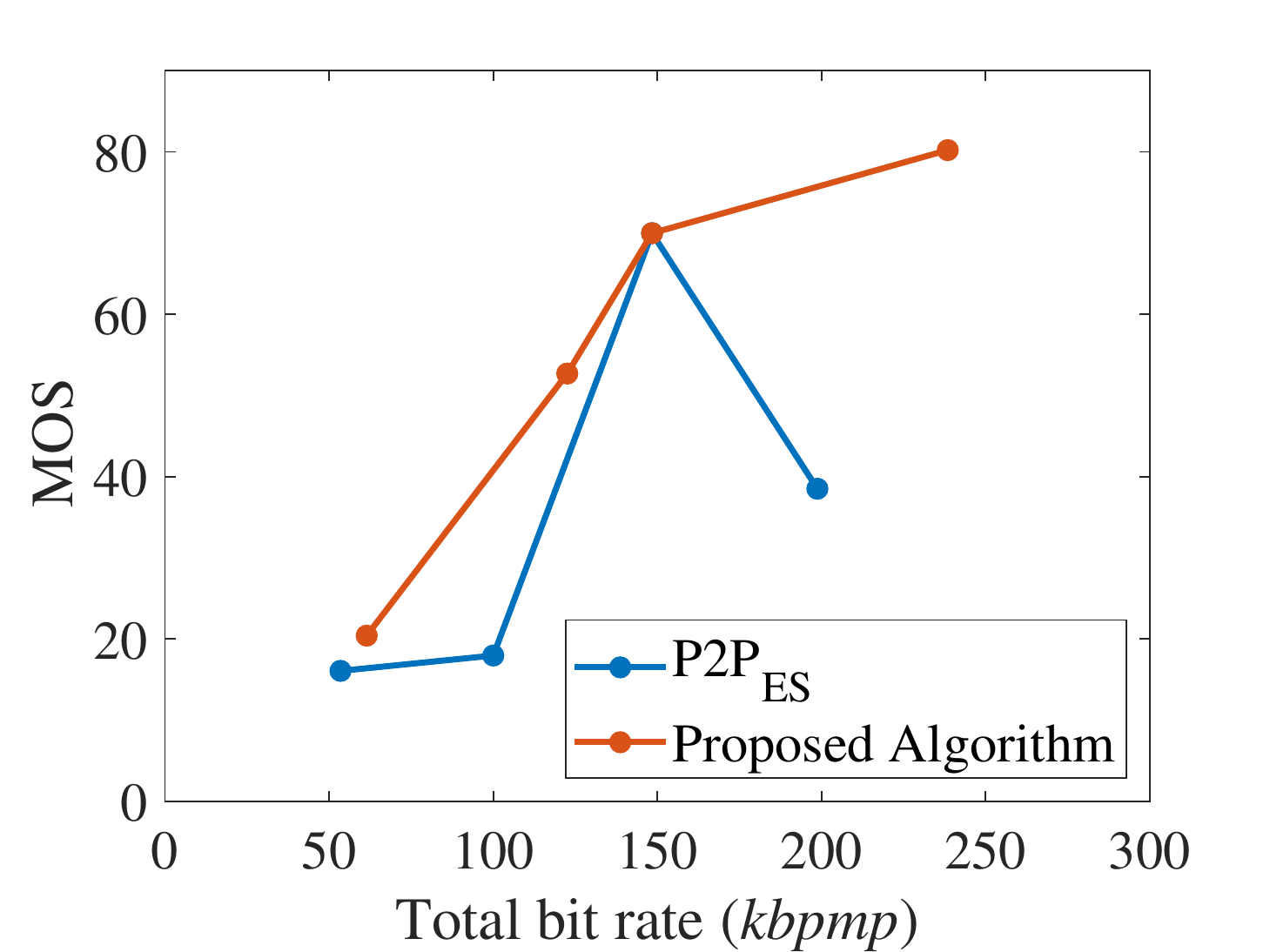}}
\subfigure[]{ \label{fig13:subfig:h}
\includegraphics[width=0.485\columnwidth]{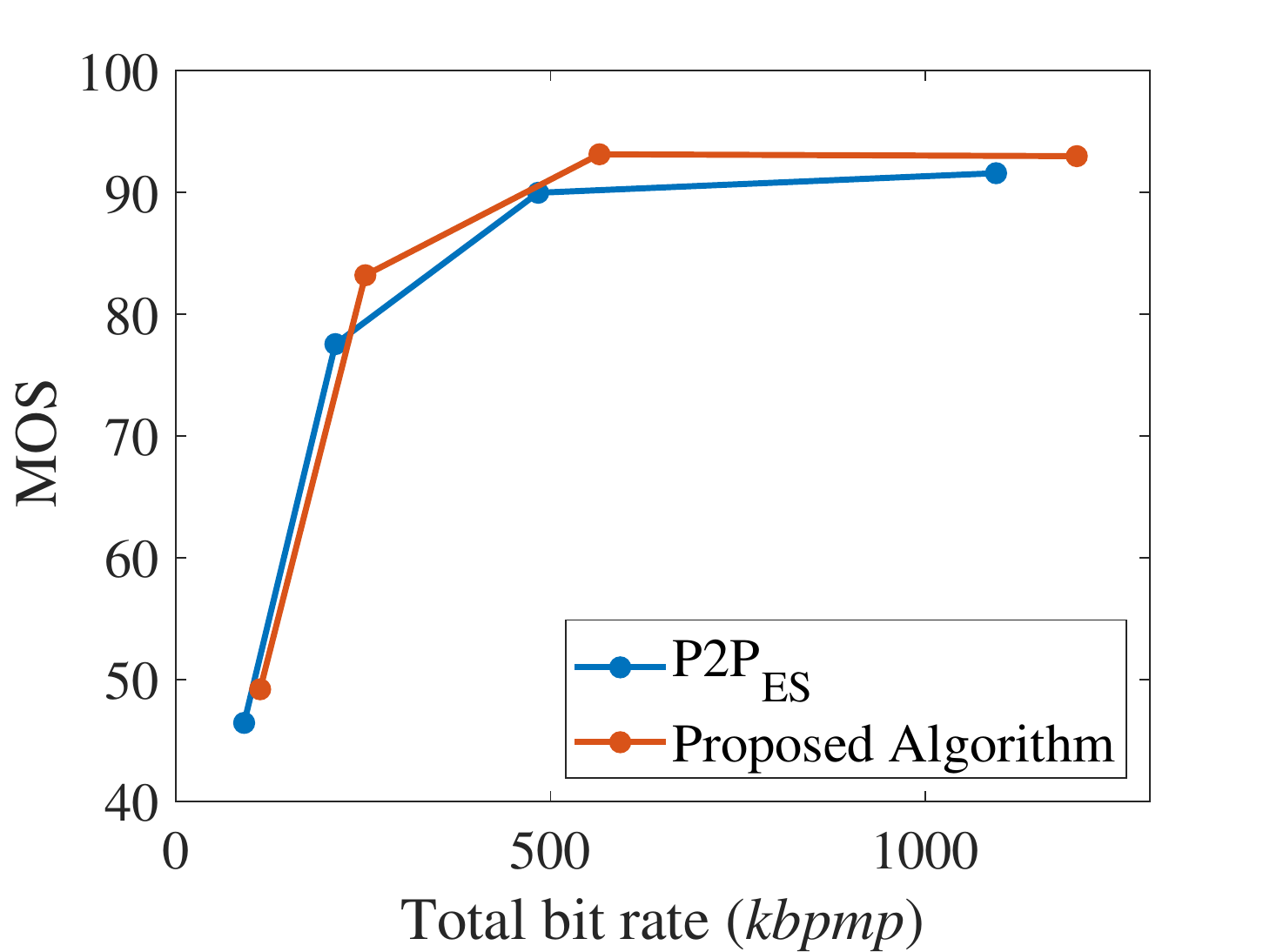}}
\caption{MOS vs. total bit rate for our rate control algorithm and
the point-to-point exhaustive search algorithm ($P2P_{ES}$). (a)
\emph{Bag}, (b) \emph{Banana}, (c) \emph{Flowerpot}, (d)
\emph{Cake}, (e) \emph{Mushroom}, (f) \emph{Puer\_tea}, (g)
\emph{Statue}, (h) \emph{Litchi}. The vertical axis shows the
average $MOS$ of all the viewers.} \label{fig13}
\end{figure*}
\begin{figure*}[t!]
\centering \subfigure[]{ \label{figsubjective:subfig:a}
\includegraphics[width=8.65cm, height=2.5cm]{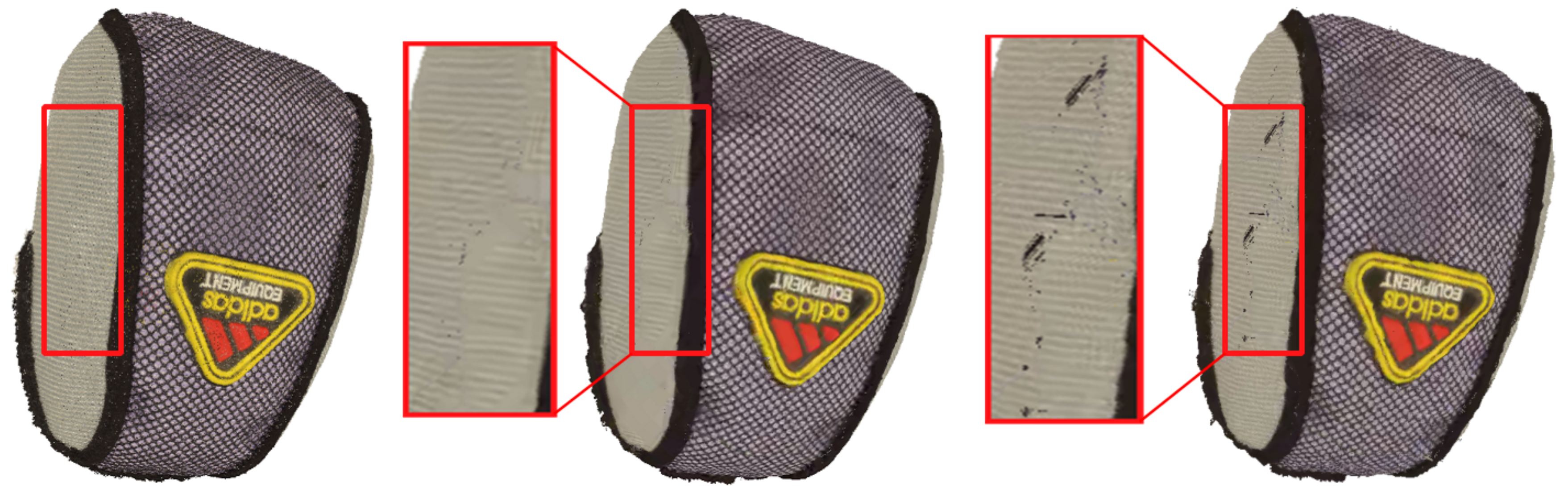}}\hspace{4mm}
\subfigure[]{ \label{ffigsubjective:subfig:b}
\includegraphics[width=8.65cm, height=2.5cm]{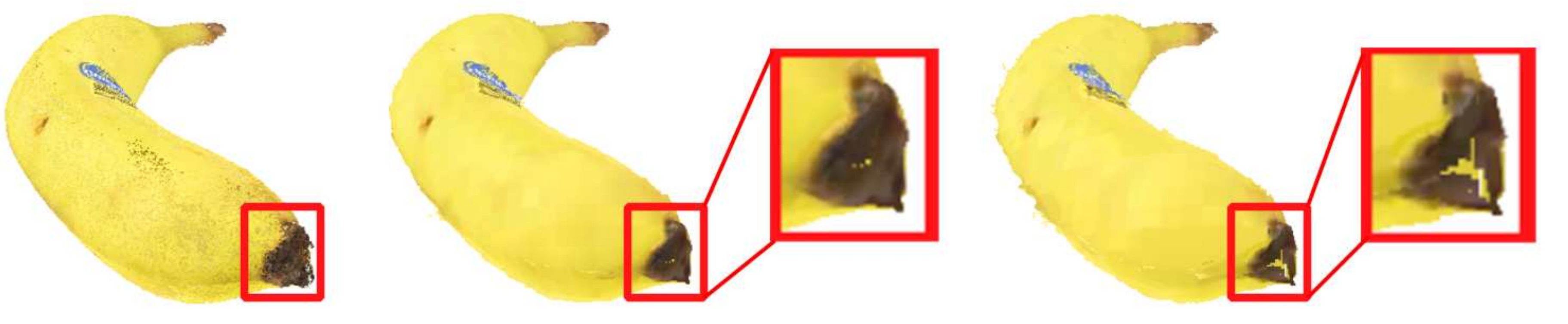}}
\subfigure[]{ \label{figsubjective:subfig:c}
\includegraphics[width=8.65cm, height=2.5cm]{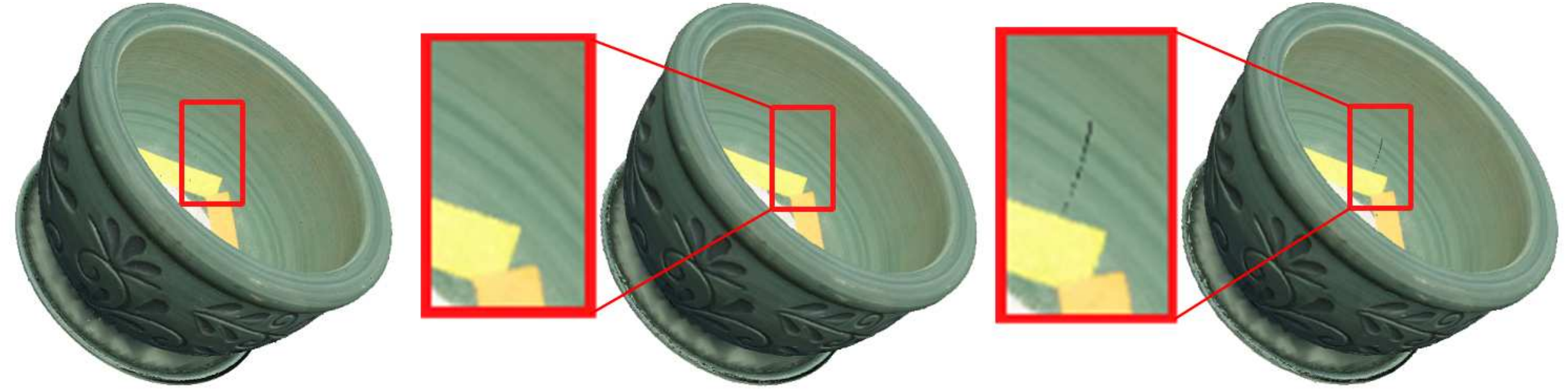}}\hspace{4mm}
\subfigure[]{ \label{figsubjective:subfig:d}
\includegraphics[width=8.65cm, height=2.5cm]{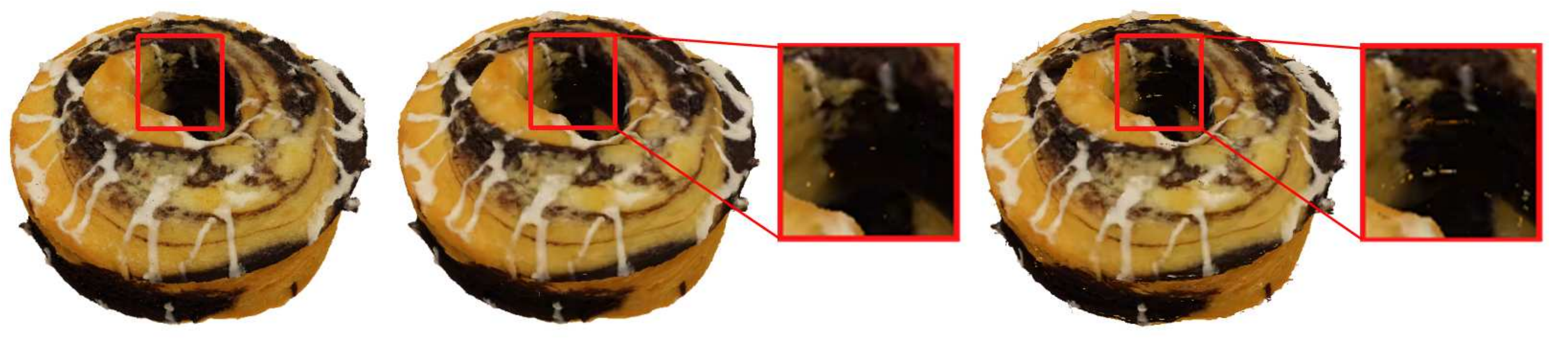}}
\subfigure[]{ \label{figsubjective:subfig:e}
\includegraphics[width=8.65cm, height=2.5cm]{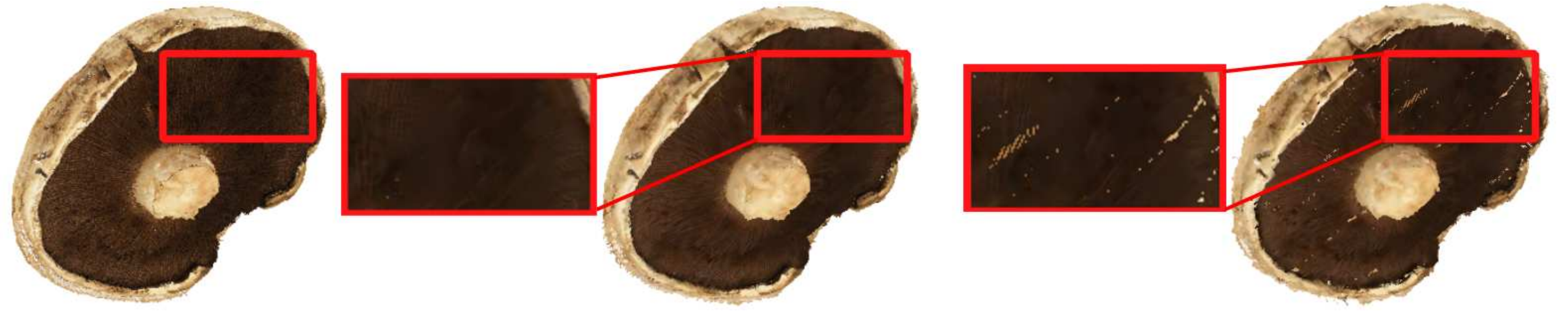}}\hspace{4mm}
\subfigure[]{ \label{figsubjective:subfig:f}
\includegraphics[width=8.65cm, height=2.5cm]{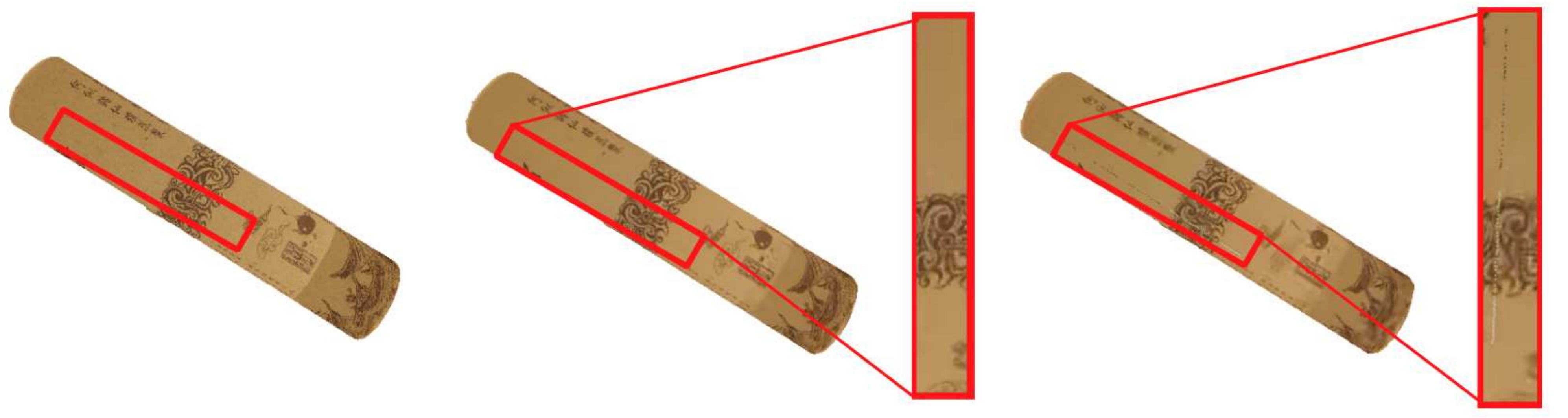}}
\subfigure[]{ \label{figsubjective:subfig:g}
\includegraphics[width=8.65cm, height=2.5cm]{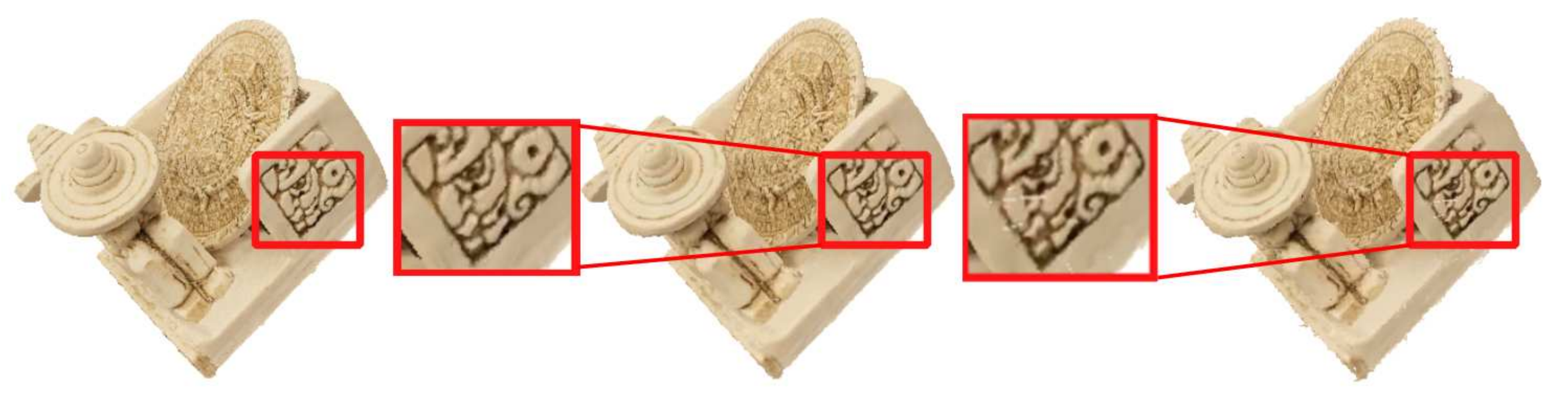}}\hspace{4mm}
\subfigure[]{ \label{figsubjective:subfig:h}
\includegraphics[width=8.65cm, height=2.5cm]{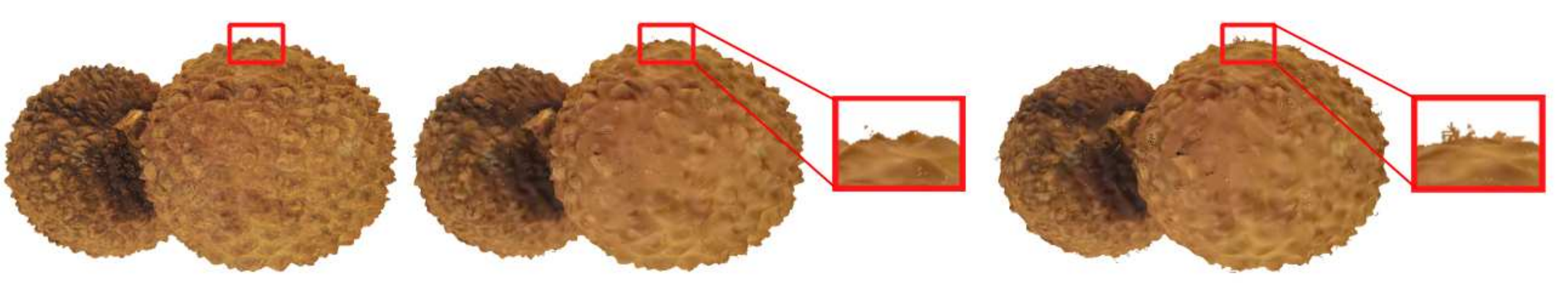}}
\caption{Perceptual quality comparison between our rate control
algorithm and $P2P_{ES}$. Left: original, Centre: proposed, Right:
$P2P_{ES}$. (a)subjective quality of \emph{Bag} with a target
bitrate of 510 kbpmp, (b)subjective quality of \emph{Banana} with a
target bitrate of 85 kbpmp, (c)subjective quality of
\emph{Flowerpot} with a target bitrate of 405 kbpmp, (d)subjective
quality of \emph{Cake} with a target bitrate of 170 kbpmp, (e)
subjective quality of \emph{Mushroom} with a target bitrate of 275
kbpmp, (f)subjective quality of \emph{Puer\_tea} with a target
bitrate of 190 kbpmp, (g)subjective quality of \emph{Statue} with a
target bitrate of 165 kbpmp, (h)subjective quality of \emph{Litchi}
with a target bitrate of 110 kbpmp.} \label{figsubjective}
\end{figure*}
The developed perceptual quality model would be of great benefit to
applications involving coding and rate control in 3DPC broadcasting
systems. In this paper, we solve the rate control problem for a
static 3DPC. Our method can also be extended to dynamic 3DPCs as
they can be seen as a sequence of successive static 3DPCs. For a
given target bitrate, we aim to find the combination of the geometry
$QP$ (corresponding to $Q_{g}$) and color $QP$ (corresponding to
$Q_{c}$) that provide the best perceptual quality. We formulate this
rate control problem as a constrained optimization problem where the
objective function is the derived perceptual quality model,
\begin{equation}
\label{eq:rdtotal2}
\begin{aligned}
\min_{(\emph{$Q_{g},Q_{c}$})} & MOS^c (Q_{g},Q_{c}) \\
\mathrm{s.t.} \quad & R_g(Q_{g}) + R_c(Q_{g},Q_{c}) \leq
\emph{$R_{T}$},
\end{aligned}
\end{equation}
where $R_g$ and $R_c$ are the geometry and color bitrate,
respectively, and  $R_T$ is the overall target bitrate. Based
on~\eqref{eq:mos100sim},~\eqref{eq:rdtotal2} and the Cauchy-based
rate model~\cite{liu2020rate}, the rate control problem can be
rewritten as
\begin{equation}
\label{eq:rdtotal}
\begin{aligned}
\min_{(\emph{$Q_{g},Q_{c}$})} & p_1 Q_g + p_2 Q_c +p_3 \\
\mathrm{s.t.} \quad & \gamma_g  Q_g^{\theta_g}+\gamma_c
Q_c^{\theta_c} \leq \emph{$R_{T}$},
\end{aligned}
\end{equation}
where $p_1$, $p_2$, and $p_3$ are the parameters of the perceptual
quality model, and $\gamma_g$, $\theta_g$, $\gamma_c$, $\theta_c$
are the parameters of the geometry and color rate models.

Together with the proposed model parameter estimation method in
Section~\ref{sec:4}, the proposed model can be embedded in the V-PCC
(TMC2) encoder to determine the optimal $Q_g$ and $Q_c$. First, the
CFGD and CBMV features of the input 3DPC are extracted, as described
in Section~\ref{sec:4}. Then by using the pre-trained matrix
$\bf{H}$ in~\eqref{predH}, the parameter vector $\bf{\hat{P}_m}$ can
be calculated. For the rate model, the parameters $\gamma_g$,
$\theta_g$, $\gamma_c$, and $\theta_c$ can be obtained by precoding
with two geometry and color quantization step pairs. Finally, with
the target bitrate $R_{T}$, the optimal $Q_{g,opt}$ and $Q_{c,opt}$
can be obtained by solving~\eqref{eq:rdtotal} using an interior
point method or another convex optimization
method~\cite{boyd2004convex}.

To assess the proposed perceptual quality model-based rate control
algorithm, we compared its performance to that of point-to-point
based exhaustive search algorithm (denoted by $P2P_{ES}$). For
$P2P_{ES}$, a 3DPC was first encoded by all the tested geometry and
color $QP$ pairs ranging from 26 to 50. Then the subset of
admissible pairs (pairs whose bitrates are smaller than or equal to
the target bitrate) was determined. Finally, the pair that gave the
highest PSNR for the Y component ($PSNR_Y$) was selected from this
subset. We focused on the Y component because it plays an important
role in visualization and in our perception of objective structure
and surface shape~\cite{palmer1999vision}. Since the texture
complexity of the tested 3DPCs are different, we set different
target bitrates for each 3DPC, as shown in
Table~\ref{tab:targetbitlist}.
\begin{table}[t!]
\centering \caption{Target bitrate in kilobits per million points
(\emph{kbpmp}) for each point cloud in the test set}
\label{tab:targetbitlist}
  \begin{tabular}{ccccc}
      \toprule
      \midrule
      Point Cloud  &$R_{T,1}$  &$R_{T,2}$  &$R_{T,3}$    &$R_{T,4}$  \\\hline
      \emph{Bag}          &170   &510   &1495   &2130     \\
      \emph{Banana}       &40  &120    &310   &850     \\
      \emph{Cake}         &110  &170   &265   &460     \\
      \emph{Flowerpot}    &75   &135    &265   &405     \\
      \emph{Litchi}       &110   &250   &565   &1200     \\
      \emph{Mushroom}     &50   &150   &220  &375    \\
      \emph{Puer\_tea}     &75   &190   &640   &1525     \\
      \emph{Statue}       &55   &105  &155   &200     \\\hline
      \bottomrule
  \end{tabular}
\end{table}
The rate-MOS curves of the proposed algorithm and $P2P_{ES}$ are
compared in Fig.~\ref{fig13}. The results demonstrate that the
proposed rate control algorithm can achieve better rate-MOS
performance than $P2P_{ES}$ with much lower complexity. Since the
value of $PSNR_Y$ in $P2P_{ES}$ is not consistent with the $MOS$,
the $MOS$s of the reconstructed 3DPCs by the $P2P_{ES}$ fluctuate
with different target bitrates. The proposed algorithm used the
proposed RR model to better predict the $MOS$s, and better
subjective quality can be achieved with given target bitrates.
Finally, Fig.~\ref{figsubjective} compares the subjective quality
between the proposed rate control algorithm and $P2P_{ES}$. We can
see that a significant subjective quality improvement can be
achieved by the proposed RR model-based rate control algorithm.

\section{Conclusion}\label{sec:6}
We proposed an RR linear quality model that accurately predicts the
perceptual quality of V-PCC compressed 3DPCs from the V-PCC geometry
and color quantization parameters. The three coefficients of our
linear model are estimated using a training set of reference 3DPCs
and two features (CFGD and CBMV) that are computed from the test
reference 3DPC. Because the number of high quality original 3DPCs
used by the MPEG PCC group is rather limited, we selected high
quality 3DPCs from the WPC dataset to conduct the subjective
experiments for static 3DPCs. The results show that the PLCC and the
SRCC between the predicted MOSs and the actual MOSs are both as high
as 0.91, indicating high accuracy of the proposed model.

Moreover, to illustrate the applications of the proposed model, we
also proposed an optimized rate control algorithm for 3DPC
compression. Benefitting from the accuracy of the proposed RR
quality model, the subjective quality of the proposed algorithm is
much better than that of $P2P_{ES}$.

In future work,  we will assess the performance of the proposed
model on the high quality 3DPCs recently provided by the MPEG PCC
group. We will also apply the proposed quality metric to
rate-distortion optimized coding and quality enhancement for 3DPCs.

\end{document}